\documentclass[%
 prd,
superscriptaddress,
twocolumn,
nofootinbib,
 amsmath,amssymb,
 aps,
]{revtex4-2}

\usepackage{float}
\usepackage{graphicx}
\usepackage{dcolumn}
\usepackage{bm}
\usepackage{mathrsfs}
\usepackage{frcursive}
\usepackage{mathtools}
\usepackage{commath}
\usepackage{color}
\usepackage[dvipsnames]{xcolor}
\usepackage[normalem]{ulem}
\newcolumntype{.}{D{.}{.}{13}}
\newcolumntype{d}[1]{D{.}{.}{#1}}
\usepackage[pdftex,breaklinks,colorlinks,
linkcolor=Blue,
citecolor=teal,
anchorcolor=red,
urlcolor=cyan,
pdfencoding=auto]{hyperref}
\usepackage{comment}
\usepackage{booktabs}
\usepackage{orcidlink}

\usepackage{array}
\newcommand{\deltaC}{\delta}
\newcommand{\deltaA}{\Delta}
\newcommand{\deltaE}{\Delta}

\def\MTOT{{M}}
\def\ETOT{E_\text{tot}}
\def\MATHRINGMTOT{\mathring{M}}
\def\MATHRINGm1{\mathring m_1}

\begin{document}

\title{Hybrid waveform model for asymmetric
spinning binaries\vspace{0.2cm} \\ 
\normalsize{
Self-force meets post-Newtonian theory}
}

\author{Loïc Honet\,\orcidlink{0009-0007-2863-6085}}
\email{loic.honet@ulb.be}
\affiliation{%
 Universit\'e Libre de Bruxelles, BLU-ULB Brussels Laboratory of the Universe, C.P. 231, B-1050 Bruxelles, Belgium
}%
\author{Adam Pound\,\orcidlink{0000-0001-9446-0638}}
 \email{A.Pound@soton.ac.uk}
\affiliation{School of Mathematical Sciences and STAG Research Centre, University of Southampton, Southampton, United Kingdom, SO17 1BJ}
\author{Geoffrey Comp\`ere\,\orcidlink{0000-0002-1977-3295}}%
 \email{geoffrey.compere@ulb.be}
\affiliation{%
 Universit\'e Libre de Bruxelles, BLU-ULB Brussels Laboratory of the Universe, C.P. 231, B-1050 Bruxelles, Belgium
}%

\date{\today}

\begin{abstract}
We develop and implement a new hybrid waveform model for quasicircular inspirals with a spinning primary and nonspinning secondary, excluding the merger and ringdown. This model, which is a core component of the more extensive \texttt{WaSABI-C} model,  consistently assembles all available first-order self-force and post-Newtonian results through a hybridization procedure without any tuning to numerical relativity, making it particularly suited for intermediate to extreme mass ratios. For almost all masses and primary spins, the resulting hybrid model significantly improves the faithfulness of both post-Newtonian and adiabatic self-force waveforms considered separately. We provide detailed comparisons with 50 simulations from the SXS catalog with mass ratios ranging from 1 to 15 and primary spins ranging from $-0.8$ to $0.8$. The hybrid model improves the median mismatch against numerical relativity waveforms by a factor of 2000 with respect to adiabatic waveforms and 40 with respect to post-Newtonian waveforms.  The mismatches are comparable to those obtained from the \texttt{SEOBNRv5EHM} model in the quasicircular limit over most of the parameter space covered by NR simulations.
\end{abstract}
                          
\maketitle

\tableofcontents

\allowdisplaybreaks

\section{\label{sec:level1}Introduction}

\subsection{Gravitational wave astronomy with asymmetric-mass systems}

The era of gravitational wave (GW) astronomy is now well underway: since the first direct detection of a GW signal~\cite{LIGOScientific:2016aoc}, the LVK (LIGO-Virgo-KAGRA) collaboration has detected more than $200$ binary coalescence candidate events~\cite{LIGOScientific:2025slb}. Many more events are yet to come at the end of the current fourth observing run in October 2025 as well as during the O5 run planned for 2027~\cite{GraceDB}. In order to analyze the signals emitted by these coalescing binaries without introducing significant biases in parameter estimation as well as to perform stringent tests of the theory of General Relativity (GR), a robust understanding of the two-body problem within GR is required~\cite{Toubiana:2024car,PhysRevD.57.4566,PhysRevD.78.124020,PhysRevD.82.024014,PhysRevD.95.104004,PhysRevResearch.2.023151}. Many modeling techniques have been developed to achieve this over the last few decades. Among them, post-Newtonian (PN)~\cite{Blanchet:1985sp,Blanchet:1992br,Blanchet:1995ez,Jaranowski:1997ky,
Goldberger:2004jt, 
Futamase:2007zz,Blanchet:2013haa,Foffa:2013qca,Damour:2014jta,Jaranowski:2015lha,Bernard:2015njp,Bernard:2016wrg,Damour:2016abl,Porto:2016pyg,Marchand:2017pir,Schafer:2018jfw,Levi:2018nxp,Foffa:2019yfl,Foffa:2021pkg,Blanchet:2023sbv,Blanchet:2023bwj} and 
post-Minkowskian (PM) theories~\cite{1979grt..conf..179.,Westpfahl:1980mk,1981GReGr..13..963B,1985ForPh..33..417W,Ledvinka:2008tk,Damour:2016gwp,Damour:2017zjx,Bini:2017xzy,Vines:2017hyw,Bjerrum-Bohr:2019kec,Kalin:2020mvi,Damour:2019lcq,Mogull:2020sak,
Buonanno:2022pgc,Travaglini:2022uwo,Bjerrum-Bohr:2022blt,Kosower:2022yvp},
self-force (SF) theory~\cite{Mino:1996nk,Poisson:2011nh,Pound:2012nt,Barack:2018yvs,Miller:2020bft,Pound:2021qin,Katz:2021yft,Wardell:2021fyy,Warburton:2021kwk,Miller:2023ers,Spiers:2023mor,Cunningham:2024dog,Mathews:2025nyb}, and  
Numerical Relativity (NR)  \cite{Shibata:1999wm,Pretorius:2004jg,Campanelli:2005dd,Baker:2005vv,Aylott:2009tn,Ajith:2012az,Hinder:2013oqa,Jani:2016wkt,Ferguson:2023vta,Healy:2017psd,Healy:2019jyf,Healy:2020vre,Healy:2022wdn,Huerta:2019oxn,Hamilton:2023qkv,Rashti:2024yoc} can be classified as \textit{first-principles} methods, as they directly solve the Einstein field equations (EFEs). In synergy with these first-principles methods, various types of fast inspiral-merger-ringdown models have been developed for data analysis: effective-one-body (EOB) models~\cite{Buonanno:1998gg,Buonanno:2000ef,Damour:2000we,Damour:2001tu,Buonanno:2005xu,Buonanno:2006ui}, which led to the families of \texttt{SEOBNR} \cite{Ramos-Buades:2021adz,Pompili:2023tna,Ramos-Buades:2023ehm,vandeMeent:2023ols,Khalil:2023kep,Leather:2025nhu} and \texttt{TEOBResumS} \cite{Damour:2014sva,Nagar:2015xqa,Nagar:2018zoe,Nagar:2019wds,Nagar:2020pcj,Riemenschneider:2021ppj,Gamba:2021ydi,Nagar:2023zxh} models, NR surrogates~\cite{Blackman:2015pia,Blackman:2017dfb,Blackman:2017pcm,Varma:2018mmi,Varma:2019csw,Williams:2019vub,Islam:2021mha,Rifat:2019ltp,Islam:2022laz}, and phenomenological models~\cite{Pan:2007nw,Ajith:2007qp,Ajith:2009bn,Santamaria:2010yb,Hannam:2013oca,Husa:2015iqa,Khan:2015jqa,London:2017bcn,Khan:2018fmp,Khan:2019kot,Estelles:2020twz}, all of which are built from the outputs of first-principles methods.

The frequency band of current ground-based detectors allows the detection of events that occur within the range $10$Hz to $10^4$Hz. Most signals within this band are emitted by the coalescence of roughly-comparable-mass binaries whose total mass reaches a few dozen solar masses~\cite{GraceDB,LIGOScientific:2021djp,LIGOScientific:2025slb}.  However, the recent detection of a $200$-solar-mass system \cite{LIGOScientific:2025rsn} points to the increasing relevance of intermediate mass black holes, the existence of which also implies greater likelihoods of events with high mass ratios $\mathring{q}$.\footnote{We use rings, as in $\mathring{q}$, to denote the initial values of evolving parameters.}  Notably, parameter estimation for this event also suffered from very large systematic biases between the five waveform models involved in the study~\cite{PhysRevResearch.1.033015,Ramos-Buades:2023ehm,PhysRevD.105.084040,PhysRevD.111.104019,PhysRevD.109.063012}. Moreover, events with mass ratios as high as $\mathring{q}\approx 9$~\cite{LIGOScientific:2020zkf} and even $\mathring q\approx 27$ (GW191219\_163120)~\cite{LIGOScientific:2021djp} have already been observed, with the latter falling outside the range of validated GW models~\cite{LIGOScientific:2021djp}. 
Such asymmetric compact binary coalescences are nowadays considered as important astrophysical sources~\cite{Bellovary:2025ris}, motivating one of LVK's observational science objectives: providing fast and accurate waveform models for asymmetric-mass-ratio binaries~\cite{LVKObsWhitePaper25} (see the task Obs-2.7.-A(i)).

Binaries with even higher $\mathring{q}$ will be even more essential in future GW astronomy. In January 2024 the European Space Agency (ESA) adopted  the LISA mission, which will mark the beginning of space-based GW interferometry~\cite{LISA:2024hlh}. LISA will probe the millihertz GW spectrum, unraveling the signals emitted by extreme-mass-ratio inspirals (EMRIs). EMRIs are binaries comprising a stellar-mass compact object of mass $m_2$ orbiting around a massive black hole of mass $\mathring m_1$. The small mass ratio $\mathring\varepsilon=1/\mathring q=m_2/{\mathring{m}_1}$ of these systems ranges in the interval $10^{-4}-10^{-6}$, which is many orders of magnitude more asymmetric than any signals detectable by the LVK observatories~\cite{GraceDB}. Due to its extreme mass ratio, an EMRI will stay months to years accumulating signal-to-noise ratio (SNR), undergoing tens or hundreds of thousands of cycles before merger or leaving the LISA band~\cite{PhysRevD.62.124021}. 

Such very long in-band signals are intrinsic to EMRIs, as the radiation-reaction timescale, over which the system loses energy through the emission of GWs, grows like the inverse of the small mass ratio $\mathring \varepsilon$. Therefore, the binary's three independent orbital frequencies $\Omega_i$~\cite{Pound:2021qin}, which evolve in the inspiral due to the emission of GWs, change at the slow rate $d\Omega_i/dt\sim \mathring\varepsilon$. Moreover, as $\Omega_i\sim 1/{\mathring{m}_1}$, the binary will spend its time in the strong-field regime while visible within the LISA band. This abundance of strong-field cycles will provide unique probes of the strong-field regime of GR~\cite{Barack:2006pq,Gair_2013,Barsanti_2022} and of the astrophysical environment in galactic cores~\cite{Amaro_Seoane_2023}. 

Intermediate-mass-ratio inspirals (IMRIs), with mass ratios $10^{-4}\leq\mathring \varepsilon\leq10^{-2}$, represent another potentially important class of high-$\mathring{q}$ of sources, for both LISA~\cite{Amaro_Seoane_2023,LISA:2024hlh} and third-generation ground-based detectors such as the Einstein Telescope (ET)~\cite{Abac:2025saz}. These systems have many of the same intrinsic characteristics as EMRIs, but they can additionally be detectable from their very early inspiral stage all the way up to merger~\cite{Chapman-Bird:2025xtd}, evolving through multiple regimes in a single detector band or across multiple bands~\cite{LISA:2024hlh,Abac:2025saz}. Furthermore, even massive black hole mergers, the loudest sources for LISA, are expected to have a wide range of mass ratios, ranging from equal mass to values of $\mathring q$ in the hundreds~\cite{Katz:2019qlu}.

All of these asymmetric binaries currently lack models with sufficient accuracy and extensiveness for high-precision GW science, with IMRIs and EMRIs particularly lying far beyond the range of validity of current models for second-generation ground-based detectors. This leaves GW modelers with a very challenging task: providing accurate, efficient, and complete waveform models across the whole range $10\lesssim\mathring q\lesssim 10^6$.

\subsection{This paper: modeling high-$\mathring{q}$ systems}

There has been substantial recent progress toward more faithful waveform models in much of the binary parameter space, but all these models have limitations in the high-$\mathring q$ regime~\cite{LISAConsortiumWaveformWorkingGroup:2023arg}. In PN theory, waveforms have been pushed to 4.5PN beyond leading order~\cite{Damour:2014jta,Blanchet:2023bwj,Trestini:2025nzr}; however, PN rapidly loses accuracy at high $\mathring q$ because the number of orbital cycles in the strong-field regime scales linearly with $\mathring q$. Links between scattering binaries and gravitationally bound systems~\cite{Damour:2016gwp,Damour:2017zjx, Bini:2019nra, Kalin:2019rwq, Kalin:2019inp, Cho:2021arx, Saketh:2021sri, Adamo:2024oxy} have also allowed PM scattering calculations to inform models of inspirals~\cite{Nagar:2024dzj, Buonanno:2024byg, Albanesi:2025txj,Dlapa:2025biy,Akpinar:2025huz}, but application of these ideas to asymmetric systems is still in a germinal stage~\cite{Gonzo:2023goe,Barack:2023oqp,Cheung:2024byb,Kosmopoulos:2023bwc,Long:2024ltn, Gonzo:2024xjk}. The SXS collaboration's catalog of NR waveforms now contains 4170
simulations, including 164 with mass ratios $\mathring q>8$~\cite{SXS:catalog,sxspython_2024,scheel2025sxscollaborationscatalogbinary}, and work on the high-$\mathring q$ regime is ongoing~\cite{Fernando:2018mov,Lousto:2020tnb,Wittek:2024pis}; however, NR is still currently limited to mass ratios $\lesssim 20$, and it is not feasible for NR to explore the whole high-$\mathring{q}$ parameter space (and effectively impossible to model EMRIs with NR) due to the quadratic scaling of NR runtime with $\mathring q$~\cite{Dhesi:2021yje}. 

In principle, the challenges of high-$\mathring q$ modeling are met by SF theory, in which the small, secondary object is treated as a source of perturbations on the spacetime of the larger, primary black hole, and the spacetime metric is consequently expanded in powers of the small mass ratio $\mathring\varepsilon$. This approach has reached recent milestones in both accuracy and efficiency. By combining a multiscale formulation of the Einstein field equations~\cite{Hinderer:2008dm,Miller:2020bft,Hughes:2021exa,Pound:2021qin,Mathews:2025nyb,Wei:2025lva} with GPU acceleration, the FastEMRIWaveforms (FEW) package~\cite{Chua:2020stf,Katz:2021yft,Speri:2023jte,Chapman-Bird:2025xtd} can generate long, LISA-length waveforms in tens of milliseconds. At the same time, the most advanced SF models have proved highly accurate for all mass ratios $\mathring q\gtrsim 10$~\cite{Wardell:2021fyy,Albertini:2022rfe,PaperII}. However, current SF models remain severely limited in their coverage of the binary parameter space, particularly for spinning and precessing systems.

It is generally accepted that, in order to meet LISA requirements, it is necessary and sufficient to go to second order in the SF expansion of the EFEs~\cite{LISAConsortiumWaveformWorkingGroup:2023arg,Arun_2022,Amaro_Seoane_2023}. This is motivated by the fact that the phase of the GW signal admits an expansion of the form~\cite{Hinderer:2008dm,Pound:2021qin} 
\begin{equation}\label{eq:accphase}
    \varphi(t,\mathring \varepsilon)=\frac{1}{\mathring\varepsilon} \varphi_{(0)}(\mathring\varepsilon t)+\varphi_{(1)}(\mathring\varepsilon t)+\mathcal{O}(\mathring\varepsilon),
\end{equation}
where the first term of the expansion is the adiabatic ($0$PA) phase and the second term is the first post-adiabatic ($1$PA) correction. The former depends on the dissipative piece of the first-order self-force (1SF), while the latter depends on the full first-order self-force as well as the dissipative piece of the second-order self-force (2SF)~\cite{Hinderer:2008dm,Miller:2020bft,Pound:2021qin}. 

Currently, the only available 1PA model is restricted to the case of nonspinning, quasicircular inspirals~\cite{Wardell:2021fyy}. 0PA models are available for generic binaries involving a spinning primary, but they are limited to weak fields and small eccentricities~\cite{Isoyama:2021jjd} or else to equatorial systems whose orbital angular momentum is aligned with the primary's spin axis~\cite{Chapman-Bird:2025xtd}. 0PA models also fall short of the necessary accuracy requirements for EMRIs. For IMRIs and other less extreme binaries, which will be observable when the two bodies are at much larger separations, even a 1PA model loses accuracy~\cite{Albertini:2022rfe}.

A recent Bayesian analysis~\cite{Burke:2023lno} confirmed that neglecting $1$PA corrections introduces significant biases on the parameter estimation for EMRIs and IMRIs. However, it also showed that these biases can be mitigated or entirely eliminated by approximating the $1$PA terms with PN data. This is the starting point of our work: \emph{to combine SF and PN results to construct a model that accurately covers the whole range of mass ratios $10\lesssim \mathring q\lesssim 10^6$ and particularly covers the spinning binaries for which there are no complete 1PA models}. 

More concretely, we seek to build a hybridized SF+PN model that achieves the following:
\begin{enumerate}
    \item To model EMRIs with sufficient accuracy for LISA, the model should be ``exact'' (accurate to 6 or more digits~\cite{Khalvati:2025znb}) in its 0PA information and should be as complete as possible in its 1PA information. Since 0PA effects~\cite{Hughes:2021exa,Fujita:2009us,Nasipak:pybhpt,TeukolskyPackage}, first-order conservative self-force effects~\cite{vandeMeent:2017bcc,Nasipak:2025tby}, and all linear-in-secondary-spin effects~\cite{Drummond:2023wqc,Grant:2024ivt, Witzany:2024ttz,Piovano:2024yks,Skoupy:2024uan,Mathews:2025nyb} can now be calculated in SF theory for generic orbital configurations around a spinning primary, completing a hybrid EMRI model for spinning binaries requires using a PN approximation to the missing second-order dissipative self-force effects.
    \item To be efficient enough for LISA data analysis and to dovetail with the prevailing EMRI modeling program, the model should take the multiscale form~\cite{Pound:2021qin,Mathews:2025nyb} that is compatible with the FEW rapid waveform-generation software package~\cite{Chua:2020stf,Katz:2021yft,Speri:2023jte,Chapman-Bird:2025xtd}. 
    \item To be sufficiently accurate for long signals that extend into the weak field, in the mass-asymmetric but non-EMRI regime $10^{-4}\lesssim\mathring\varepsilon \lesssim 0.1$~\cite{Chapman-Bird:2025xtd}, the model must contain terms beyond 1PA order~\cite{Albertini:2022rfe}. More generally, for the purpose of achieving high accuracy over the broadest possible range of signals, all available PN information should be included.
    \item Following the principle of parsimony, we also aspire to keep the model as conceptually simple as possible and built entirely from first principles, with no calibration to NR data.
\end{enumerate}

\begin{figure*}[!htb]
{\centering
\includegraphics[width=.99\textwidth]{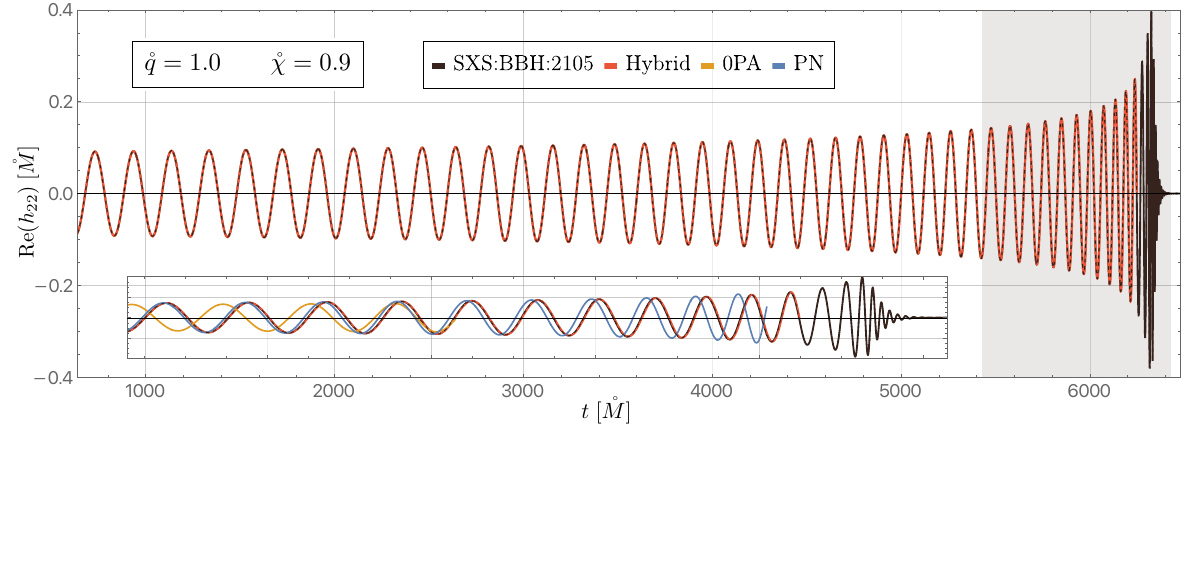}}\vspace{-2.3cm}
\caption{Self-force/post-Newtonian hybrid waveform (in red) and NR waveform SXS:BBH:2105 (in black)~\cite{Boyle:2019kee} for a quasicircular binary with primary spin ${\mathring\chi}=0.9$ and mass ratio $\mathring q=1$. The inset zooms in on the shaded gray region close to the merger. The hybrid model is described in the core of this article. We also display $0$PA and 4PN waveforms for comparison (in orange and blue, respectively), aligned with the NR waveform at the same (early) reference time as the hybrid waveform.
}\label{WF:2105}
\end{figure*}

In this paper, we develop a model achieving each of these objectives in the case of a nonspinning secondary object on a quasicircular orbit around a spinning primary black hole. Our model is restricted to the inspiral regime, but it could be extended to the merger-ringdown regime using the framework in Refs.~\cite{Kuchler:2024esj,Kuchler:2025hwx}. When applied for mass ratios $\mathring q\leq 15$, we find that our hybrid model matches NR inspiral waveforms far more accurately than either the 0PA or PN models taken individually. We find excellent numerical agreement with SXS simulations even at comparable mass ratios, as illustrated in Fig.~\ref{WF:2105} for an equal-mass, rapidly spinning binary. 

Our model is not unique in combining information from the weak-field (PN) and small-mass-ratio (SF) limits. EOB dynamics is designed to capture the test-mass limit by construction~\cite{Buonanno:1998gg}, and there has been a long history of incorporating SF information into EOB models~\cite{Damour:2009sm,Barack:2010ny,Akcay:2012ea,Bini:2013rfa,Bini:2014zxa,Bernuzzi:2014owa,Akcay:2015pjz,Bini:2015xua,Bini:2015bfb,Kavanagh:2017wot,Nagar:2018zoe,Nagar:2019wrt,Antonelli:2019fmq,vandeMeent:2023ols,Leather:2025nhu}. Models such as \texttt{ENIGMA} and \texttt{ESIGMA} have also incorporated both PN and SF results~\cite{Huerta:2017kez,Paul:2024ujx}. This paradigm of combining results from different limits and techniques is perhaps best exemplified by the ``tutti frutti'' approach~\cite{Bini:2019nra}. Moreover, EOB models specifically tailored to the small-mass-ratio limit have also been constructed~\cite{Damour:2007xr, Damour:2009sm, Yunes:2010zj, Taracchini:2014zpa, Bini:2019nra, Nagar:2022fep, Albanesi:2023bgi,Leather:2025nhu}.

We see our model as complementary to these other models. While they incorporate test-body and SF information, they are largely rooted in PN theory and generally utilize PN expansions of SF results. As emphasized in our enumerated list of objectives above, our approach prioritizes using exact SF information wherever available and seamlessly slotting into the mainstream EMRI modeling framework. 

Like the multiscale approach as a whole, our formulation (i) is modular, immediately improvable as PN and SF data advances, and (ii) will ultimately enable rapid generation of long  waveforms for generic, eccentric, precessing binaries through seamless integration with the FEW package. We hence expect that our approach will provide accurate, efficient models of IMRIs and serve as accurate stand-ins for EMRI models until complete 1PA results are available.

In the companion Letter~\cite{PaperIV}, we further extend our model to include additional SF information, and we provide a more thorough accuracy benchmarking against both NR and other models.

\subsection{Outline}

The plan of the paper is as follows. In Sec.~\ref{sec:waveforms}, we provide a brief overview of the conventions generally in use in NR, PN, and SF communities. We provide a common dictionary between them and  settle on a common set of parameters. In Sec.~\ref{sec:flux-balance}, we derive, for a nonspinning secondary orbiting a spinning primary, the explicit form of the forcing functions of the phase space parameters written in terms of energy and angular momentum fluxes as well as the binding energy of the system. In Sec.~\ref{sec:hybridization}, we explain the hybridization procedure and the quantities that are being hybridized in our waveform model. In Sec.~\ref{sec:implementation}, we describe the numerical implementation of our hybrid model, as well as three other models we shall compare to: the 0PA and the PN model corresponding to the (state-of-the-art) truncation that we consider, as well as 0PA4PN, a variant of our model that utilizes less SF information. We make explicit how the multiscale expansion enables rapid waveform generation through the segregation between the offline step (computations that can be done prior to waveform generation) and the online step (computations that should be performed on the fly). In Sec.~\ref{sec:comparison}, we assess the quality of our hybrid waveform model by comparing it to NR simulations. We explore the dephasing of the waveform against the NR simulations as well as their mismatches (using the AdVirgo+ sensitivity curve). Finally, in Sec.~\ref{sec:outlook}, we wrap up the discussion on the comparisons with other waveform models and discuss perspectives.

\subsection{Notation and nomenclature}

We adopt geometric units with $G=c=1$. We then express each quantity's dimensions using powers of the binary's total mass $\mathring M$ at reference time $\mathring t$, as in the label ``$t\; [\mathring M]$'' in Fig.~\ref{WF:2105}. Dimensionless quantities are indicated with ``[/]''.

Following common (but opaque) practice, we use the following terminology:
\begin{itemize}
    \item an $n$PN term in a quantity $f$ is a term of \emph{relative} order $x^{n/2}$ in the binary dynamics, where $1/x$ is an invariant measure of the orbital separation (defined below). These orders are always counted relative to the leading power of $x$ in the quantity~$f$, with the exception of the fluxes at the horizon, which are counted relative to the leading power of $x$ of the corresponding flux at infinity.
    \item an $n$SF term in a quantity $f$ is a term calculated from the $n$th-order metric perturbation $h^{(n)}_{\alpha\beta}$ in an expansion of the metric of the form $g_{\alpha\beta} + \varepsilon h^{(1)}_{\alpha\beta} + \varepsilon^2 h^{(2)}_{\alpha\beta} + \ldots$ This is equivalent to an \emph{absolute} order counting in powers of $\varepsilon$.
    \item an $n$PA term in a quantity $f$ is a term of \emph{relative} order $\varepsilon^n$ in the binary dynamics. This counting is analogous to $n$PN counting.
\end{itemize}

Finally, we refer to slowly evolving binary variables, such as masses and frequencies, as parameters, which make up the parameter space. We refer to the full set of evolving binary variables, including those that evolve on the orbital time scale, as ``phase space coordinates'', which make up the binary's phase space (restricted by phase-space reductions such as quasicircularity, spin alignment, etc.). 

\section{Setup and conventions}\label{sec:waveforms}

We consider the quasicircular compact binary inspiral of a secondary body of small mass $m_2$ and a primary body of larger mass $m_1(t)$. Here we allow the primary's mass to evolve with observation time $t$, assuming a time foliation that extends to future null infinity and penetrates the primary's event horizon. In principle, the secondary's mass also evolves due to GW absorption, but this effect is strongly suppressed in both PN and PA counting~\cite{Poisson:2004cw}; for completeness, we account for it in the companion letter~\cite{PaperIV}, but for simplicity, we neglect it in this paper. In general, the primary mass's evolution is also a small effect~\cite{Hughes:2018qxz,Warburton:2024xnr,PaperII}, but we account for it because it becomes increasingly important close to merger, especially for prograde, high-spin systems. 

We define the total mass as $\MTOT(t)=m_2+m_1(t)$, the symmetric mass ratio as $\nu(t)=m_1(t)m_2/(m_1(t)+m_2)^2$ and the primary's dimensionless spin as $\chi(t)=J_1(t)/m_1^2(t)$. In the companion letter~\cite{PaperIV} we include the spin of the secondary, but we neglect it here. We can invert the mass relationships to express the primary and secondary masses in terms of the symmetric mass ratio and total mass as  
\begin{align}\label{Mmp}
m_1 &= \MTOT\frac{1+\sqrt{1-4\nu}}{2}, \\    
m_2 &= \MTOT\frac{1-\sqrt{1-4\nu}}{2}.
\end{align}
The observable waveform at sky position $(R,\Theta,\Phi)$ and observation time $t$ is decomposed in spin-weighted spherical harmonics in the standard fashion as
\begin{align}
 R(  h_+-i h_\times) = \sum_{\ell =0}^\infty \sum_{m=-\ell}^\ell  \, h_{\ell m}(t)\,  \mbox{}_{-2}Y^{\ell m}(\Theta,\Phi).
\end{align}
We perform the following decomposition of each $(\ell,m)$ mode of the wavestrain:
\begin{equation}\label{eq:hu}
    h_{\ell m}(t)=\MTOT(t) \hat h_{\ell m}(t)e^{-i m \psi_{\ell m}(t)},
\end{equation}
where, by convention, the evolving total mass of the binary $\MTOT$ has been factored out, and the dimensionless mode amplitude $\hat h_{\ell m}\equiv |h_{\ell m}/\MTOT|$ is  real-valued. Hence, all variations of complex amplitudes have been promoted to variations of the waveform phase, and $\psi_{\ell m}(t)$ corresponds to the full GW phase. We define 
\begin{equation}
x \equiv  \omega^{2/3}
\end{equation}
from the halved, dimensionless GW frequency of the $(2,2)$ mode, 
\begin{equation}
\omega(t) \equiv \MTOT(t) d\psi_{22}/dt.
\end{equation}
Note our definition of $\hat h_{\ell m}$ differs from that of~\cite{Warburton:2024xnr}. Also note that, via Kepler's law, $x$ is a (gauge-invariant) proxy for the binary's orbital separation.

Let us now contrast the PN and SF waveforms. The GW modes in the PN expansion are written in terms of a single waveform phase, $\psi = \psi_{22}$, as opposed to one phase per mode:
\begin{align}\label{eq:h}
   \!\!\!\! h_{\ell m}^\text{PN}(t) &= \MTOT(\mathring t) \hat h_{\ell m}^{\text{PN}}(t)e^{-i m \psi(t)},
\end{align}
where $\hat h_{22}^{\text{PN}}$ is real-valued by convention, while other mode amplitudes $\hat h_{\ell m}^{\text{PN}}$ are complex. The waveform is normalized with respect to the total mass at an initial time~ $\mathring t$. The real-valued amplitude $\hat h_{22}$ at 4PN order is provided in Eq. (20) of~\cite{Warburton:2024xnr} following the earlier results of~\cite{Henry:2022ccf,Blanchet:2023bwj,Blanchet:2023sbv}. For our hybridization, we reexpand Eq.~\eqref{eq:h} at fixed dynamical total mass $\MTOT(t)$. Since the evolution of the mass is a $5$PN effect~\cite{Warburton:2024xnr} and the amplitudes are precise at $4$PN, this only amounts to replacing occurrences of $\MTOT(\mathring t)$ by $\MTOT(t)$.

For completeness, we note the halved GW phase of the $(2,2)$ mode, $\psi$, is also equal to the bare PN orbital phase $ \phi_\text{PN}$ up to a 1.5PN (absolute, or 4PN relative) tail correction given by 
\begin{align}\label{psi22}
\psi \equiv \phi_\text{PN} -2\Omega_\text{PN} {\MTOT(t)} \log \left(\frac{\Omega_\text{PN}}{\Omega_0}\right)+O(\text{5PN}),
\end{align}
where $\Omega_\text{PN}=\frac{d\phi_\text{PN}}{dt}$  is the bare (dimensionful) orbital frequency. Taking a time derivative, and using the leading 2.5PN result for the radiation-reaction of the orbital frequency, one finds the dimensionless quantity $x$ is related to the PN orbital parameter $y_{\text{PN}} \equiv [\MTOT(t)\Omega_\text{PN}(t)]^{2/3} $ as 
\begin{align}\label{xxorb}
y_{\text{PN}}=x\left\{ 1+ \frac{192}{5}\nu x^4 \left[ \log \left( \frac{x}{x_0} \right)+\frac{2}{3}\right] +O(\text{5PN})\right\}. 
\end{align}
with $x_0=(\MTOT \Omega_0)^{2/3}$.  In these expressions, $\Omega_0$ is a free parameter in the choice of time slicing between the binary and null infinity. To avoid this gauge ambiguity in linking near and far zones, we only work with the GW frequency parameter $x$ in what follows. 

In the SF expansion, the GW modes are instead expanded in terms of the secondary's orbital phase $\phi_p(t)$~\cite{Warburton:2024xnr}:
\begin{align}\label{SFh}
   \!\!\!\! h_{\ell m}(t) = \MATHRINGMTOT \tilde  h^\text{SF}_{\ell m}(t)  e^{-i m \phi_p(t)},
\end{align}
where we use a subscript $p$ because the secondary is treated as a point particle in the SF limit. Here all the dimensionless amplitudes are complex, and their SF expansion is 
\begin{equation}
    \tilde h^\text{SF}_{\ell m}=\mathring\nu \tilde h^{(1)}_{\ell m}+\mathring\nu^2 \tilde h^{(2)}_{\ell m}+O(\mathring\nu^3),
\end{equation}
with $\mathring\MTOT$ and $\mathring\nu$ denoting the total mass and symmetric mass ratio at time $\mathring t$. The secondary's  (dimensionless) orbital frequency is $\hat \Omega=\MATHRINGm1\Omega_{\text{SF}}=\MATHRINGm1  d\phi_p/dt$, where here $t$ on the secondary's trajectory is required to reduce, in the test-mass limit, to the Boyer-Lindquist time of the Kerr background of the primary. Finally, the SF orbital parameter $y$ is defined  as
\begin{align}\label{defy}
y_{\text{SF}}(t) \equiv  (M\Omega_{\rm SF})^{2/3} = \left[ \frac{\MTOT(t)}{\MATHRINGm1 }\hat \Omega(t) \right]^{2/3}. 
\end{align}
Note this differs from the parameter $\hat \Omega^{2/3}$, which is often denoted $y$ in the literature~\cite{Tiec:2014lba}. Also note that while $\Omega_{\rm SF}$ and $\Omega_{\rm PN}$ are generally treated as identical in the literature, we distinguish them because each develops a gauge-dependent link to the invariant, observable waveform frequency at 4PN and 1PA.

We stress that $\phi_p$ is not the GW phase, as a modulation of the imaginary part of $\tilde h^{ (n)}_{\ell m}$ will produce a modulation of the waveform phase. In order to make the phase explicit, we follow Ref.~\cite{Warburton:2024xnr} in bringing Eq.~\eqref{SFh} into the form of Eq.~\eqref{eq:hu}. 

We first decompose the complex amplitude into a phase and a real amplitude as $\tilde h_{\ell m}^\text{SF}=\hat h_{\ell m}^{\text{SF}}e^{-i m \deltaC \phi_{\ell m}}$ where $m\deltaC \phi_{\ell m}=- \text{arctan}( \text{Im} \tilde h_{\ell m}^\text{SF}/\text{Re} \tilde h_{\ell m}^\text{SF})$. The total waveform phase can then be related to the SF orbital phase as 
\begin{equation}\label{psi22SF}
\psi_{\ell m}=\phi_p+\deltaC \phi_{\ell m}. 
\end{equation}
Taking this equation minus the corresponding equation for the 22 mode, we can write the phases as 
\begin{align}\label{eq:deltapsilm}
   \psi_{\ell m} = \psi + \deltaC \phi_{\ell m}-\deltaC \phi_{22} \equiv \psi + \deltaC \psi_{\ell m}. 
\end{align}
in terms of phase shifts $\deltaC \psi_{\ell m}$ that are determined by the complex amplitudes. 

Next, we take a time derivative of the 22 mode phase~\eqref{psi22SF} to obtain
\begin{align}
\omega =\MTOT \Omega_{\text{SF}}
+\nu  F_{(0)}^{\hat \Omega}(\MTOT\Omega_{\text{SF}}
,\chi) \partial_{\hat \Omega} \delta \phi_{22}+\mathcal{O}(\nu^2),     \label{omega}
\end{align}
where $\frac{d\hat \Omega}{dt}=\frac{\mathring \nu}{\MATHRINGMTOT}F_{(0)}^{\hat \Omega}(\hat\Omega,\mathring\chi)+\mathcal{O}(\mathring\nu^2)$ is the rate of change of the secondary's orbital frequency. In Eq. \eqref{omega} we have already reexpressed the functions in terms of the evolving parameters. As reviewed in the next section,  $F_{(0)}^{\hat \Omega}$ can be expressed in terms of the fluxes of energy at infinity, $\mathcal{F}^\infty_{(0)}(\hat \Omega,\mathring \chi)$, and at the  primary's horizon, $\mathcal{F}^\mathcal{H}_{(0)}(\hat \Omega,\mathring \chi)$:  
\begin{equation}\label{FOmega0PA}
    F_{(0)}^{\hat \Omega}(\hat \Omega,\mathring \chi)=\frac{3}{\hat \Omega U_{(0)}^3 D
    }\left( \mathcal{F}^\infty_{(0)}+\mathcal{F}^\mathcal{H}_{(0)}\right).
\end{equation}
Here
\begin{align}
    U_{(0)}(\hat \Omega,\mathring \chi) &=1/\sqrt{(1-{\mathring\chi}\hat  \Omega) \left[1+{\mathring\chi}\hat \Omega-3 {\hat \Omega}^{\frac{2}{3}} ({1-{\mathring\chi} \hat \Omega})^{\frac{1}{3}}\right]}\label{U0}
\end{align}
is the leading-order (0PA) term in the (inverse) redshift factor $dt_p/d\tau$ along the particle's trajectory. The quantity
\begin{multline}
    D(\hat \Omega,\mathring \chi)\equiv -3{\mathring\chi}^2-6(\hat \Omega^{-1}-{\mathring\chi})^{2/3}\\
    +7{\mathring\chi}(\hat \Omega^{-1}-{\mathring\chi})^{1/3} +\hat \Omega^{-1}(\hat \Omega^{-1}-{\mathring\chi})^{1/3}\label{eqD}
\end{multline}
is the leading-order term in $\partial E_p/\partial\hat\Omega$, up to the factor $U^3_{(0)}\hat\Omega/3$ in Eq.~\eqref{FOmega0PA}, where $E_p$ is the particle's energy. 

Substituting Eq. \eqref{FOmega0PA} into \eqref{omega} and expanding to linear order in the symmetric mass ratio, we find the analogue of Eq.~\eqref{xxorb}:
\begin{equation}
    y_{\text{SF}}=x \left( 1+\nu \deltaC x \right),\label{ytox}
\end{equation}
where 
\begin{align}
  \deltaC x &=  \frac{2}{3}\frac{\mathcal{F}^{\infty\text{SF}}_{(0)}+\mathcal{F}^{\mathcal H\text{SF}}_{(0)}}{x^{7/2}U_{(0)}^3D|\tilde h_{22}^{(1)}|^2} \label{delta x SF}\\*
   &\quad\times \left[\text{Im}\left(\partial_x \tilde h_{22}^{(1)} \right)\text{Re}\left(\tilde h_{22}^{(1)}\right)-\text{Im}\left( \tilde h_{22}^{(1)} \right)\text{Re}\left(\partial_x\tilde h_{22}^{(1)}\right)\right]\nonumber
\end{align}
can be evaluated as a function of $x$. Here $\mathcal{F}^{\infty\text{SF}}_{(0)}(x,\chi)\equiv\mathcal{F}^{\infty}_{(0)}(x^{3/2},\chi)$ and $\mathcal{F}^{\mathcal H\text{SF}}_{(0)}(x,\chi)\equiv\mathcal{F}^{\mathcal H}_{(0)}(x^{3/2},\chi)$.  
Equations~\eqref{ytox} and \eqref{delta x SF} generalize Eqs.~(36)--(37) of~\cite{Warburton:2024xnr} to the case of a spinning, Kerr primary. They are important as many SF quantities are  computed on a grid of $\hat \Omega$ values, but should be consistently reexpanded onto a grid of $\omega$ (or $x$) values in order to be hybridised with their related PN quantities. We have checked numerically that $\deltaC x ={\cal O}(4\text{PN})$, in analogy with the PN relationship~\eqref{xxorb}.

In this paper we work at leading order in the SF amplitudes, where $\tilde h_{\ell m}=\mathring\nu \tilde h_{\ell m}^{(1)}(\hat \Omega , \mathring \chi)+\mathcal{O}(\mathring\nu^2)$. Here $ \mathring \chi$ is the background dimensionless primary spin $\mathring \chi=\mathring J_1/\MATHRINGm1^2$, which is time independent. After reexpansion, we have 
\begin{equation}    
\MATHRINGMTOT\tilde  h^\text{SF}_{\ell m}(t)=\nu \MTOT\tilde h_{\ell m}^{(1)\text{SF}}(x, \chi)+\mathcal{O}(\nu^2),
\end{equation}
where $\tilde h_{\ell m}^{(1)\text{SF}}(x, \chi)=\tilde h_{\ell m}^{(1)}(x^{3/2}, \chi)$. 

\section{Dynamics from flux-balance laws}\label{sec:flux-balance}

Assuming a quasicircular inspiral with a primary spin but no secondary spin, the dynamics of the system is entirely captured by the dynamical evolution of the GW frequency parameter $x=\omega^{2/3}$, the primary's dimensionless spin $\chi$, the symmetric mass ratio $\nu$, and the total mass $\MTOT$. 

As noted above, we neglect the time dependence of the secondary's mass; we return to the justification for that at the end of the section. Given constant $m_2$, the evolution equation for $\MTOT$ becomes redundant with the evolution equation for $\nu$, and one can deduce that
\begin{equation}\label{Mtotconstr}
    \MTOT=\MATHRINGMTOT\frac{1-\sqrt{1-4\mathring\nu}}{1-\sqrt{1-4\nu}},
\end{equation} 
where $\MATHRINGMTOT$ and $\mathring\nu$ are the initial total mass and symmetric mass ratio, respectively. The nontrivial evolution equations then necessarily take the form%
\begin{subequations}\label{FBeom}%
\begin{align}
\frac{d\psi}{dt}&=\omega/\MTOT, \\ 
    \frac{d\omega}{dt}&=F_\omega(\omega^{2/3}, \chi, \nu,\MTOT),\\
    \frac{d\chi}{dt}&=F_\chi(\omega^{2/3}, \chi, \nu,\MTOT),\\
    \frac{d\nu}{dt}&=F_\nu(\omega^{2/3}, \chi, \nu,\MTOT),
\end{align}
\end{subequations}
for some forcing functions $F_a(x,\chi,\nu,\MTOT)$, with $a=\{\omega,\chi,\nu\}$. We purposely use a basis $(x,\chi,\nu)$ of dimensionless parameters so that all dependence on $\MTOT$ appears as an overall factor in any dimensionful quantity. 

Using flux-balance laws, we can write the  forcing functions $F_a$ in terms of (i) the binding energy $E(x,\chi,\nu,\MTOT)$ and (ii) the fluxes of GW energy at null infinity, $\mathcal F^\infty = dE^{GW}/dt \vert_{\mathcal I}$, and at the horizon, $\mathcal F^{\mathcal H} = dE^{GW}/dt\vert_{\mathcal H}$. For this purpose we introduce the total flux 
\begin{equation}
\mathcal F = \mathcal F^\infty+\mathcal F^{\mathcal H},
\end{equation} 
and we define the binding energy in terms of the Bondi energy $\ETOT$ at future null infinity as 
\begin{equation}
E(x, \chi, \nu,\MTOT) = \ETOT(x, \chi, \nu,\MTOT)-\MTOT.
\end{equation}
Note that $E = \MTOT \hat E(x,\chi,\nu)$, where $\hat E$ is dimensionless. 

We first obtain the forcing functions for the masses and spin, $F_\nu$ and $F_\chi$. The primary's mass and spin evolve according to the energy and angular momentum flux through its horizon, 
\begin{align}
\frac{dm_1}{dt} &=\mathcal F^{\mathcal H},\label{dm1dt}\\
\frac{dJ_1}{dt} &=\mathcal G^{\mathcal H}.
\end{align}
For a quasicircular orbit, these are related by $\Omega\,\mathcal G^{\mathcal H} = \mathcal F^{\mathcal H}$~\cite{Teukolsky:1974yv,Poisson:2004cw}, which we rewrite as
\begin{equation}\label{G from F}
    \mathcal{G}^\mathcal{H} = \frac{\MTOT}{y^{3/2}}\mathcal{F}^\mathcal{H}
\end{equation}
using Eq.~\eqref{defy}. Explicitly computing the time derivative of $\chi=J_1/m_1^2$, we find
\begin{equation}\label{Fchisymb}
    F_\chi=\frac{4\left(\mathcal{G}^\mathcal{H}-\MTOT\mathcal{F}^\mathcal{H}\left(1+\Delta\right)\chi\right)}{\MTOT^2\left(1+\Delta\right)^2},
\end{equation}
and similarly for the time derivative of $\nu=m_1 m_2 /\MTOT^2$: 
\begin{equation}\label{Fnusymb}
    F_\nu=-\frac{\left(1-\Delta\right)\Delta}{2\MTOT}\mathcal{F}^\mathcal{H}.
\end{equation}
For compactness we have introduced the normalized difference between the two masses,
\begin{equation}
    \Delta \equiv \frac{m_1-m_2}{\MTOT} = \sqrt{1-4\nu}.
\end{equation}

Finally, we express the forcing function $F_\omega$ in terms of the binding energy and fluxes. Taking a time derivative of the total energy of the binary, using the Bondi mass-loss formula $\mathcal F^\infty
= - d\ETOT/dt$~\cite{Madler:2016xju}, applying the chain rule, and rearranging, we find
\begin{align}\label{eq:Etot1}
F_\omega &= -\frac{\mathcal{F} + \frac{\partial E}{\partial \chi}F_\chi + \frac{\partial E}{\partial \nu}F_\nu +\frac{\partial E}{\partial \MTOT}\mathcal{F}^\mathcal{H}}{\frac{\partial E}{\partial x}\frac{\partial x}{\partial \omega}}.
\end{align}
Substituting Eqs. \eqref{Fchisymb} and \eqref{Fnusymb} into Eq. \eqref{eq:Etot1} yields our final result:
\begin{widetext}
\begin{equation}\label{Fomegasymb}
        F_\omega=-\frac{\mathcal{F}+\left[\frac{\partial E}{\partial \MTOT}-\frac{4\chi}{\MTOT\left(1+\Delta\right)}\frac{\partial E}{\partial \chi}-\frac{\left(1-\Delta\right)}{2\MTOT}\Delta\frac{\partial E}{\partial\nu}\right]\mathcal{F}^\mathcal{H}+\frac{4}{\MTOT^2\left(1+\Delta\right)^2}\frac{\partial E}{\partial\chi}\mathcal{G}^\mathcal{H}}{\frac{\partial E}{\partial x}\frac{\partial x}{\partial \omega}}.
\end{equation}
\end{widetext}

Finally, we comment on the impact of including the evolution of $m_1$ but neglecting the evolution of $m_2$. The evolution of each mass affects the waveform phasing in two ways: an indirect contribution through the energy balance that governs $d\omega/dt=F_\omega$; and a direct contribution by evolving the arguments of the forcing function~$F_\omega$. 

In the case of the primary mass, the flux through the horizon has the same small-$\nu$ scaling as the flux to infinity, ${\cal F}^{\cal H}= {\cal O}(\nu^2)$. Hence, its \emph{indirect} effect enters at leading, 0PA order in the SF counting, through its contribution to $\cal F$ in Eq.~\eqref{Fomegasymb}. On the other hand, this scaling of ${\cal F}^{\cal H}$ shows that $m_1$ evolves by an amount $\delta m_1={\cal O}(\nu)$ over the radiation-reaction time scale. Consequently, it alters $F_\omega$ by an amount of order $\delta m_1\frac{\partial F_\omega}{\partial m_1}$, suppressing it by one order in $\nu$ relative to the leading, 0PA forcing term; hence, $m_1$'s evolution has a 1PA \emph{direct} effect on the phasing. 
In PN counting, the flux scales as ${\cal F}^{\cal H}={\cal O}(x^{15/2})$, a 2.5PN contribution relative to the leading quadrupole formula ${\cal F}^{\infty}={\cal O}(x^5)$; see Eqs.~\eqref{FH4PN} and~\eqref{F4.5PN}. Hence, the indirect contribution is 2.5PN. As in the SF counting, the direct contribution from the change in mass $\delta m_1 = \int \frac{{\cal F}^{\cal H}}{dx/dt} dx = {\cal O}(x^{7/2})$ is one order smaller, at 3.5PN, where we have used $dx/dt={\cal O}(x^5)$. If the primary were nonspinning, these PN scalings would be suppressed by 1.5PN order; again see Eq.~\eqref{FH4PN}.

The effect of the secondary's mass evolution is substantially smaller. If the secondary were spinning, its mass would evolve at a rate $dm_2/dt={\cal O}(\nu^5)$~\cite{Poisson:2004cw}. This would add an order-$\nu^5$ term, ${\cal F}^{{\cal H}_2}$, to the total flux ${\cal F}$, three orders smaller in $\nu$ than the leading, 0PA flux. Hence, it would represent a 3PA indirect contribution to the phasing. Unlike in the case of $m_1$, the direct effect of the change in $m_2$ is not suppressed by an additional order relative to the indirect effect; both effects enter at 3PA order. We can glean this by noting $m_2$ evolves by an amount $\delta m_2=\int {\cal F}^{{\cal H}_2}dt = {\cal O}(\nu^4)$ over the radiation-reaction time scale, altering $F_\omega$ by an amount of order $\delta m_2 \frac{\partial F_\omega}{\partial m_2}$. Since $F_\omega={\cal O}(\nu)$ and the partial derivative lowers the power of $\nu$ by one, we see $\delta m_2 \frac{\partial F_\omega}{\partial m_2}$ is of order $\nu^4$, which is a 3PA term in $d\omega/dt$. 

All of this is in the case of a spinning secondary. Since we consider a nonspinning secondary in this paper, the neglected effect is another order smaller in $\nu$~\cite{Poisson:2004cw}, making it 4PA. Its PN scalings are the same as for the primary, meaning in the nonspinning case it is additionally suppressed by 4PN orders relative to the leading-order flux to infinity. This combined SF and PN suppression makes it negligible for the model considered here.

\section{Hybridization}\label{sec:hybridization}

In this section we detail our hybridization procedure between PN and SF quantities. At its core, our method is a straightforward application of the standard method of combining two asymptotic expansions: for any function that is expanded in two ways, one builds a composite expansion by adding together the two expansions and subtracting doubly counted common terms. However, there is considerable flexibility in choosing \emph{which} functions are expanded and hybridized in this way and \emph{which} variables they are expanded in. 

Our choices are motivated by the following criteria:
\begin{enumerate}
    \item As far as possible, we formulate the waveform evolution in terms of gauge-invariant functions of gauge-invariant variables. This ensures variables and functions of them represent the same quantities in both expansions. While cross-validation and cross-pollination between PN and SF results is often carried out in terms of orbital frequencies to eliminate gauge ambiguity~\cite{Damour:2009sm,Tiec:2014lba,Barack:2018yvs}, even these quantities become gauge-dependent at 4PN and 2SF, and comparisons at that order require working in terms of invariant \emph{waveform} frequencies~\cite{Warburton:2024xnr}. This motivates us to adopt the waveform phase $\psi$ and frequency $\omega$ as our core phase-space variables.  
    \item We adopt variables $\nu$ and $M$ in favor of $m_1$ and $m_2$ as this choice enforces symmetries of the system under interchange of the two masses. It is also well known to improve the accuracy of SF expansions in the comparable-mass regime~\cite{LeTiec:2011ab,Le_Tiec_2012,Tiec:2014lba,Nagar:2013sga,Ramos-Buades:2022lgf,Warburton:2021kwk,Wardell:2021fyy}.
    \item We seek to preserve pole structure when it arises from an exact feature of the system. In particular, we avoid expanding the fraction in Eq.~\eqref{Fomegasymb} in either the PN or SF limits. By leaving the fraction intact, we preserve the feature that the denominator vanishes, and the forcing function $F_\omega$ blows up, at the innermost stable circular orbit (ISCO)   in the quasicircular limit. A PN expansion of the fraction dissolves this pole, while an SF expansion of it gives it exaggerated strength at each successive order~\cite{Kuchler:2024esj,Albertini:2022rfe}. As shown in Ref.~\cite{Burke:2023lno}, simple resummations of PN based on this pole structure can dramatically improve PN's accuracy; and as shown in the companion paper~\cite{PaperII}, avoiding an SF expansion of it provides a marked improvement of SF waveform models for comparable masses.  
\end{enumerate}
We return to the last point in Sec.~\ref{subsubsec:models} below. 

With these guiding principles, we define our hybrid model from the hybridization of the energy flux $\mathcal F^\infty$ at infinity, the energy and angular momentum fluxes  $\mathcal F^{\mathcal H}$ and  $\mathcal G^{\mathcal H}$ at the horizon, the binding energy $E$, the real amplitudes $\hat h_{\ell m}$, and the phase shifts $\deltaC\psi_{\ell m}$ with respect to the $(2,2)$ mode phase $\psi$. The forcing functions $F_a$, which govern the dynamical evolution of the waveform through Eq.~\eqref{FBeom}, are then given in terms of the hybridized quantities by the functionals \eqref{Fchisymb}, \eqref{Fnusymb}, and \eqref{Fomegasymb}, where $\mathcal F^\infty$, $\mathcal F^{\mathcal H}$, $\mathcal G^{\mathcal H}$, and $E$ are replaced with the corresponding hybrids $\mathcal F^\infty_H$, $\mathcal F^{\mathcal H}_H$, $\mathcal G^{\mathcal H}_{H}$, and $E_H$ defined below. The $(\ell,m)$ waveform modes are defined from \eqref{eq:hu} where the amplitudes $\hat h_{\ell m}$ and phases $\psi_{\ell m}$ defined from \eqref{eq:deltapsilm} are replaced with the corresponding hybrid quantities $\hat h^H_{\ell m}$ and $\psi+\deltaC \psi^H_{\ell m}$. While countless other constructions are possible, we have not found any simple alternatives that improve the accuracy of our model. 

Avoiding re-expansions of the forcing functions~\eqref{Fchisymb}, \eqref{Fnusymb}, and \eqref{Fomegasymb}, particularly our avoidance of expanding the fraction in Eq.~\eqref{Fomegasymb}, are the only resummations we employ. All our other operations are transformations on the binary's phase space (after which we always fully re-expand and truncate sums) and standard compositions of asymptotic series. Since Eqs.~\eqref{Fchisymb}, \eqref{Fnusymb}, and \eqref{Fomegasymb} are exact results in GR (in the quasicircular limit), this resummation is aimed at preserving as much of the exact dynamics as possible.

In the rest of this section, we derive the hybrid expressions for the fluxes, binding energy, and complex amplitudes.

\subsection{Overview}

Post-Newtonian expansions are typically power series in the PN parameter $y_{\rm PN}^{1/2}$ at fixed symmetric mass ratio $\mathring \nu$, total mass $\mathring M$, and spin $\mathring a_1=\MATHRINGm1{\mathring \chi}$; SF expansions are typically power series in the mass ratio $\mathring{\varepsilon}=m_2/\MATHRINGm1$ at fixed dimensionless orbital frequency $\hat\Omega=\MATHRINGm1\Omega$, primary mass $\mathring m_1$, and spin $\mathring a_1 = \MATHRINGm1{\mathring \chi}$. In order to write down an SF+PN hybrid quantity, one has to reexpand the PN and SF quantities in terms of the same set of parameters, which we choose to be $(x,\chi,\nu,M)$. 

We first align our normalizations between PN and SF expansions by setting $m_1(\mathring t)=\MATHRINGm1$, $J_1(\mathring t) = \mathring J_1$ at a reference time $\mathring t$, typically at the start of the considered inspiral phase. We then express both expansions in terms of the evolving parameters $(x,\chi,\nu,M)$. To see how this works, consider the SF expansion, in which 1PA terms (including 1SF conservative quantities such as the 1SF binding energy) depend on the (dimensionless) corrections $\deltaC \hat m_1 \equiv (m_1 - \mathring m_1)/m_2$ and $\deltaC \hat \chi_1 \equiv (\chi - \mathring\chi)/\mathring\varepsilon$~\cite{Miller:2020bft}, leading to expressions such as 
\begin{equation}
E = m_2\left[\hat E_{(0)}(\hat\Omega,\mathring\chi) + \mathring\varepsilon \hat E_{(1)}(\hat\Omega,\mathring \chi,\deltaC \hat m_1, \deltaC \hat\chi) +\mathcal{O}(\mathring\varepsilon^2)\right].       
\end{equation}
Prior to hybridization, we re-expand this in terms of the physical, evolving mass and spin parameters $m_1$ and $\chi$. However, one always finds, for gauge-invariant quantities, that the dependence on $\deltaC \hat m_1$ and $\deltaC \hat \chi$ is of the form $\deltaC \hat m_1\frac{\partial\hat E_{(0)}}{\partial\hat m_1}$ and $\deltaC \hat \chi\frac{\partial\hat E_{(0)}}{\partial {\hat \chi}}$ --- that is, equivalent to the effect of slightly perturbing the background parameters $\mathring m_1$ and $\mathring\chi$ in the 0SF functions. Hence, when we rewrite $E$ using $\mathring m_1 = m_1(1 - \varepsilon\deltaC \hat m_1)$ and $\chi = \mathring\chi -\varepsilon\deltaC \hat \chi$, the dependence on $\delta \hat m_1$ and $\delta\hat\chi$ immediately cancels. As a consequence, in the SF expansions of interest here, one can simply neglect $\delta\hat m_1$ and $\delta\chi$ and replace $\mathring m_1$ and $\mathring\chi$ with the evolving variables $m_1$ and $\chi$. It is then straightforward to re-express $(m_1\Omega,\chi,m_1,m_2)$ in terms of $(x,\chi,\nu,M)$ and re-expand in powers of $\nu$, as illustrated in the previous section. Analogous steps apply for PN expansions. For quantities such as the flux, which has no 0SF term, this procedure solely amounts to replacing $\mathring m_1\Omega$ by $x^{3/2}$, $\mathring m_1$ by $M$, and $\mathring\varepsilon$ by $\nu$.

Once functions are written in terms of the dynamical variables $(x,\chi,\nu,M)$ both in PN and SF expansions, we use the total mass $\MTOT$ to set the scale of dimensionful quantities. All dimensionful functions of $(x,\chi,\nu,M)$ can then be written as a power of $M$ multiplying a dimensionless function of $(x,\chi,\nu)$.

For a given dimensionless function $f(x,{\chi}, \nu)$, which depends upon the (time evolving) spin ${\chi}$ of the primary black hole and the (time evolving) symmetric mass ratio $\nu$, one can perform a PN expansion of the quantity $f$ as a double Taylor expansion, 
\begin{align}\label{fPN}
 f(x,{\chi},\nu)= x^{N/2} \sum_{l=0}^\infty \sum_{n=\underline{n}(l)}^\infty f^{\text{PN}}_{nl}( \nu , \log(x)) x^{n/2}{\chi}^l,
\end{align}
where $x^{N/2}$ is the leading PN behaviour of $f$ and $\underline{n}(l)$ is the minimal value of $n$ of the spin$^l$ sector of the quantity $f$. As in the PN literature, we shall refer to the $l=0,1,2,3$ terms as the nonspinning, spin-orbit (SO), spin-spin (SS) and spin-spin-spin (SSS) sectors, respectively. For all quantities considered, the non-spinning sector starts at 0PN order, $\underline{n}(0)=0$, except for the energy flux at the horizon where $\underline{n}(0)=3$ but $\underline{n}(1)=0$; see Eq. \eqref{FH4PN}.  The coefficients of the series expansion $f_{nl}^\text{PN}( \nu)$ depend on $ \nu$ and also possibly contain $\log(x)$ terms. The spin-induced quadrupole and other spin-induced multipoles of the primary body are included as part of the truncated PN expansion. The spin-induced multipoles of the primary black hole are also included as the 0PA expansion is performed around the Kerr black hole. We ignore the secondary spin and its higher multipole moments in both the SF and PN expansions.

Alternatively, one can as well perform an SF expansion of the same quantity $f$:
\begin{align}\label{fSF}
 f(x,{\chi}, \nu)=  \nu^{K} \sum_{k=0}^\infty \nu^k f^{\text{SF}}_{(k)}(x , \chi),
\end{align}
where $\nu^K$ is the leading SF behaviour of the quantity $f$. We emphasize that this SF expansion is in terms of the dynamically evolving parameters $\nu,\chi$, which requires a reexpansion of the standard SF quantities written in terms of $\mathring \nu,\mathring \chi$. We will denote with a superscript $\mbox{}^\text{SF}$ self-force expansions in terms of dynamically evolving parameters. 

Since both expansions are from first principles, the PN expansion of the SF series yields the same asymptotic expansion as the SF expansion of the PN series~\cite{Warburton:2024xnr}. This yields the asymptotic expansion of the function $f$ as a triple Taylor series,
\begin{align}\label{fSFPN}
 f(x,{\chi},\nu)=  \nu^{K}x^{N/2} \sum_{l=0}^\infty\sum_{k=0}^\infty \sum_{n=\underline{n}(l)}^\infty f^{\text{SF$\vert$PN}}_{(k)nl} \nu^k x^{n/2}{\chi}^l,
\end{align}
where $f^{\text{SF$\vert$PN}}_{(k)nl}$ can be obtained from either the SF expansion of Eq. \eqref{fPN} or the PN expansion of Eq. \eqref{fSF}. 

We define the SF+PN hybrid expansion of the quantity $f$ to order $(n_\text{max},l_\text{max},k_\text{max})$ as the sum of the two expansions minus their common terms:
\begin{align}
& x^{N/2} \sum_{l=0}^{l_\text{max}} \sum_{n=\underline{n}(l)}^{n_\text{max}}f^{\text{PN}}_{nl}( \nu , \log(x)) x^{n/2}{\chi}^l + \nu^{K} \sum_{k=0}^{k_\text{max}} \nu^k f^{\text{SF}}_{(k)}(x , \chi)\nonumber\\*
 &-\nu^{K}x^{N/2} \sum_{l=0}^{l_\text{max}}\sum_{k=0}^{k_\text{max}}\sum_{n=\underline{n}(l)}^{n_\text{max}} f^{\text{SF$\vert$PN}}_{(k)nl} \nu^k x^{n/2}{\chi}^l. \label{fSFPN hybrid}
\end{align}
The last sum cancels the doubly counted terms. 

Equation~\eqref{fSFPN hybrid} is a standard composite expansion.
By design, it is uniformly accurate through orders $x^{(N+n_{\rm max})/2}$ and $\nu^{K+k_{\rm max}}$ over the whole $(x,\nu)$ plane.  It is expected a priori, though not certain, that the most accurate such hybrid is obtained by truncating each series at the highest available orders in the expansions.

\subsection{Energy and angular momentum fluxes}

The forcing terms \eqref{Fchisymb}, \eqref{Fnusymb}, and \eqref{Fomegasymb} are the main ingredients of the hybrid model as well as the PN-only and SF-only models. They all depend upon four different functions on parameter space: three fluxes $\mathcal{F}^\infty$, $\mathcal{F}^\mathcal{H}$, and $\mathcal{G}^\mathcal{H}$, and the binding energy $E$. 

Regarding the fluxes $\mathcal{F}^\infty$, $\mathcal{F}^\mathcal{H}$, and $\mathcal{G}^\mathcal{H}$ from a nonspinning secondary compact body orbiting a spinning primary, only first-order SF computations have been computed in the literature:
\begin{align}
\mathcal{F}^\infty&= \nu^{2} \mathcal{F}^{\infty \text{SF}}_{(0)}(x,\chi)+\mathcal{O}(\nu^3),\label{fluxSF}\\
\mathcal{F}^\mathcal{H}&= \nu^{2} \mathcal{F}^{\mathcal{H} \text{SF}}_{(0)}(x,\chi)+\mathcal{O}(\nu^3),\label{fluxSFH}\\
\mathcal{G}^\mathcal{H}&= \nu^{2} \mathcal{G}^{\mathcal{H} \text{SF}}_{(0)}(x,\chi)+\mathcal{O}(\nu^3)\label{fluxSFHJ}.
\end{align}
Results for these fluxes are now standard~\cite{Taracchini:2014zpa,Barack:2018yvs} and are readily available from the Black Hole Perturbation Toolkit's \texttt{Teukolsky} package~\cite{TeukolskyPackage}, for example. We review their computation in Appendix~\ref{sec:ChebyshevInterpolation}. 

The energy fluxes have also been independently obtained in PN theory. In the nonspinning sector, the energy flux at infinity has been recently computed in PN through $\mathcal{O}(x^{19/2})$ ($4.5$PN accuracy)~\cite{Blanchet:2023bwj,Blanchet:2023sbv}, building on Refs.~\cite{Marchand:2020fpt,Larrouturou:2021dma,Larrouturou:2021gqo,Blanchet:2022vsm,Henry:2021cek,Trestini:2022tot,Trestini:2023wwg}; and through $\mathcal{O}(x^{18/2})$  ($4$PN) in the spin-orbit (SO)~\cite{Marsat:2013caa,Blanchet:2006gy,Bohe:2013cla,Cho:2022syn}, spin-spin (SS)~\cite{Bohe:2015ana,Cho:2021mqw,Cho:2022syn}, and cubic-in-spin (SSS) sectors~\cite{Marsat:2014xea}. The horizon fluxes are known including spin effects through $\mathcal{O}(x^{18/2})$ (relative $1.5$PN accuracy compared to the leading-PN horizon flux)~\cite{Tagoshi:1997jy,Alvi:2001mx,Porto:2007qi,Chatziioannou:2012gq,Saketh:2022xjb}. The angular momentum fluxes are known to the same PN orders via Eq.~\eqref{G from F}. We summarize these PN results in more detail in Appendix~\ref{app:PN}.

Therefore, the overall SF+PN hybrid fluxes at null infinity and at the horizon are given by the formula \eqref{fSFPN hybrid} as
\begin{align}
\mathcal{F}_H^\infty &= \nu^{2} \mathcal{F}^{{\infty\text{SF}}}_{(0)}(x,\chi)+x^{5}\sum_{n=0}^{9}\mathcal{F}^{\infty\text{PN}}_{n0}(\nu)x^{n/2}\nonumber\\
& +{\chi} x^{13/2}\sum_{n=0}^{5}\mathcal{F}^{\infty\text{PN}}_{n1}(\nu)x^{n/2}+{\chi}^2 x^{7}\sum_{n=0}^{4}\mathcal{F}^{\infty\text{PN}}_{n2}(\nu)x^{n/2} \nonumber\\
& +{\chi}^3 x^{17/2}\mathcal{F}^{\infty\text{PN}}_{03}(\nu)
-\nu^2\Biggl(x^{5}\sum_{n=0}^{9}\mathcal{F}^{\infty\text{SF$\vert$PN}}_{(0)n0}x^{n/2} \nonumber\\
& +{\chi} x^{13/2}\sum_{n=0}^{5}\mathcal{F}^{\infty\text{SF$\vert$PN}}_{(0)n1}x^{n/2}+{\chi}^2 x^{7}\sum_{n=0}^{4}\mathcal{F}^{\infty\text{SF$\vert$PN}}_{(0)n2}x^{n/2}\nonumber\\
&\quad +{\chi}^3 x^{17/2}\mathcal{F}^{\infty\text{SF$\vert$PN}}_{(0)03}\Biggr),\label{FluxHyb}
\end{align}
and
\begin{align}
\mathcal{F}_H^\mathcal{H}&=\nu^2\mathcal{F}^{\mathcal{H}\text{SF}}_{(0)}(x,\chi)+x^{15/2}\sum_{n=0}^{3}\mathcal{F}^{\mathcal{H}\text{PN}}_{n}({\chi},\nu)x^{n/2}\nonumber\\*
&\quad -\nu^2 x^{15/2}\sum_{n=0}^{3}\mathcal{F}^{\mathcal{H}\text{SF$\vert$PN}}_{(0)n}({\chi})x^{n/2},\label{fH}\\
\mathcal{G}_{H}^\mathcal{H}&=\nu^2\mathcal{G}^{\mathcal{H}\text{SF}}_{(0)}(x,\chi)+x^{15/2}\sum_{n=0}^{3}\mathcal{G}^{\mathcal{H}\text{PN}}_{n}({\chi},\nu)x^{n/2}\nonumber\\*
&\quad -\nu^2 x^{15/2}\sum_{n=0}^{3}\mathcal{G}^{\mathcal{H}\text{SF$\vert$PN}}_{(0)n}({\chi})x^{n/2},\label{fHJ}
\end{align}
where the subscript ``$H$'' stands for ``Hybrid''. All required PN data to build these expressions can be found on the data repository PNpedia~\cite{PNpedia}.

\subsection{Binding energy}\label{sec:binding energy}

The binding energy is substantially subtler than the flux to infinity.\footnote{In principle, fluxes through the primary's horizon have the same subtleties as we describe for the binding energy, but because these fluxes are already a small effect, subtleties in their definitions have not yet proved to be relevant.} While the latter is computed from waveform amplitudes at future null infinity, the binding energy is usually computed from a local, \emph{mechanical} energy describing the two-body system in the near zone. Our evolution equation~\eqref{Fomegasymb} was derived under the assumption that the binding energy is defined directly from the Bondi mass, whose relationship with the local mechanical energy is highly nontrivial~\cite{Pound:2019lzj}. 

At leading SF order, and at low PN orders, $E$ is not influenced by these subtleties. At leading SF order, it is the geodesic binding energy (i.e., geodesic energy minus inertial secondary mass $m_2$) as a function of the evolving dimensionless spin $\chi=\chi(t)$. At low PN orders, it is identical to the local mechanical energy.

However, recent work in Ref.~\cite{Trestini:2025nzr} has shown that the two quantities differ by a Schott term that arises at 4PN order. Forthcoming work will show how this Schott term extends to the fully relativistic 1SF energy~\cite{GrantInPrep}. 

Since the Schott term has not yet been implemented, here we follow Ref.~\cite{Wardell:2021fyy} in simply replacing the Bondi mass with the binary's mechanical energy. This is known to be a relatively small error~\cite{Pound:2019lzj} in a quantity that is already subleading (1SF beyond geodesic). However, there is some evidence it could still be appreciable~\cite{Albertini:2022rfe}. We will incorporate the Schott term in future work.

The 1SF mechanical energy was first defined from the first law of binary mechanics~\cite{LeTiec:2011ab,Le_Tiec_2012,Blanchet:2012at,LeTiec:2015kgg,Fujita:2016igj}, after which it was used to inform EOB~\cite{Barausse:2011dq} and to fix the 4PN conservative dynamics~\cite{Damour:2014jta}. More recently, it was shown to have a more traditional interpretation as the on-shell value of the mechanical Hamiltonian~\cite{Lewis:2025ydo}. It can be computed from the Detweiler redshift invariant $z(x,{\chi},\nu)$~\cite{Detweiler:2008ft,Barack:2011ed,Shah:2012gu,Fujita:2016igj,Nasipak:2025tby}, as illustrated in Ref.~\cite{Isoyama:2014mja}, for example. The redshift invariant is known both from a PN expansion and an SF expansion~\cite{Bini:2014nfa}, with the form
\begin{align}
 z &= z_{(0)}^\text{SF}(x ,  \chi)+\nu z^\text{SF}_{(1)}(x,  \chi)+ {\cal O}( \nu^2)\nonumber \\* \label{defz}
 &= 1+\left[ -\frac{3}{4}\left(1+\Delta\right)+\frac{\nu}{2} \right]x+{\cal O}(x^2)
\end{align} 
at the first two orders. Here, $z_{(0)}^\text{SF}(x, \chi)$ is given by $z_{(0)}(x,\chi)=1/U_{(0)}(x^{3/2},\chi)$, with $U_{(0)}$  given in Eq.~\eqref{U0} and with $\hat{\Omega}$ replaced by $x^{3/2}$ and $\mathring\chi$ by $\chi$. The 1SF term, $z_{(0)}^\text{SF}(x, \chi)$, has been computed extensively for quasicircular binaries in the literature~\cite{Barack:2018yvs}; see, for example, Refs.~\cite{Detweiler:2008ft,Sago:2008id,Shah:2010bi,Shah:2012gu,Bini:2015xua,Kavanagh:2016idg}.

The interpolation and reexpansion procedure for the 1SF redshift and binding energy are once again described in Appendix~\ref{sec:ChebyshevInterpolation}. On the PN side, the PN expansion is performed in both the nonspinning and spinning sectors through $\mathcal{O}(x^{5})$ (4PN)~\cite{Damour:2014jta,Marchand:2017pir,Foffa:2019yfl,Cho:2022syn,Bohe:2013cla,Boh__2013,Bohe:2015ana,Marsat:2014xea}. 

The overall SF+PN hybrid binding energy hence reads 
\begin{widetext}
\begin{align}\label{eq:EHybrid}
E_H= \nu& \MTOT \hat E^{\text{geo}}(x,\chi)+\nu^2 \MTOT  E^\text{SF}_{(1)}(x,\chi)+x \MTOT\sum_{n=0}^{8}E^\text{PN}_{n0}(\nu)x^{n/2}+{\chi} x^{5/2}\MTOT\sum_{n=0}^{4}E^\text{PN}_{n1}(\nu)x^{n/2}+{\chi}^2x^3\MTOT\sum_{n=0}^{4}E^\text{PN}_{n2}(\nu)x^{n/2}\nonumber\\*
&\hspace{-1cm}+{\chi}^3x^{9/2}\MTOT E^\text{PN}_{03}(\nu)
-\nu \MTOT\left(x\sum_{n=0}^{8}E^\text{SF$\vert$PN}_{(0)n0}x^{n/2}+{\chi} x^{5/2}\sum_{n=0}^{4}E^\text{SF$\vert$PN}_{(0)n1}x^{n/2}+{\chi}^2x^3\sum_{n=0}^{4}E^\text{SF$\vert$PN}_{(0)n2}x^{n/2}+{\chi}^3x^{9/2} E^\text{SF$\vert$PN}_{(0)03}x^{n/2}\right)\nonumber\\*
&\hspace{-1cm}-\nu^2 \MTOT\left(x\sum_{n=0}^{8}E^\text{SF$\vert$PN}_{(1)n0}x^{n/2}+{\chi} x^{5/2}\sum_{n=0}^{4}E^\text{SF$\vert$PN}_{(1)n1}x^{n/2}+{\chi}^2x^3\sum_{n=0}^{3}E^\text{SF$\vert$PN}_{(1)n2}x^{n/2}+{\chi}^3x^{9/2} E^\text{SF$\vert$PN}_{(1)03}x^{n/2}\right),
\end{align}
\end{widetext}
noting that $x^5$ terms identically vanish in the SO and SSS sectors.
Here
\begin{align}\label{Egeo}
\hat E^{\text{geo}}(x,\chi)=-1+\frac{1-2 xf_\chi^{1/3}}{\sqrt{f_\chi\left(1+x^{3/2}{\chi} -3xf_\chi^{1/3}\right)}}
\end{align}
with $f_\chi\equiv 1-x^{3/2}\chi$. The 1SF corrections to the energy and redshift are related as~\cite{Isoyama:2014mja}
\begin{align}
E^\text{SF}_{(1)}(x)= &\frac{1}{2} z^\text{SF}_{(1)}(x,\chi)- \frac{1}{3} x \partial_x z^{\text{SF}}_{(1)}(x,\chi)\nonumber\\
&\hspace{-1cm}+\hat E^\text{geo}(x,\chi)+x\left(-\frac{2}{3}+\deltaC x-\frac{2\delta \MTOT}{3 \MTOT}{}\right)\partial_x \hat E^\text{geo}(x,\chi)\nonumber\\
&\hspace{-1cm} -\delta \chi \partial_\chi E^\text{geo}(x,\chi), \label{eq:E1reexp}
\end{align}
with
\begin{align}
    &\delta \MTOT=\frac{\MTOT-\MATHRINGMTOT}{\nu},
    &\delta\chi=\frac{\chi-\mathring\chi}{\nu}.
\end{align}
Here the last four terms come from reexpanding the geodesic binding energy $\varepsilon m_1 \hat E^\text{geo}$ into the variables $x$, $\chi$, $\nu$, and $\MTOT$; see Appendix \ref{sec:ChebyshevInterpolation}. 

\subsection{Amplitudes}

Let us finally discuss the hybridization of the complex amplitudes. Recall that in our conventions, the mode amplitudes $\hat h_{\ell m}$ as defined in Eq. \eqref{eq:hu} are real-valued. This allows for an unambiguous definition of what contributes to the waveform amplitude and what contributes to the waveform phase.  In the SF data we use, the modes $m' \neq 0$ are instead expanded as 
\begin{align}\label{waveform}
 R\, h_{\ell' m'} = \MTOT\frac{2}{(m' \Omega)^2} C^{\text{SF}}_{\ell' m'}e^{-i m \phi_p},    
\end{align}
where primed mode indices refer to the basis of spheroidal harmonics $\mbox{}_{-2}S_{\ell' m'}$. We follow standard practice by re-expressing the SF amplitudes in a spherical harmonic basis~\cite{Chapman-Bird:2025xtd}, starting from a grid of SF spheroidal-harmonic amplitudes, using the procedure described in~\cite{Cook:2014cta} to convert the amplitudes between the two bases at each node of the grid. The details of the procedure are relegated to Appendix~\ref{sec:ChebyshevInterpolation}. As a result, the numerical functions obtained from the 1SF expansion are put in a spherical-harmonic expansion of the form
\begin{align}\label{waveform2}
 R\, h_{\ell m} = \MTOT\frac{2}{(m \Omega)^2} C^{\text{SF}}_{\ell m}e^{-i m \phi_p}.    
\end{align}

In the basis of spin-weighted spherical harmonics, one can now write $\tilde h_{\ell m}^{(1)}=\frac{2}{(m\Omega)^2}C^{\text{SF}}_{\ell m,(0)}$ and finally go to the convention where the mode amplitudes are real, $\tilde h_{\ell m}^{(1)}=\hat h_{\ell m}^{\text{SF},(1)}e^{-im\deltaC\phi_{\ell m}^{(1)}}$. The real amplitudes $\hat h_{\ell m}^{\text{SF},(1)}$ and phase shifts $\deltaC \phi_{\ell m}^{(1)}$ are given by
\begin{align}
    \hat h_{\ell m}^{\text{SF},(1)} & =\frac{2}{(m\Omega)^2}\left|C^{\text{SF}}_{\ell m,(0)}\right|, \\
    \deltaC \phi_{\ell m}^{\text{SF},(1)} &= -\frac{1}{m}\text{arg}\left( C^{\text{SF}}_{\ell m,(0)}\right).
\end{align}

The quantities we hybridize with PN are, first, the real mode amplitude $\hat h_{\ell m}^{\text{SF},(1)}$ and, second, the quantity $\deltaC\psi_{\ell m}^{\text{SF},(1)}=\deltaC\phi_{\ell m}^{\text{SF},(1)}-\deltaC\phi_{22}^{\text{SF},(1)}$, as it has been defined in Eq. \eqref{eq:deltapsilm}. Matching the physical waveforms, we therefore build the hybrid mode amplitudes as
\begin{widetext}
\begin{align}\label{hybridamp}
\hat h_{\ell m}^{H}&= \alpha_{\ell m}(\nu)\nu \hat h_{\ell m}^{\text{SF},(1)}+x\sum_{n=0}^8\hat{h}_{\ell m,n0}^\text{PN}x^{n/2}+x^{5/2}\sum_{n=0}^
4\hat{h}_{\ell m,n1}^\text{PN}\chi x^{n/2}+x^3\sum_{n=0}^3\hat{h}_{\ell m,n2}^\text{PN}\chi^2 x^{n/2}+x^{9/2}\hat{h}_{\ell m,03}^\text{PN}\chi^3\nonumber\\*
&\hspace{0.3cm}-\alpha_{\ell m}(\nu)\nu \left(x\sum_{n=0}^8\hat{h}_{\ell m,(0)n0}^\text{SF$\vert$PN}x^{n/2}+x^{5/2}\sum_{n=0}^
4\hat{h}_{\ell m,(0)n1}^\text{SF$\vert$PN}\chi x^{n/2}+x^3\sum_{n=0}^3\hat{h}_{\ell m,(0)n2}^\text{SF$\vert$PN}\chi^2 x^{n/2}+x^{9/2}\hat{h}_{\ell m,(0)03}^\text{SF$\vert$PN}\chi^3\right), 
\end{align}
\end{widetext} 
and
\begin{align}
\deltaC \psi_{\ell m}^H& = \nu{^0} \deltaC \psi_{\ell m}^{\text{SF},(1)} + \deltaC \psi_{\ell m}^{\text{PN}} - \nu^{0} \left( \deltaC \psi_{\ell m,(0)}^\text{SF$\vert$PN} \right). \label{hybridamp2}
\end{align}

In Eq.~\eqref{hybridamp}, we have resummed the SF $\ell m \neq 22$ mode amplitudes by multiplying the SF amplitude by the mass ratio dependent factor $\alpha_{\ell m}(\nu)=1+O(\nu)$ that appears in the leading PN term $\nu^{-1}\hat{h}_{\ell m,00}^\text{PN}$. We have in particular $\alpha_{22}(\nu)=1$, $\alpha_{21}(\nu)=\alpha_{33}(\nu)=\alpha_{31}(\nu)=\sqrt{1-4\nu}$ and $\alpha_{32}(\nu)=1-3\nu$. The coefficients $\hat{h}_{\ell m,(0)n l}^\text{SF$\vert$PN}$ are defined as Eq.~\eqref{fSFPN} for the function $\hat h_{\ell m}/(\nu \alpha_{\ell m}(\nu))$.  All PN data used in this paper are summarized in Appendix~\ref{app:PN}.

\section{Numerical implementation}\label{sec:implementation}

We now summarize the implementation of each of the three models we can construct using the scheme outlined above: a 0PA model, a PN model, and a hybrid model.

\subsection{Models}\label{subsubsec:models}

The central model of this paper is the SF+PN hybrid model. Starting from the energy flux-balance law (and horizon flux-balance laws), we have derived in Sec.~\ref{sec:flux-balance} the system of coupled ODEs that rules the evolution of the binary, Eqs. \eqref{Fchisymb}, \eqref{Fnusymb}, and \eqref{Fomegasymb}. The forcing terms of those ODEs depend upon four quantities: the two energy fluxes $\mathcal{F}^\infty$ and $\mathcal{F}^\mathcal{H}$, the angular momentum flux down the primary's horizon, $\mathcal{G}^\mathcal{H}$, and the binding energy $E$. We have described in Sec.~\ref{sec:hybridization} how those four quantities can be hybridized using SF and PN information. We have also described how one can hybridize the waveform mode amplitudes using Eqs. \eqref{hybridamp} and \eqref{hybridamp2}. All those hybrid quantities are  put together in a model that we shall call the hybrid model, and refer to as $H$ when labeling expressions. In order to assess the accuracy improvement of the hybrid model as compared to both of its constituents, we also introduce the standard $0$PA model, referred to as $0$PA, and a PN model, referred to as PN. 

As emphasized earlier, keeping the fraction in Eq.~\eqref{Fomegasymb} unexpanded leads to a better comparison with NR. Hence, we keep it as it is when generating waveforms. In this sense, not expanding the fraction of Eq. \eqref{Fomegasymb} in powers of $\nu$ may be viewed as a resummation of the waveform frequency evolution. On the other hand, since we always re-expand our hybridized functions when changing  parameters from (${\mathring\varepsilon}$, $\mathring\chi$, $\MATHRINGm1$) to ($\nu$, $\chi$, $\MTOT$), that transformation represents a reexpansion rather than a resummation.  We do stress that the PN model we define just below is effectively a resummed PN model, as we similarly do not expand the fraction of \eqref{Fomegasymb} in powers of $x$. 

In this section, we describe what pieces of information each of the models contains and how we numerically implement them. We postpone the comparison with NR to Sec.~ \ref{sec:comparison}.

\bigskip

\noindent\textbf{0PA model: } This is used as a benchmark for our hybrid model at small $\nu$, where the 0PA part of the driving force is dominant. This model corresponds to only including the $0$PA contributions in the quantities $\mathcal{F}^\infty$, $\mathcal{F}^\mathcal{H}$, $\mathcal{G}^\mathcal{H}$, and $E$ and keeping only the leading-order-in-$\nu$ term in the forcing functions \eqref{Fomegasymb}, \eqref{Fchisymb}, and \eqref{Fnusymb}. All the terms in the numerator of Eq. \eqref{Fomegasymb} that are proportional to a derivative of $E$ are $\mathcal{O}(\nu)$, which is subdominant and hence neglected at $0$PA. We then end up exactly with
\begin{equation}    F^{\rm 0PA}_\omega\left(x, \chi\right)=\frac{\nu}{\MTOT} F_{(0)}^{\hat \Omega}\left(\omega,\chi \right)=\frac{\nu}{\MTOT} \frac{\mathcal F^{\infty\text{SF}}_{(0)}+\mathcal F^{\mathcal H\text{SF}}_{(0)}}{\frac{\partial \hat E^\text{geo}(\omega^{2/3},\chi)}{\partial\omega}}.\label{F0PA}
\end{equation}
Here, $F_{(0)}^{\hat \Omega}\left(\omega,\chi\right)$ is defined in Eq. \eqref{FOmega0PA} but with $\hat \Omega$ replaced by  $\omega$ and $\mathring \chi$ replaced by $\chi$. This approximation to $F_\omega$ has a residual error of order $\mathcal{O}(\nu^{2})$ but is exact in $\omega$ and $\chi$, away from the unperturbed ISCO frequency $\Omega_\star$. It has a simple pole at the unperturbed ISCO, where the function $D(\omega)$ vanishes. We emphasize the reason why we can replace $\hat \Omega$ by $\omega$ in the forcing terms is that $\hat \Omega=\omega+\mathcal{O}(\nu)$; the corrections from replacing the orbital frequency with half the $(2,2)$ mode waveform frequency is a 1PA effect in the SF expansion. The mode amplitudes we use in this model are simply the ones from a first-order Teukolsky computation, reexpanded as a function of the parameters $x$ and $\chi$,
\begin{align}
	\hat h^{\rm 0PA}_{\ell m}(x,\chi)=\nu\frac{2}{m^2x^3} \left|C^{\rm SF}_{\ell m,(0)}(x,\chi)\right|.
\end{align}

\bigskip

\noindent 
\textbf{PN model:} This is our second model that serves as a benchmark. For this model, we only use PN information to build the waveform frequency forcing function $F^{\rm PN}_\omega$. We build the four quantities $\mathcal{F}_{\rm PN}^\infty$, $\mathcal{F}_{\rm PN}^\mathcal{H}$, $\mathcal{G}_{\rm PN}^\mathcal{H}$, and $E_{\rm PN}$ from their PN expansions and substitute those expressions into the forcing functions \eqref{Fchisymb}, \eqref{Fnusymb}, and \eqref{Fomegasymb}. The result for $F^{\rm PN}_\omega$ will include all powers of $\nu$ but exhibit errors already at the leading order $\mathcal{O}(\nu)$ in the small-$\nu$ limit (i.e., 0PA), as the infinite set of PN orders would be necessary to obtain the complete 0PA term. Because the forcing term is a quotient of PN series, it is effectively resummed.  The mode amplitudes we use for this model are the PN mode amplitudes $\hat h^{\rm PN}_{\ell m}(x)$.  In short, our PN model is analogous to a TaylorT1 model \cite{Damour:2000zb} with 4PN (resp. 4.5PN) information for the binding energy (resp. energy flux at infinity), with horizon fluxes added and where the main expansion variable is the waveform frequency parameter $x$. 

\bigskip

\noindent 
\textbf{Hybrid model:} For this model we use all four hybrid quantities $\mathcal{F}^\infty_H$, $\mathcal{F}^\mathcal{H}_H$, $\mathcal{G}^\mathcal{H}_{H}$, and $E_H$ as defined in Eqs. \eqref{FluxHyb}, \eqref{fH}, \eqref{fHJ}, and \eqref{eq:EHybrid} and substitute those expressions into the forcing terms \eqref{Fchisymb}, \eqref{Fnusymb}, and \eqref{Fomegasymb}. Even though we do not expand the fraction in Eq. \eqref{Fomegasymb} in powers of $\nu$ when numerically generating waveforms, it is still of interest to do so when assessing convergence properties of the hybrid model. Doing so, one obtains 
\begin{equation}\label{SFresumexpansion}
    F^H_\omega=\nu \left[ F^H_{\omega (0)}(x,\chi)+ \sum_{n=1}^\infty \nu^n  F_{\omega (n)}^{H}(x,\chi)\right].
\end{equation}
By construction,  $F^H_{\omega(0)}=F_{\omega (0)}$; the 0PA term is exact. But the $1$PA term, $F_{\omega (1)}^{H}$, is incomplete. It contains the exact $1$PA corrections to the binding energy (modulo the Schott term described earlier), but it only contains a PN approximation of the 2SF (1PA) flux; see Eq. \eqref{FluxHyb}. Hence, for small $\nu$ it formally incurs a 1PA error, though as we shall see, this error is small. The largest error in the 2SF  flux occurs in the spin sectors at $4.5$PN relative to the leading PN term in the nonspinning sector. Hence, the errors we make in the $1$PA term $F_{\omega (1)}^{H}$, as compared to a full 2SF calculation are a relative $4.5$PN error linear and quadratic in primary spin, and an $\mathcal{O}({\chi}^4)$ error, as we neglected higher than SSS interactions. Using the leading PN behavior of Eq.~\eqref{Fomegasymb}, one can therefore write the full $1$PA waveform-frequency forcing function as
\begin{multline}
    F_{\omega (1)}(x,{\chi}) =F_{\omega (1)}^{H}(x,{\chi})+x^{\frac{11}{2}}\left[O(x^{5})+O({\chi} x^{9/2})\right. \\  \left.+O({\chi}^2 x^{9/2})+O({\chi}^3 x^{9/2})+O(\chi^4x^4)\right],
\end{multline}
where we have noted that the $\chi^4$ terms begin at 4PN.  
Although the error is formally 1PA, the model behaves very differently from the $0$PA model, as it includes the PN information contained in the energy flux up to relative $4$PN (and up to $4.5$PN in the nonspinning sector) as well as the full 1SF piece of the binding energy. In addition, it also contains higher-PA PN corrections to the fluxes and binding energy, as described in Sec. \ref{sec:hybridization}. The mode amplitudes we use for this model are the $\hat h^{H}_{\ell m}(x)$ as they are defined in Eq. \eqref{hybridamp}. 

\bigskip

One of the main features of keeping the forcing term $F_\omega$ unexpanded, as in Eq. \eqref{Fomegasymb}, is the following: in the expansion~\eqref{SFresumexpansion}, we observe that each term $F_{\omega (n)}^{H}$ has a pole of the form $1/D^{\frac{1}{4}(1+3(-1)^n+10n)}$, as can be derived from the asymptotic matching with the transition to plunge; see, e.g., Sec.~5 of~\cite{Kuchler:2024esj} or Table~I of~\cite{Albertini:2022rfe}, with $D(\Omega)$ defined in Eq.~\eqref{eqD}. Those poles are all located precisely at the primary's ISCO frequency $\Omega_\star$. Keeping the fraction unexpanded in Eq.~\eqref{Fomegasymb} effectively resums the expansion \eqref{SFresumexpansion}. As alluded to earlier, this reduces the strength of the divergence \emph{at} the ISCO frequency but also shifts the location \emph{of} the ISCO frequency. Indeed, the ISCO is reached at the breakdown of the two-timescale expansion, which itself happens when the evolution equations of the inspiral break down. When fully expanding the forcing terms as in Eq. \eqref{SFresumexpansion}, the evolution equations \eqref{FBeom} break down when each individual $F_{\omega (n)}^{H}$ blows up to infinity, which occurs exactly where $D(\omega=\Omega_\star)=0$. However, when keeping the fraction unexpanded as in Eq. \eqref{Fomegasymb}, the breakdown occurs where $\partial_\omega E_H=0$. Let us limit our analysis to $1$SF corrections, $E_H=\nu \MTOT\left[\hat E^\text{geo}+\nu E_{(1)}^\text{SF}+\mathcal{O}{(\nu^2)}\right]$. We solve $\partial_\omega E_H\left[\Omega_\star(1+\nu C_\Omega)\right]=0$ for the ISCO shift $C_\Omega$ at linear order in $\nu$, and find that
\begin{align}
    C_\Omega=1+\frac{1}{2}\frac{\partial_\omega^2z_{(1)}\left(\Omega_\star\right)}{\partial_\omega^2\hat E^\text{geo}\left(\Omega_\star\right)},
\end{align}
which reproduces Eq. (16) of~\cite{Isoyama:2014mja}. We have checked numerically that our ISCO shift agrees with Table I of~\cite{Isoyama:2014mja}.

\bigskip

\noindent
\textbf{0PA4PN model:} This model is nearly identical to the hybrid model, with the only difference that  the 1SF binding energy is not used, i.e. the binding energy is only the hybridization of the geodesic energy with the 4PN energy. This model will be used to assess the importance of using the numerical 1SF binding energy.  

\subsection{Offline and online computations}\label{subsec:WFgen}

We stressed in the Introduction that modern SF waveforms can be generated rapidly due to the offline-online split that the multiscale expansion enables. By design, the hybrid waveform enjoys precisely that same split of waveform generation between an offline and an online step. The offline step consists of solving the field equations to (i) express the waveform as a function of the binary's phase-space variables---$(\phi,\Omega_{\rm PN},\chi,\nu,M)$ or $(\phi_p,\Omega_{\rm SF},\chi,\nu,M)$ and then $(\psi,\omega,\chi,\nu,M)$---(ii) compute the fluxes and binding energy as functions of the binary parameters---$(\Omega_{\rm PN},\chi,\nu,M)$ or $(\Omega_{\rm SF},\chi,\nu,M)$ and then $(\omega,\chi,\nu,M)$. The online step then consists of generating a waveform by obtaining a trajectory in phase space and summing the waveform modes. This online stage can be carried out ``on the fly'' from the precomputed offline results.

In our context, the offline stage largely reduces to gathering existing results or generating data using existing codes. Concretely, for each of the three models above, the two steps consist of the following:

\bigskip

\noindent \textbf{Offline step}
\begin{itemize}
    \item Obtain a Chebyshev interpolant of the energy fluxes to infinity, $\mathcal{F}_{(0)}^\infty(x,\chi)$, and down the horizon, $		\mathcal{F}_{(0)}^\mathcal{H}(x,\chi)$, as described in Appendix~\ref{sec:ECheb}.
    \item Obtain a Chebyshev interpolant of the redshift invariant $z_{(1)}(x,\chi)$ as described in Appendix \ref{app:1b}, and from it compute the 1SF (1PA) correction to the binding energy using Eq.~\eqref{eq:E1reexp}.
    \item Obtain a Chebyshev interpolant of the mode amplitudes $C^{\rm SF}_{\ell m,(0)}$, as described in Appendix \ref{sec:AmpChebyshev}.
    \item Code all PN expressions (fluxes, binding energy, and amplitudes) for a quasicircular inspiral.
    \item Build the forcing functions $F_\omega^X(x,\chi,\nu,\MTOT)$, $F_\chi^X(x,\chi,\nu,\MTOT)$, and  $F_\nu^X(x,\chi,\nu,\MTOT)$ as well as the amplitudes $\hat h^X_{\ell m}\left(x(t),\chi(t),\nu(t)\right)$ and phase shifts $\deltaC \psi^X_{\ell m}\left(x(t),\chi(t),\nu(t)\right)$ for each model $X\in\{H,0\text{PA},\text{PN}\}$. 
\end{itemize}

\smallskip

\noindent 
\textbf{Online step}
\begin{itemize}
    \item For a given model $X\in\{H,0\text{PA},\text{PN}\}$ and a given set of initial values $\mathring \chi$, $\mathring \nu$, $\mathring \omega < \Omega_\star(\mathring \chi)$, and $\MATHRINGMTOT$, solve the system of coupled ordinary differential equations
    \begin{equation}\label{eq:ODE}
    \begin{cases}
        d\psi/dt &=\omega/\MTOT,\\
        d\omega/dt &=F_\omega^X(x,\chi,\nu,\MTOT),\\ 
        d\chi/dt &= F_\chi^X(x,\chi,\nu,\MTOT), \\
        d\nu/dt &= F_\nu^X(x,\chi,\nu,\MTOT),
    \end{cases}
    \end{equation}
    with the initial condition $\psi(\mathring t)=\mathring\psi$, $\omega(\mathring t)=\mathring \omega $, $\chi(\mathring t) = \mathring\chi$, $\nu(\mathring t)={\mathring\nu}$ with $\MTOT$ given by Eq. \eqref{Mtotconstr} and $x=\omega^{2/3}$.
    \item Once the (2,2) mode waveform phase $\psi(t)$ and the waveform frequency $\omega(t)$ are known, we can compute the $(\ell ,m)$-mode of the wavestrain with\\
    
\begin{align}
h^X_{\ell m}&=  \MTOT (t)\hat h^X_{\ell m}\left(x(t),\chi(t),\nu(t)\right)  \nonumber \\ 
&\qquad \times e^{-i m (\psi_{\ell m}(t)+\deltaC \psi^X_{\ell m}(t))}.
\end{align}
Note that, even though we do not label the phase space trajectory explicitly with a superscript $X$, the solutions $x(t)$, $\chi(t)$, and $\nu(t)$, as well as $\MTOT(t)$ are implicitly model dependent, as the forcing terms in Eqs. \eqref{eq:ODE} do depend upon model $X$.
\item Finally, when comparing to NR, one should rather normalize all quantities with respect to the reference total mass $\mathring M$. Hence we rescale the Boyer-Lindquist time as $t\rightarrow \frac{1}{\MATHRINGMTOT}t$ and the $(\ell,m)$ mode of the wavestrain as $h^X_{\ell m}\rightarrow\frac{\MATHRINGMTOT}{\MTOT}h^X_{\ell m}$.
\end{itemize}

\section{Comparison of waveforms with numerical relativity and other models}\label{sec:comparison}

In this section, we compare our hybrid SF+PN waveform model to NR waveforms accessible from the SXS catalog~\cite{Lovelace:2011nu,Lovelace:2010ne,Mroue:2013xna,Mroue:2012kv,Lovelace:2014twa,Kumar:2015tha,Lovelace:2016uwp,Abbott:2016nmj,Hinder:2013oqa,Abbott:2016apu,Chu:2015kft,Abbott:2016wiq,Varma:2018mmi,Varma:2018aht,Varma:2019csw,Varma:2020bon,Islam:2021mha,Ma:2021znq}. We also compare against the $0$PA and PN models as benchmarks. These comparisons, in Secs.~\ref{subsec:comparison} and \ref{subsec:accuracy}, span the parameter range ${\mathring\chi}\in [-1,1]$ for a variety of mass ratios. To perform such comparisons, one first has to align the models with the NR waveform, a procedure described in Sec.~\ref{subsec:WFalig}.

\subsection{Waveform alignment scheme}\label{subsec:WFalig}
 The procedure we use  to align a waveform $h_{\ell m}^X$ from model $X$ to an NR waveform $h_{\ell m}^\text{NR}$ goes as follows:
\begin{enumerate}
    \item We fix a time interval $[t_0,t_0+T]$ in which to align both waveforms. We choose the time interval such that it starts after the relaxation time of the NR simulation in order to avoid initial boundary condition effects. In the following, we fix $t_0$ and $T$ in two different ways for two different batches of comparisons. The first batch of comparisons, which we shall refer to as ``Fixed$\chi$", compare waveforms that have the same value of the primary spin $\mathring\chi$ but distinct values of the symmetric mass ratio $\mathring\nu$. This set  allows us to assess the convergence properties of the hybrid model as the mass ratio decreases, \textit{ceteris paribus}. For the second batch of comparisons, referred to as ``$\Omega_0$", we  only consider the performance of the hybrid model against each individual NR simulation, comparing both waveforms across the maximal time window (i.e., starting from the lowest possible frequency after junk radiation has faded away) that the NR waveform provides.
    \item Given that some residual eccentricity or junk radiations might remain present even after relaxation time, finding the times $t_X$ and $t_\text{NR}$ at which the model waveform $X$ and the NR waveform reach a chosen frequency does not provide an accurate waveform alignment scheme. 
    
    Instead, we perform the following minimization procedure. We find the time shift $\deltaA t$ that minimizes the square error function 
    \begin{align}\label{eq:SE}
      \;\;\;\;\,\;\; SE(\deltaA t)\equiv\int_{t_0}^{t_0+T} dt \left[\omega_\text{NR}(t) -\omega_X(t + \deltaA t)\right]^2
    \end{align}
    between the model waveform $X$ and the NR waveform. The error function is typically a parabola with minimum, which allows for a fast minimization scheme~\cite{Boyle:2008ge}. In practice, we use a ternary search algorithm for finding the minimum of $SE(\deltaA t)$. 
    
    Given that residual eccentricities are small, our first guess for the time shift $t_X-t_\text{NR}$ serves as a good guess for localizing the minimum. We then look for a minimum in the time shift interval $[(t_X-t_\text{NR})-t_-,(t_X-t_\text{NR})-t_+]$, where $t_-$ and $t_+$ are two alignment parameters that define the time window upon which we look for a minimum of the square error \eqref{eq:SE}. When the algorithm has converged to a given precision that we chose to be $\sim10^{-3}$, we store the value of $\deltaA t$ that achieves the minimum error, $\deltaA t_\text{min}$. 
    
    \item To align the waveforms in phase, we compute the waveform dephasing $\deltaE\varphi_\text{min}$ from the average over the $[t_0,t_0+T]$ interval of the $(2,2)$ mode dephasing:
    \begin{align}
\;\;\;\;\;\;\;\deltaE\varphi_\text{min} \equiv \frac{-1}{T}\, \int_{t_0}^{t_0+T} dt \,  \text{arg}\left( \frac{h_{22}^X(t+\deltaA t_\text{min})}{h_{22}^\text{NR}(t)} \right).    
    \end{align}
    \item We finally align the waveforms by performing a time translation and phase shift of the complex waveform of the model $X$,  
    \begin{equation}\label{eq:halign}
h_{\ell m}^X(t)\rightarrow h_{\ell m}^X(t+\deltaA t_\text{min})e^{-i\deltaA\varphi_\text{min}}.
    \end{equation}
    In terms of the waveform phase, this amounts to 
    \begin{align}\label{eq:psialign}
        \psi^X_{\ell m}(t) \mapsto \psi^X_{\ell m}(t+\deltaA t_\text{min}) + \frac{\deltaA\varphi_\text{min}}{m} .  
    \end{align}
\end{enumerate}
Compared to~\cite{Boyle:2008ge}, our minimization procedure corresponds to performing a least-square error algorithm on the waveform frequencies, and not on the waveform phase. After some numerical checks, we find both procedures yield very similar values for $\deltaA t_\text{min}$ and $\deltaA\varphi_\text{min}$, for all comparisons considered in this paper.

\subsection{Waveform comparisons}\label{subsec:comparison}

We now carry out our comparisons. Using the python package \texttt{sxs}~\cite{sxspython_2024} and its second catalog~\cite{SXS:catalog}, we selected simulations without eccentricity, inclination nor secondary spin, and spanning the largest available range of primary spins ${\mathring\chi}$ and mass ratios $\mathring q=1/{\mathring\varepsilon}$. The set of simulations we use is displayed in the bottom-right panel of Fig.~\ref{fig:Spaghet} and listed in Table \ref{tab:mismatchdephasing}, arranged into groups of fixed $\mathring q$ and varying ${\mathring\chi}$ or vice versa. To assess the accuracy of our hybrid model, we use three different indicators: qualitative comparisons, waveform dephasing, and mismatches.

\begin{table*}[htbp]
\makebox[\textwidth][c]{
    \begin{tabular}{lcccccccccc}
\toprule
      &    	&	&	\multicolumn{2}{c}{Hybrid} & \multicolumn{2}{c}{$0$PA} & \multicolumn{2}{c}{PN} \\
\cmidrule(lr){4-5}\cmidrule(lr){6-7}\cmidrule(lr){8-9}
\multicolumn{1}{c}{ID} 
& 	$\mathring q[/]$
& ${\mathring\chi}[/]$		
&{$\mathcal{M}^H [/]$}	
&	\multicolumn{1}{c}{$\deltaE\psi^H_{22}(t_\star)[{\text{rad}}]$}
&{$\mathcal{M}^{0\text{PA}} [/]$}	
&	\multicolumn{1}{c}{$\deltaE\psi^{0\text{PA}}_{22}(t_\star)[{\text{rad}}]$}
&{$\mathcal{M}^\text{PN} [/]$}	
&	\multicolumn{1}{c}{$\deltaE\psi^\text{PN}_{22}(t_\star)[{\text{rad}}]$}  \\ \midrule
  SXS:BBH:0461~\cite{Blackman:2017dfb,Boyle:2019kee}		
  &	\multicolumn{1}{c}{$2$}		
  &	\multicolumn{1}{c}{$-0.6$}	
  &	\multicolumn{1}{c}{$6.7\cdot 10^{-7}$}	
  &	\multicolumn{1}{c}{$-1.1\cdot 10^{-1}$}
  &	\multicolumn{1}{c}{$1.6\cdot 10^{-2}$}	
  &	\multicolumn{1}{c}{$-9.7\cdot 10^{0}$}
  &	\multicolumn{1}{c}{$1.3\cdot 10^{-5}$}	
  &	\multicolumn{1}{c}{$-4.8\cdot 10^{-1}$}
  \\
  SXS:BBH:2115~\cite{Chu:2015kft,Boyle:2019kee}	
  &\multicolumn{1}{c}{$2$}
  &	\multicolumn{1}{c}{$-0.3$}
  &	\multicolumn{1}{c}{$1.8\cdot 10^{-5}$}
  &	\multicolumn{1}{c}{$-4.5\cdot 10^{-1}$}
  &	\multicolumn{1}{c}{$3.3\cdot 10^{-2}$}
  &	\multicolumn{1}{c}{$-1.5\cdot 10^{1}$}
  &	\multicolumn{1}{c}{$3.0\cdot 10^{-5}$}
  &	\multicolumn{1}{c}{$-1.2\cdot 10^{0}$}
  \\
  SXS:BBH:2425~\cite{SXS:catalog}	
  &	\multicolumn{1}{c}{$2$}	
  &	\multicolumn{1}{c}{$0$}
  &	\multicolumn{1}{c}{$3.7\cdot 10^{-6}$}
  &	\multicolumn{1}{c}{$1.5\cdot 10^{-1}$}
  &	\multicolumn{1}{c}{$7.9\cdot 10^{-3}$}	
  & \multicolumn{1}{c}{$-8.8\cdot 10^{0}$}
  &	\multicolumn{1}{c}{$7.2\cdot 10^{-5}$}	
  &	\multicolumn{1}{c}{$-1.4\cdot 10^{0}$}
  \\
  SXS:BBH:2124~\cite{Chu:2015kft,Boyle:2019kee}	
  &	\multicolumn{1}{c}{$2$}
  &	\multicolumn{1}{c}{$0.3$}
  &	\multicolumn{1}{c}{$5.4\cdot 10^{-6}$}	
  &	\multicolumn{1}{c}{$-4.1\cdot 10^{-1}$}
  &	\multicolumn{1}{c}{$1.1\cdot 10^{-2}$}
  &	\multicolumn{1}{c}{$-1.2\cdot 10^{1}$}
  &	\multicolumn{1}{c}{$1.4\cdot 10^{-4}$}
  &	\multicolumn{1}{c}{$-3.1\cdot 10^{0}$}
  \\
  SXS:BBH:2129~\cite{Chu:2015kft,Boyle:2019kee}	
  &	\multicolumn{1}{c}{$2$}
  &	\multicolumn{1}{c}{$0.6$}
  &	\multicolumn{1}{c}{$8.7\cdot 10^{-6}$}	
  &	\multicolumn{1}{c}{$-5.2\cdot 10^{-2}$}
  &	\multicolumn{1}{c}{$4.0\cdot 10^{-4}$}	
  &	\multicolumn{1}{c}{-$4.6\cdot 10^{0}$}
  &	\multicolumn{1}{c}{$3.1\cdot 10^{-4}$}	
  &	\multicolumn{1}{c}{$-4.3\cdot 10^{0}$}
  \\
\midrule
  SXS:BBH:2179~\cite{Varma:2019csw}	
  &	\multicolumn{1}{c}{$6$}
  &	\multicolumn{1}{c}{$-0.8$}
  &	\multicolumn{1}{c}{$4.1\cdot 10^{-6}$}	
  &	\multicolumn{1}{c}{$-4.0\cdot 10^{-1}$}
  &	\multicolumn{1}{c}{$8.4\cdot 10^{-3}$}	
  &	\multicolumn{1}{c}{$-7.4\cdot 10^{0}$}
  &	\multicolumn{1}{c}{$4.7\cdot 10^{-5}$}	
  &	\multicolumn{1}{c}{$4.2\cdot 10^{-1}$}
  \\
  SXS:BBH:2208~\cite{Varma:2019csw}	
  &	\multicolumn{1}{c}{$6$}	
  &	\multicolumn{1}{c}{$-0.4$}
  &	\multicolumn{1}{c}{$2.1\cdot 10^{-6}$}	
  &	\multicolumn{1}{c}{$-5.1\cdot 10^{-1}$}
  &	\multicolumn{1}{c}{$1.0\cdot 10^{-2}$}	
  &	\multicolumn{1}{c}{$-8.4\cdot 10^{0}$}
  &	\multicolumn{1}{c}{$8.2\cdot 10^{-6}$}	
  &	\multicolumn{1}{c}{$9.8\cdot 10^{-3}$}
  \\
  SXS:BBH:3630~\cite{SXS:catalog}	
  &	\multicolumn{1}{c}{$6$}
  &	\multicolumn{1}{c}{$0$}	
  &	\multicolumn{1}{c}{$1.7\cdot 10^{-6}$}	
  &	\multicolumn{1}{c}{$-5.2\cdot 10^{-1}$}
  &	\multicolumn{1}{c}{$1.2\cdot 10^{-2}$}	
  &	\multicolumn{1}{c}{$-1.0\cdot 10^{1}$}
  &	\multicolumn{1}{c}{$1.1\cdot 10^{-4}$}	
  &	\multicolumn{1}{c}{$-1.4\cdot 10^{0}$}
  \\
  SXS:BBH:2185~\cite{Varma:2019csw}	
  &	\multicolumn{1}{c}{$6$}
  &	\multicolumn{1}{c}{$0.4$}
  &	\multicolumn{1}{c}{$6.8\cdot 10^{-6}$}	
  &	\multicolumn{1}{c}{$-4.7\cdot 10^{-1}$}
  &	\multicolumn{1}{c}{$5.5\cdot 10^{-3}$}	
  &	\multicolumn{1}{c}{$-8.8\cdot 10^{0}$}
  &	\multicolumn{1}{c}{$9.1\cdot 10^{-4}$}	
  &	\multicolumn{1}{c}{$-4.8\cdot 10^{0}$}
  \\
  SXS:BBH:1437~\cite{Varma:2018mmi,Boyle:2019kee}	
  &	\multicolumn{1}{c}{$6$}	
  &	\multicolumn{1}{c}{$0.8$}	
  &	\multicolumn{1}{c}{$2.1\cdot 10^{-4}$}	
  &	\multicolumn{1}{c}{$-3.1\cdot 10^{0}$}
  &	\multicolumn{1}{c}{$7.2\cdot 10^{-5}$}	
  &	\multicolumn{1}{c}{$-2.9\cdot 10^{0}$}
  &	\multicolumn{1}{c}{$6.4\cdot 10^{-3}$}		
  &	\multicolumn{1}{c}{$-1.0\cdot 10^{1}$}
  \\
\midrule
  SXS:BBH:2478~\cite{Yoo:2022erv}	
  &	\multicolumn{1}{c}{$14$}	
  &	\multicolumn{1}{c}{$-0.5$}	
  &	\multicolumn{1}{c}{$1.6\cdot 10^{-6}$}	
  &	\multicolumn{1}{c}{$-2.8\cdot 10^{-1}$}
  &	\multicolumn{1}{c}{$7.0\cdot 10^{-3}$}	
  &	\multicolumn{1}{c}{$-6.7.\cdot 10^{0}$} 
  &	\multicolumn{1}{c}{$3.6\cdot 10^{-4}$}	
  &	\multicolumn{1}{c}{$1.7\cdot 10^{0}$}
  \\
  SXS:BBH:2479~\cite{Yoo:2022erv}	
  &	\multicolumn{1}{c}{$14$}	
  &	\multicolumn{1}{c}{$-0.25$}	
  &	\multicolumn{1}{c}{$5.1\cdot 10^{-6}$}	
  &	\multicolumn{1}{c}{$-1.2\cdot 10^{-2}$}
  &	\multicolumn{1}{c}{$1.0\cdot 10^{-2}$}	
  &	\multicolumn{1}{c}{$-7.4\cdot 10^{0}$}
  &	\multicolumn{1}{c}{$2.1\cdot 10^{-4}$}	
  &	\multicolumn{1}{c}{$1.4\cdot 10^{0}$}
  \\
  SXS:BBH:2480~\cite{Yoo:2022erv}	
  &	\multicolumn{1}{c}{$14$}	
  &	\multicolumn{1}{c}{$0$}	
  &	\multicolumn{1}{c}{$7.4\cdot 10^{-6}$}	
  &	\multicolumn{1}{c}{$-7.8\cdot 10^{-1}$}
  &	\multicolumn{1}{c}{$1.0\cdot 10^{-2}$}	
  &	\multicolumn{1}{c}{$-8.7\cdot 10^{0}$}
  &	\multicolumn{1}{c}{$1.0\cdot 10^{-4}$}	
  &	\multicolumn{1}{c}{$-1.3\cdot 10^{0}$}
  \\
  SXS:BBH:2481~\cite{Yoo:2022erv}	
  &	\multicolumn{1}{c}{$14$}	
  &	\multicolumn{1}{c}{$0.25$}
  &	\multicolumn{1}{c}{$2.3\cdot 10^{-6}$}	
  &	\multicolumn{1}{c}{$-5.9\cdot 10^{-1}$}
  &	\multicolumn{1}{c}{$9.5\cdot 10^{-3}$}	
  &	\multicolumn{1}{c}{$-8.8\cdot 10^{0}$}
  &	\multicolumn{1}{c}{$1.8\cdot 10^{-3}$}	
  &	\multicolumn{1}{c}{$-4.6\cdot 10^{0}$}
  \\
  SXS:BBH:2482~\cite{Yoo:2022erv}	
  &	\multicolumn{1}{c}{$14$}		
  &	\multicolumn{1}{c}{$0.5$}
  &	\multicolumn{1}{c}{$5.8\cdot 10^{-6}$}	
  &	\multicolumn{1}{c}{$-4.3\cdot 10^{-1}$}
  &	\multicolumn{1}{c}{$3.6\cdot 10^{-3}$}	
  &	\multicolumn{1}{c}{$-7.3\cdot 10^{0}$}
  &	\multicolumn{1}{c}{$6.1\cdot 10^{-3}$}	
  &	\multicolumn{1}{c}{$-9.2\cdot 10^{0}$}
  \\
\midrule
  SXS:BBH:2463~\cite{Yoo:2022erv}	
  &	\multicolumn{1}{c}{$15$}	
  &	\multicolumn{1}{c}{$-0.5$}	
  &	\multicolumn{1}{c}{$3.7\cdot 10^{-6}$}	
  &	\multicolumn{1}{c}{$-4.5\cdot 10^{-1}$}
  &	\multicolumn{1}{c}{$6.7\cdot 10^{-3}$}
  &	\multicolumn{1}{c}{$-6.7\cdot 10^{0}$}
  & \multicolumn{1}{c}{$3.6\cdot 10^{-4}$}
  &	\multicolumn{1}{c}{$1.7\cdot 10^{0}$}
  \\
  SXS:BBH:2473~\cite{Yoo:2022erv}	
  &	\multicolumn{1}{c}{$15$}	
  &	\multicolumn{1}{c}{$-0.35$}	
  &	\multicolumn{1}{c}{$1.1\cdot 10^{-5}$}
  &	\multicolumn{1}{c}{$-5.9\cdot 10^{-1}$}	
  &	\multicolumn{1}{c}{$6.5\cdot 10^{-3}$}
  &	\multicolumn{1}{c}{$-7.0\cdot 10^{0}$}
  &	\multicolumn{1}{c}{$2.2\cdot 10^{-4}$}
  &	\multicolumn{1}{c}{$1.5\cdot 10^{0}$}
  \\
  SXS:BBH:2477~\cite{Yoo:2022erv}	
  &	\multicolumn{1}{c}{$15$}	
  &	\multicolumn{1}{c}{$0$}
  &	\multicolumn{1}{c}{$2.7\cdot 10^{-6}$}
  &	\multicolumn{1}{c}{$-1.9\cdot 10^{-1}$}
  &	\multicolumn{1}{c}{$9.9\cdot 10^{-3}$}
  &	\multicolumn{1}{c}{$-8.3\cdot 10^{0}$}
  &	\multicolumn{1}{c}{$4.0\cdot 10^{-5}$}
  &	\multicolumn{1}{c}{$-6.2\cdot 10^{-1}$}
  \\
  SXS:BBH:2465~\cite{Yoo:2022erv}	
  &	\multicolumn{1}{c}{$15$}	
  &	\multicolumn{1}{c}{$0.31$}
  & \multicolumn{1}{c}{$1.0\cdot 10^{-5}$}
  &	\multicolumn{1}{c}{$-8.6\cdot 10^{-2}$}	
  & \multicolumn{1}{c}{$6.6\cdot 10^{-3}$}
  &	\multicolumn{1}{c}{$-8.0\cdot 10^{0}$}
  & \multicolumn{1}{c}{$2.4\cdot 10^{-3}$}
  &	\multicolumn{1}{c}{$-5.4\cdot 10^{0}$}
  \\
  SXS:BBH:2464~\cite{Yoo:2022erv}	
  &	\multicolumn{1}{c}{$15$}	
  &	\multicolumn{1}{c}{$0.5$}
  &	\multicolumn{1}{c}{$5.1\cdot 10^{-7}$}
  &	\multicolumn{1}{c}{$-2.5\cdot 10^{-1}$}
  &	\multicolumn{1}{c}{$4.1\cdot 10^{-3}$}
  &	\multicolumn{1}{c}{$-7.3\cdot 10^{0}$}
  &	\multicolumn{1}{c}{$7.6\cdot 10^{-3}$}
  &	\multicolumn{1}{c}{$-9.6\cdot 10^{0}$}
  \\
\midrule
  SXS:BBH:2328~\cite{SXS:catalog}	
  &	\multicolumn{1}{c}{$1$}	
  &	\multicolumn{1}{c}{$-0.5$}	
  &	\multicolumn{1}{c}{$3.5\cdot 10^{-5}$}
  &	\multicolumn{1}{c}{$1.2\cdot 10^{0}$}	
  &	\multicolumn{1}{c}{$8.0\cdot 10^{-2}$}
  &	\multicolumn{1}{c}{$-2.4\cdot 10^{1}$}
  & \multicolumn{1}{c}{$2.0\cdot 10^{-5}$}
  &	\multicolumn{1}{c}{$-6.0\cdot 10^{-1}$}
  \\
  SXS:BBH:2358~\cite{SXS:catalog}	
  &	\multicolumn{1}{c}{$3$}	
  &	\multicolumn{1}{c}{$-0.5$}	
  & \multicolumn{1}{c}{$2.0\cdot 10^{-6}$}
  &	\multicolumn{1}{c}{$8.9\cdot 10^{-1}$}
  & \multicolumn{1}{c}{$6.5\cdot 10^{-2}$}
  &	\multicolumn{1}{c}{$-2.1\cdot 10^{1}$}
  & \multicolumn{1}{c}{$2.8\cdot 10^{-5}$}
  &	\multicolumn{1}{c}{$-1.8\cdot 10^{-1}$}
  \\
  SXS:BBH:0060~\cite{Mroue:2013xna,Boyle:2019kee}	
  &	\multicolumn{1}{c}{$5$}	
  &	\multicolumn{1}{c}{$-0.5$}
  &	\multicolumn{1}{c}{$4.5\cdot 10^{-5}$}
  &	\multicolumn{1}{c}{$-2.2\cdot 10^{-1}$}	
  &	\multicolumn{1}{c}{$2.9\cdot 10^{-2}$}
  &	\multicolumn{1}{c}{$-1.2\cdot 10^{1}$}	
  &	\multicolumn{1}{c}{$5.0\cdot 10^{-5}$}
  &	\multicolumn{1}{c}{$-1.8\cdot 10^{-1}$}
  \\
  SXS:BBH:3623~\cite{SXS:catalog}	
  &	\multicolumn{1}{c}{$8$}	
  &	\multicolumn{1}{c}{$-0.5$}
  &	\multicolumn{1}{c}{$1.4\cdot 10^{-6}$}
  &	\multicolumn{1}{c}{$-6.4\cdot 10^{-3}$}	
  &	\multicolumn{1}{c}{$8.2\cdot 10^{-3}$}
  &	\multicolumn{1}{c}{$-7.5\cdot 10^{0}$}
  & \multicolumn{1}{c}{$5.8\cdot 10^{-5}$}
  &	\multicolumn{1}{c}{$7.4\cdot 10^{-1}$}
  \\
  SXS:BBH:2474~\cite{Yoo:2022erv}	
  &	\multicolumn{1}{c}{$13$}	
  &	\multicolumn{1}{c}{$-0.5$}
  &	\multicolumn{1}{c}{$1.9\cdot 10^{-5}$}
  &	\multicolumn{1}{c}{$-2.9\cdot 10^{-1}$}
  &	\multicolumn{1}{c}{$8.9\cdot 10^{-3}$}
  &	\multicolumn{1}{c}{$-7.2\cdot 10^{0}$}
  &	\multicolumn{1}{c}{$2.0\cdot 10^{-4}$}
  &	\multicolumn{1}{c}{$1.5\cdot 10^{0}$}
  \\
  SXS:BBH:2478~\cite{Yoo:2022erv}	
  &	\multicolumn{1}{c}{$14$}	
  &	\multicolumn{1}{c}{$-0.5$}
  &	\multicolumn{1}{c}{$1.6\cdot 10^{-6}$}
  &	\multicolumn{1}{c}{$-2.8\cdot 10^{-1}$}
  &	\multicolumn{1}{c}{$7.0\cdot 10^{-3}$}
  &	\multicolumn{1}{c}{$-6.7\cdot 10^{0}$} 
  & \multicolumn{1}{c}{$3.6\cdot 10^{-4}$}
  &	\multicolumn{1}{c}{$1.7\cdot 10^{0}$}
  \\
  SXS:BBH:2463~\cite{Yoo:2022erv}	
  &	\multicolumn{1}{c}{$15$}	
  &	\multicolumn{1}{c}{$-0.5$}
  &	\multicolumn{1}{c}{$3.7\cdot 10^{-6}$}
  &	\multicolumn{1}{c}{$-4.5\cdot 10^{-1}$}	
  &	\multicolumn{1}{c}{$6.7\cdot 10^{-3}$}
  &	\multicolumn{1}{c}{$-6.7\cdot 10^{0}$}
  & \multicolumn{1}{c}{$3.6\cdot 10^{-3}$}
  &	\multicolumn{1}{c}{$1.7\cdot 10^{0}$}
  \\
\midrule 
  SXS:BBH:1165~\cite{Boyle:2019kee}	
  &	\multicolumn{1}{c}{$2$}	
  &	\multicolumn{1}{c}{$0$}	
  &	\multicolumn{1}{c}{$2.0\cdot 10^{-5}$}
  &	\multicolumn{1}{c}{$1.4\cdot 10^{0}$}
  &	\multicolumn{1}{c}{$7.6\cdot 10^{-2}$}
  &	\multicolumn{1}{c}{$-2.8\cdot 10^{1}$}
  & \multicolumn{1}{c}{$5.1\cdot 10^{-5}$}
  &	\multicolumn{1}{c}{$-1.2\cdot 10^{0}$}
  \\
  SXS:BBH:0259~\cite{Chu:2015kft,Boyle:2019kee}	
  &	\multicolumn{1}{c}{$3.5$}	
  &	\multicolumn{1}{c}{$0$}	
  &	\multicolumn{1}{c}{$9.1\cdot 10^{-6}$}
  &	\multicolumn{1}{c}{$2.2\cdot 10^{0}$}
  &	\multicolumn{1}{c}{$3.8\cdot 10^{-2}$}
  &	\multicolumn{1}{c}{$-1.8\cdot 10^{1}$}
  &	\multicolumn{1}{c}{$8.3\cdot 10^{-5}$}
  &	\multicolumn{1}{c}{$-1.3\cdot 10^{-1}$}
  \\
  SXS:BBH:1906~\cite{Varma:2019csw,Boyle:2019kee}	
  &	\multicolumn{1}{c}{$4$}	
  &	\multicolumn{1}{c}{$0$}
  &	\multicolumn{1}{c}{$1.3\cdot 10^{-6}$}
  &	\multicolumn{1}{c}{$-3.1\cdot 10^{-1}$}
  & \multicolumn{1}{c}{$1.4\cdot 10^{-2}$}
  &	\multicolumn{1}{c}{$-1.1\cdot 10^{1}$}
  & \multicolumn{1}{c}{$1.1\cdot 10^{-4}$}
  &	\multicolumn{1}{c}{$-1.8\cdot 10^{0}$}
  \\
  SXS:BBH:1220~\cite{Blackman:2015pia,Boyle:2019kee}	
  &	\multicolumn{1}{c}{$4$}	
  &	\multicolumn{1}{c}{$0$}
  &	\multicolumn{1}{c}{$4.9\cdot 10^{-6}$}
  &	\multicolumn{1}{c}{$1.3\cdot 10^{-1}$}	
  &	\multicolumn{1}{c}{$2.4\cdot 10^{-2}$}
  &	\multicolumn{1}{c}{$-1.5\cdot 10^{1}$}
  &	\multicolumn{1}{c}{$9.1\cdot 10^{-5}$}
  &	\multicolumn{1}{c}{$-1.6\cdot 10^{0}$}
  \\
    SXS:BBH:3630~\cite{SXS:catalog}	
  &	\multicolumn{1}{c}{$6$}	
  &	\multicolumn{1}{c}{$0$}
  &	\multicolumn{1}{c}{$1.7\cdot 10^{-6}$}
  &	\multicolumn{1}{c}{$-5.2\cdot 10^{-1}$}
  &	\multicolumn{1}{c}{$1.2\cdot 10^{-2}$}
  &	\multicolumn{1}{c}{$-1.0\cdot 10^{1}$}
  & \multicolumn{1}{c}{$1.1\cdot 10^{-4}$}
  &	\multicolumn{1}{c}{$-1.4\cdot 10^{0}$}
  \\[.33em]
    SXS:BBH:0298~\cite{Chu:2015kft,Boyle:2019kee}	
  &	\multicolumn{1}{c}{$7$}	
  &	\multicolumn{1}{c}{$0$}
  &	\multicolumn{1}{c}{$7.4\cdot 10^{-7}$}
  &	\multicolumn{1}{c}{$9.2\cdot 10^{-2}$}
  &	\multicolumn{1}{c}{$1.1\cdot 10^{-2}$}
  &	\multicolumn{1}{c}{$-8.3\cdot 10^{0}$}
  & \multicolumn{1}{c}{$1.2\cdot 10^{-4}$}
  &	\multicolumn{1}{c}{$-9.7\cdot 10^{-1}$}
  \\
    SXS:BBH:0299~\cite{Chu:2015kft,Boyle:2019kee}	
  &	\multicolumn{1}{c}{$7.5$}	
  &	\multicolumn{1}{c}{$0$}
  &	\multicolumn{1}{c}{$1.3\cdot 10^{-6}$}
  &	\multicolumn{1}{c}{$2.7\cdot 10^{-1}$}
  &	\multicolumn{1}{c}{$1.1\cdot 10^{-2}$}
  &	\multicolumn{1}{c}{$-8.4\cdot 10^{0}$}
  & \multicolumn{1}{c}{$1.2\cdot 10^{-4}$}
  &	\multicolumn{1}{c}{$-8.1\cdot 10^{-1}$}
  \\
    SXS:BBH:0300~\cite{Chu:2015kft,Boyle:2019kee}	
  &	\multicolumn{1}{c}{$8.5$}	
  &	\multicolumn{1}{c}{$0$}
  &	\multicolumn{1}{c}{$1.4\cdot 10^{-6}$}
  &	\multicolumn{1}{c}{$3.1\cdot 10^{-1}$}
  &	\multicolumn{1}{c}{$1.1\cdot 10^{-2}$}
  &	\multicolumn{1}{c}{$-8.3\cdot 10^{0}$}
  & \multicolumn{1}{c}{$1.2\cdot 10^{-4}$}
  &	\multicolumn{1}{c}{$-7.7\cdot 10^{-1}$}
  \\
    SXS:BBH:0199~\cite{Blackman:2015pia,Boyle:2019kee}	
  &	\multicolumn{1}{c}{$8.7$}	
  &	\multicolumn{1}{c}{$0$}
  &	\multicolumn{1}{c}{$7.0\cdot 10^{-6}$}
  &	\multicolumn{1}{c}{$-8.2\cdot 10^{-2}$}
  &	\multicolumn{1}{c}{$1.1\cdot 10^{-2}$}
  &	\multicolumn{1}{c}{$-8.9\cdot 10^{0}$}
  & \multicolumn{1}{c}{$1.5\cdot 10^{-4}$}
  &	\multicolumn{1}{c}{$-1.0\cdot 10^{0}$}
  \\
    SXS:BBH:0301~\cite{Chu:2015kft,Boyle:2019kee}	
  &	\multicolumn{1}{c}{$9$}	
  &	\multicolumn{1}{c}{$0$}
  &	\multicolumn{1}{c}{$2.2\cdot 10^{-6}$}
  &	\multicolumn{1}{c}{$-2.4\cdot 10^{-1}$}
  &	\multicolumn{1}{c}{$5.2\cdot 10^{-3}$}
  &	\multicolumn{1}{c}{$-7.0\cdot 10^{0}$}
  & \multicolumn{1}{c}{$6.1\cdot 10^{-5}$}
  &	\multicolumn{1}{c}{$-1.0\cdot 10^{0}$}
  \\[.33em]
        SXS:BBH:1107~\cite{Boyle:2019kee}	
  &	\multicolumn{1}{c}{$10$}	
  &	\multicolumn{1}{c}{$0$}
  &	\multicolumn{1}{c}{$9.4\cdot 10^{-6}$}
  &	\multicolumn{1}{c}{$-9.2\cdot 10^{-1}$}
  &	\multicolumn{1}{c}{$1.8\cdot 10^{-2}$}
  &	\multicolumn{1}{c}{$-1.2\cdot 10^{1}$}
  & \multicolumn{1}{c}{$1.8\cdot 10^{-4}$}
  &	\multicolumn{1}{c}{$-2.0\cdot 10^{0}$}
  \\
        SXS:BBH:0303~\cite{Chu:2015kft,Boyle:2019kee}	
  &	\multicolumn{1}{c}{$10$}	
  &	\multicolumn{1}{c}{$0$}
  &	\multicolumn{1}{c}{$9.4\cdot 10^{-7}$}
  &	\multicolumn{1}{c}{$-3.1\cdot 10^{-1}$}
  &	\multicolumn{1}{c}{$4.9\cdot 10^{-3}$}
  &	\multicolumn{1}{c}{$-6.9\cdot 10^{0}$}
  & \multicolumn{1}{c}{$5.3\cdot 10^{-5}$}
  &	\multicolumn{1}{c}{$-9.6\cdot 10^{-1}$}
  \\
        SXS:BBH:2480~\cite{Yoo:2022erv}	
  &	\multicolumn{1}{c}{$14$}	
  &	\multicolumn{1}{c}{$0$}
  &	\multicolumn{1}{c}{$7.4\cdot 10^{-6}$}
  &	\multicolumn{1}{c}{$-7.8\cdot 10^{-1}$}
  &	\multicolumn{1}{c}{$1.0\cdot 10^{-2}$}
  &	\multicolumn{1}{c}{$-8.7\cdot 10^{0}$}
  & \multicolumn{1}{c}{$1.1\cdot 10^{-4}$}
  &	\multicolumn{1}{c}{$-1.3\cdot 10^{0}$}
  \\
        SXS:BBH:2477~\cite{Yoo:2022erv}	
  &	\multicolumn{1}{c}{$15$}	
  &	\multicolumn{1}{c}{$0$}
  &	\multicolumn{1}{c}{$2.7\cdot 10^{-6}$}
  &	\multicolumn{1}{c}{$-1.9\cdot 10^{-1}$}
  &	\multicolumn{1}{c}{$9.9\cdot 10^{-3}$}
  &	\multicolumn{1}{c}{$-8.3\cdot 10^{0}$}
  & \multicolumn{1}{c}{$4.0\cdot 10^{-5}$}
  &	\multicolumn{1}{c}{$-6.2\cdot 10^{-1}$}
  \\
\midrule 
    SXS:BBH:2329~\cite{SXS:catalog}	
  &	\multicolumn{1}{c}{$1$}	
  &	\multicolumn{1}{c}{$0.5$}
  &	\multicolumn{1}{c}{$1.1\cdot 10^{-5}$}
  &	\multicolumn{1}{c}{$1.5\cdot 10^{0}$}
  &	\multicolumn{1}{c}{$2.6\cdot 10^{-4}$}
  &	\multicolumn{1}{c}{$-1.7\cdot 10^{0}$}
  & \multicolumn{1}{c}{$1.1\cdot 10^{-4}$}
  &	\multicolumn{1}{c}{$-2.7\cdot 10^{0}$}
  \\
  SXS:BBH:2427~\cite{SXS:catalog}	
  &	\multicolumn{1}{c}{$3$}	
  &	\multicolumn{1}{c}{$0.5$}	
  &	\multicolumn{1}{c}{$3.5\cdot 10^{-6}$}
  &	\multicolumn{1}{c}{$-8.6\cdot 10^{-1}$}
  &	\multicolumn{1}{c}{$1.0\cdot 10^{-2}$}
  &	\multicolumn{1}{c}{$-1.4\cdot 10^{1}$}
  &	\multicolumn{1}{c}{$6.1\cdot 10^{-4}$}
  &	\multicolumn{1}{c}{$-5.6\cdot 10^{0}$}
  \\
  SXS:BBH:2385~\cite{SXS:catalog}	
  &	\multicolumn{1}{c}{$5$}	
  &	\multicolumn{1}{c}{$0.5$}
  &	\multicolumn{1}{c}{$4.4\cdot 10^{-6}$}	
  &	\multicolumn{1}{c}{$-1.1\cdot 10^{-1}$}
  &	\multicolumn{1}{c}{$4.2\cdot 10^{-3}$}
  &	\multicolumn{1}{c}{$-8.0\cdot 10^{0}$}
  &	\multicolumn{1}{c}{$1.2\cdot 10^{-3}$}
  &	\multicolumn{1}{c}{$-5.2\cdot 10^{0}$}
  \\
  SXS:BBH:0065~\cite{Mroue:2013xna,Boyle:2019kee}	
  &	\multicolumn{1}{c}{$8$}	
  &	\multicolumn{1}{c}{$0.5$}
  &	\multicolumn{1}{c}{$4.7\cdot 10^{-5}$}
  &	\multicolumn{1}{c}{$-1.2\cdot 10^{0}$}
  & \multicolumn{1}{c}{$6.9\cdot 10^{-3}$}
  &	\multicolumn{1}{c}{$-1.1\cdot 10^{1}$}
  & \multicolumn{1}{c}{$2.8\cdot 10^{-3}$}
  &	\multicolumn{1}{c}{$-8.1\cdot 10^{0}$}
  \\
  SXS:BBH:2469~\cite{Yoo:2022erv}	
  &	\multicolumn{1}{c}{$10$}	
  &	\multicolumn{1}{c}{$0.5$}
  &	\multicolumn{1}{c}{$7.3\cdot 10^{-6}$}
  &	\multicolumn{1}{c}{$-2.5\cdot 10^{-1}$}	
  &	\multicolumn{1}{c}{$4.3\cdot 10^{-3}$}
  &	\multicolumn{1}{c}{$-8.2\cdot 10^{0}$}
  &	\multicolumn{1}{c}{$3.6\cdot 10^{-3}$}
  &	\multicolumn{1}{c}{$-7.8\cdot 10^{0}$}
  \\[.33em]
    SXS:BBH:2467~\cite{Yoo:2022erv}	
  &	\multicolumn{1}{c}{$11.5$}	
  &	\multicolumn{1}{c}{$0.5$}
  &	\multicolumn{1}{c}{$1.0\cdot 10^{-5}$}
  &	\multicolumn{1}{c}{$-1.9\cdot 10^{-1}$}
  &	\multicolumn{1}{c}{$4.4\cdot 10^{-3}$}
  &	\multicolumn{1}{c}{$-7.7\cdot 10^{0}$}
  & \multicolumn{1}{c}{$5.0\cdot 10^{-3}$}
  &	\multicolumn{1}{c}{$-8.3\cdot 10^{0}$}
  \\
    SXS:BBH:2476~\cite{Yoo:2022erv}	
  &	\multicolumn{1}{c}{$13.2$}	
  &	\multicolumn{1}{c}{$0.5$}
  &	\multicolumn{1}{c}{$1.4\cdot 10^{-6}$}
  &	\multicolumn{1}{c}{$-7.7\cdot 10^{-1}$}
  &	\multicolumn{1}{c}{$4.5\cdot 10^{-3}$}
  &	\multicolumn{1}{c}{$-8.1\cdot 10^{0}$}
  & \multicolumn{1}{c}{$6.4\cdot 10^{-3}$}
  &	\multicolumn{1}{c}{$-9.4\cdot 10^{0}$}
  \\
    SXS:BBH:2482~\cite{Yoo:2022erv}	
  &	\multicolumn{1}{c}{$14$}	
  &	\multicolumn{1}{c}{$0.5$}
  &	\multicolumn{1}{c}{$5.8\cdot 10^{-6}$}
  &	\multicolumn{1}{c}{$-4.3\cdot 10^{-1}$}
  &	\multicolumn{1}{c}{$3.6\cdot 10^{-3}$}
  &	\multicolumn{1}{c}{$-7.3\cdot 10^{0}$}
  & \multicolumn{1}{c}{$6.1\cdot 10^{-3}$}
  &	\multicolumn{1}{c}{$-9.2\cdot 10^{0}$}
  \\
    SXS:BBH:2464~\cite{Yoo:2022erv}	
  &	\multicolumn{1}{c}{$15$}	
  &	\multicolumn{1}{c}{$0.5$}
  &	\multicolumn{1}{c}{$5.1\cdot 10^{-7}$}
  &	\multicolumn{1}{c}{$-2.5\cdot 10^{-1}$}
  &	\multicolumn{1}{c}{$4.1\cdot 10^{-3}$}
  &	\multicolumn{1}{c}{$-7.3\cdot 10^{0}$}
  & \multicolumn{1}{c}{$7.6\cdot 10^{-3}$}
  &	\multicolumn{1}{c}{$-9.6\cdot 10^{0}$}
  \\
\bottomrule
\end{tabular}%
}
\caption{\label{tab:mismatchdephasing}Mismatches $\cal M$ and dephasings $\deltaE\psi_{22}$ of the models $X\in\{H,\text{0PA},\text{PN}\}$ against SXS simulations, as calculated from the maximum value of the mismatch~\eqref{eq:mismatch} in the range $\MTOT\in[10M_\odot,300M_\odot]$ and from Eq.~\eqref{eq:dephasing}. In contrast to Figs.~\ref{fig:dephasingprograde}--\ref{fig:dephasingSchwarzschild}, here the dephasing is computed on a fixed (dimensionless) frequency interval (the ``$\Omega_0$'' comparison described in the text).}
\end{table*}%

\begin{figure*}[tp]
\includegraphics[width=.95\textwidth,trim={0 10pt 0 0}]{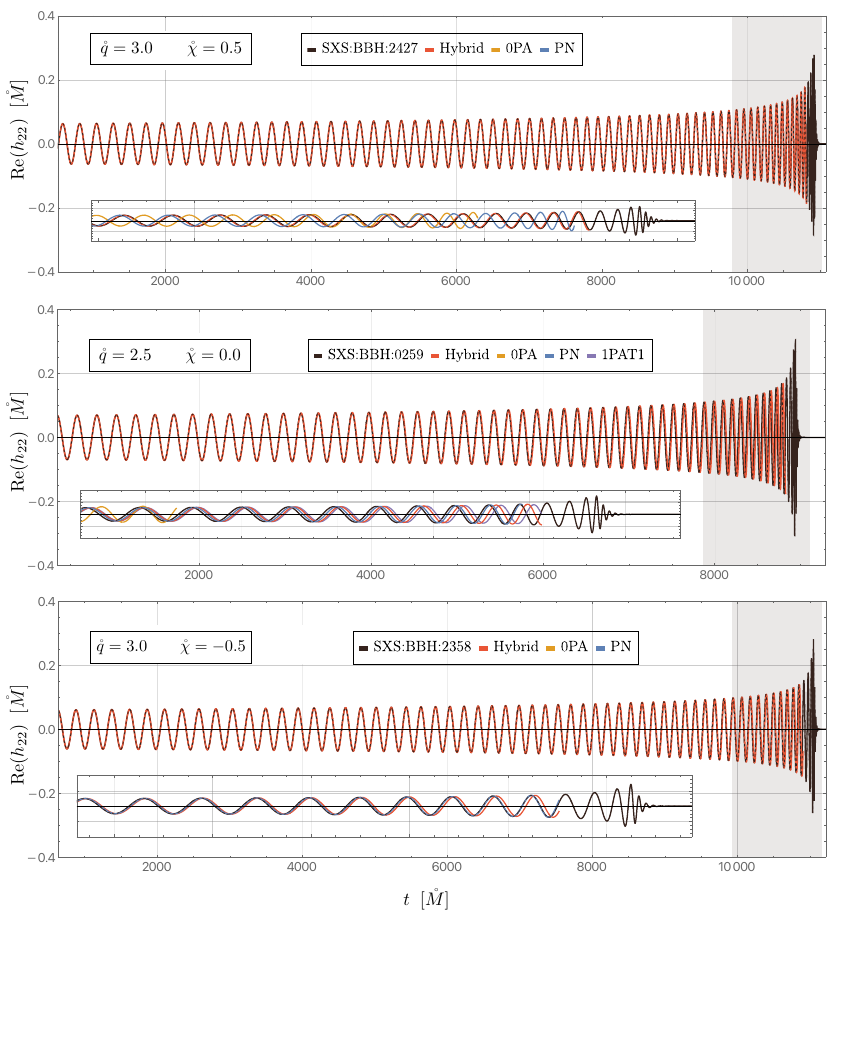}\vspace{-2.7cm}
\caption{\label{WF:Prograde}$(\ell,m)=(2,2)$ mode of the gravitational waveform from three comparable-mass binary configurations with differing primary spins. Numerical relativity (in black), hybrid (red), $0$PA (orange), PN (blue), and \texttt{1PAT1} (purple) waveforms are displayed. The references of the SXS simulations are listed in Table \ref{tab:mismatchdephasing}.}
\end{figure*}

\begin{figure*}[tp]
\includegraphics[width=.95\textwidth,trim={0 15pt 0 0}]{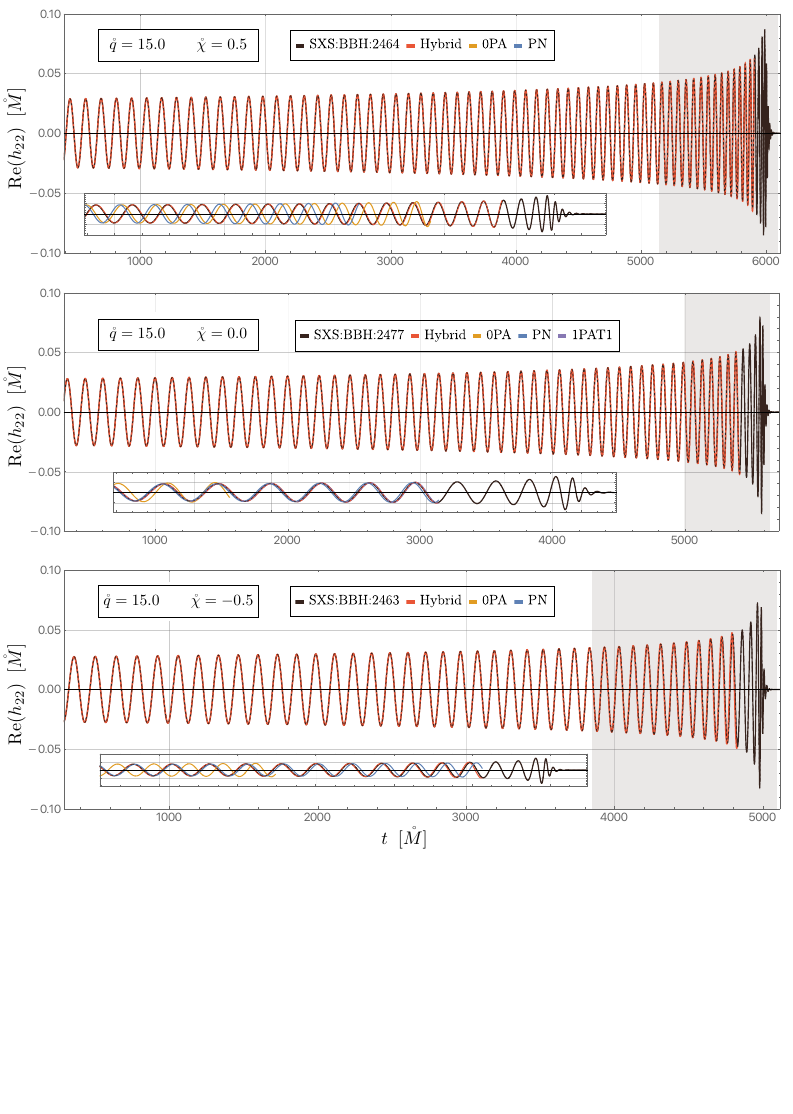}\vspace{-5cm}
\caption{\label{WF:Schw}$(\ell,m)=(2,2)$ mode of the gravitational waveform from three $\mathring q=15$ binary configurations with differing primary spins. Numerical relativity (in black), hybrid (red), $0$PA (orange), PN (blue), and \texttt{1PAT1} (purple) waveforms are displayed. The references of the SXS simulations are listed in Table \ref{tab:mismatchdephasing}.}
\end{figure*}

\subsubsection{Qualitative comparison}\label{subsubsec:qualitative}

We align the model $X$ waveform against the NR waveform using the alignment procedure of Sec.~\ref{subsec:WFalig}, for the three models $X\in \{H,0\text{PA},\text{PN}\}$ defined in Sec.~\ref{subsubsec:models}. A sample of such aligned time-domain waveforms are shown in Figs.~\ref{WF:Prograde} and~\ref{WF:Schw}. We show the waveforms for the dominant $(2,2)$ mode for comparable-mass binaries ($\mathring{q}=2.5$ or 3) in Fig.~\ref{WF:Prograde} and for asymmetric binaries with $\mathring{q}=15$ in Fig.~\ref{WF:Schw}. In both cases we show waveforms with primary spin $\mathring\chi=+0.5$ (prograde), 0, and $-0.5$ (retrograde). In the nonspinning case, we include \texttt{1PAT1} waveforms for comparison. While we only display the $(2,2)$ mode, we find the accuracy of the hybrid model is essentially the same for other modes.

The general trends we observe are the following: First, the adiabatic ($0$PA) waveforms dephase very quickly during the inspiral, accumulating a few radians of error before reaching the ISCO, where the two-timescale expansion breaks down. This low accuracy is apparent for all mass ratios we consider, as is to be expected from the small-mass-ratio expansion of the accumulated waveform phase in Eq.~\eqref{eq:accphase}: a 0PA model incurs a mass-ratio-independent phase error (when considering different mass ratios on a fixed interval of dimensionless frequency~\cite{Albertini:2022rfe}). 

Second, for comparable masses, the PN waveforms typically remain in phase with the NR waveform until the very late stage of the inspiral. In some rare cases, the PN waveform even performs better than the hybrid model. However, for binaries with large prograde spins, which evolve to much smaller separations before merger, PN can fare poorly even for comparable masses, as exhibited in the top panel of Fig.~\ref{WF:Prograde}. Importantly in our context, the PN waveform also worsens considerably for more asymmetric binaries. This is again to be expected from the small-mass-ratio expansion of the phase in Eq.~\eqref{eq:accphase}: a PN model will always lack high-PN terms in the 0PA forcing terms of the evolution equations~\eqref{eq:ODE}, resulting in a phase error that grows linearly with $\mathring q$. However, this degradation is significantly less severe than for lower-order PN waveforms; compare Fig.~\ref{WF:Schw} here to Fig.~2 of the Supplementary Material in Ref.~\cite{Wardell:2021fyy}. In addition to benefiting from the inclusion of 4PN terms, the PN waveforms might also be more accurate due our resummed form of the PN waveform generation. 

Finally, the hybrid model also remains very well in phase with the NR waveforms until the late inspiral stage. In particular, for $\mathring q\gtrsim6$, the hybrid model consistently outperforms the PN and $0$PA models, for all values of the primary spin.
Moreover, compared to the PN model, the waveform dephasings do not worsen as we go to more asymmetric binaries. This is also to be expected: the dynamics of the hybrid model is complete at 0PA order but not at 1PA order. Therefore, the dephasing should go to a constant, as for the $0$PA model, as we move to more asymmetric binaries. However, the value of this constant residual $\varphi_{(1)}(\mathring\varepsilon t)$ is dramatically lower for the hybrid than for the $0$PA model. We address this question quantitatively in the next subsection.

\subsubsection{Waveform dephasing}\label{sec:wfdephasing}

Once again, we align the model $X$ waveforms against the NR waveforms using the alignment procedure of Sec.~\ref{subsec:WFalig}, for the three models $X\in \{H,0\text{PA},\text{PN}\}$ defined in Sec.~\ref{subsubsec:models}. To lighten the notation, we drop $\deltaA t_\text{min}$ and $\deltaA\varphi_\text{min}$ from  the expressions in Eqs.~\eqref{eq:halign} and~\eqref{eq:psialign} and refer to the aligned waveform and phase as $h_{\ell m}^X$ and $\psi_{\ell m}^X$. We compute the $(2,2)$-mode waveform phase error $\deltaE\psi_{22}^X(t)$ as a function of time as 
\begin{equation}\label{eq:dephasing}
\deltaE\psi_{22}^X(t)=\left[\psi^X_{22}(t)-\psi^{NR}_{22}(t)\right]-\left[\psi^X_{22}(t_0)-\psi^{NR}_{22}(t_0)\right],
\end{equation}
where we have set the dephasing to be $0$ at the initial time $t_0$. 

\begin{figure}[tp]
 \center
 \includegraphics[width=\columnwidth]{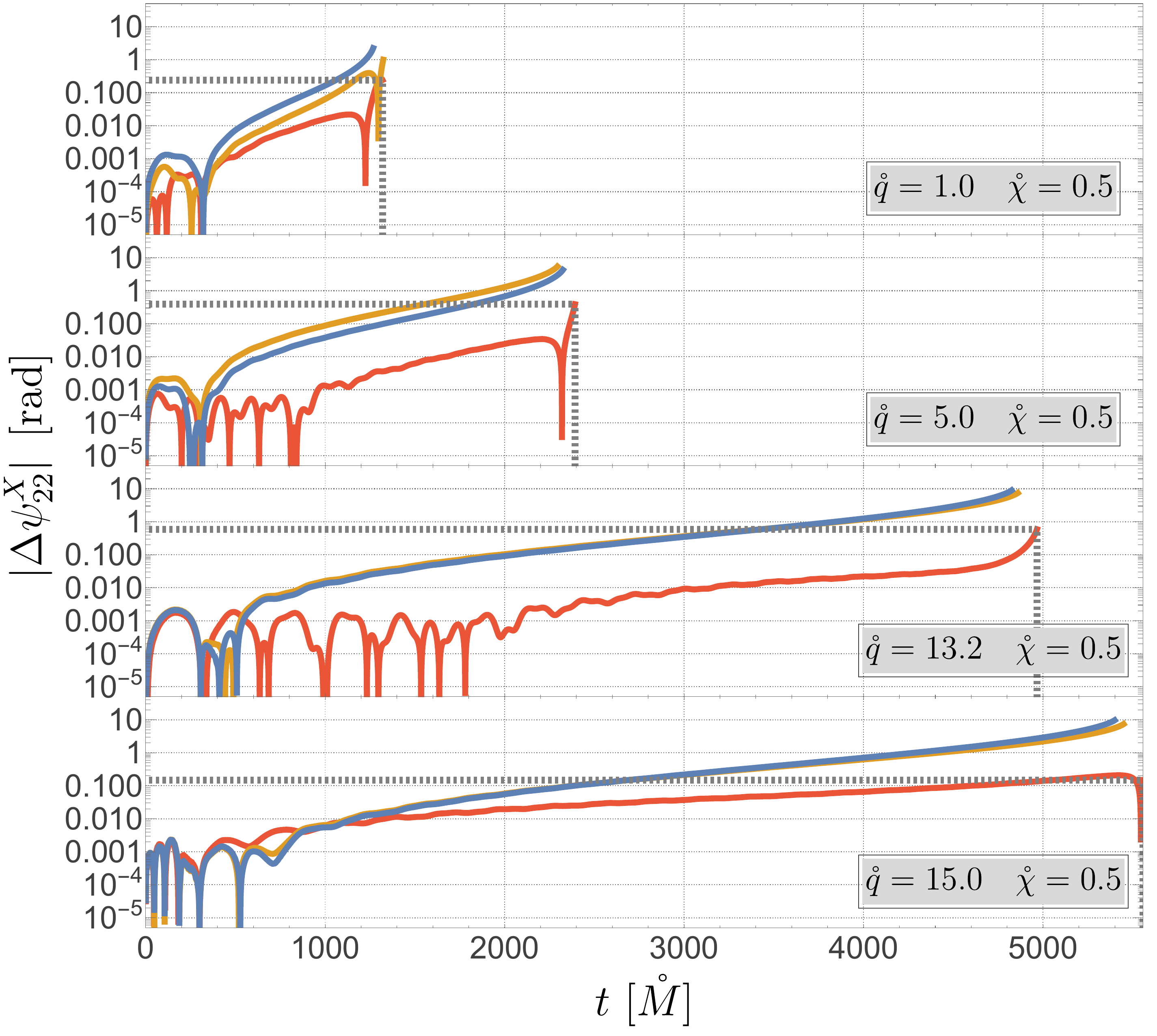}
 \caption{\label{fig:dephasingprograde}Accumulated phase error, as defined through Eq.~\eqref{eq:dephasing}, of the hybrid (red), PN (blue), and 0PA (orange) waveform models against NR simulations for a set of prograde quasicircular orbits with primary spin $\mathring \chi=+0.5$. The mass ratio ranges from $\mathring q=1$ (top panel) to $\mathring q=15$ (bottom panel). The grey horizontal lines indicate the value of the phase error accumulated when the hybrid model reaches the primary ISCO frequency: $\deltaE\psi_{22}^H(t_\star)$ where $\omega(t_\star) = \Omega_\star(\mathring \chi)$. The SXS simulations used in this figure are those with $\mathring \chi=0.5$ in Table~\ref{tab:mismatchdephasing}.} 
\end{figure}
\begin{figure}[tp]
 \center
 \includegraphics[width=\columnwidth]{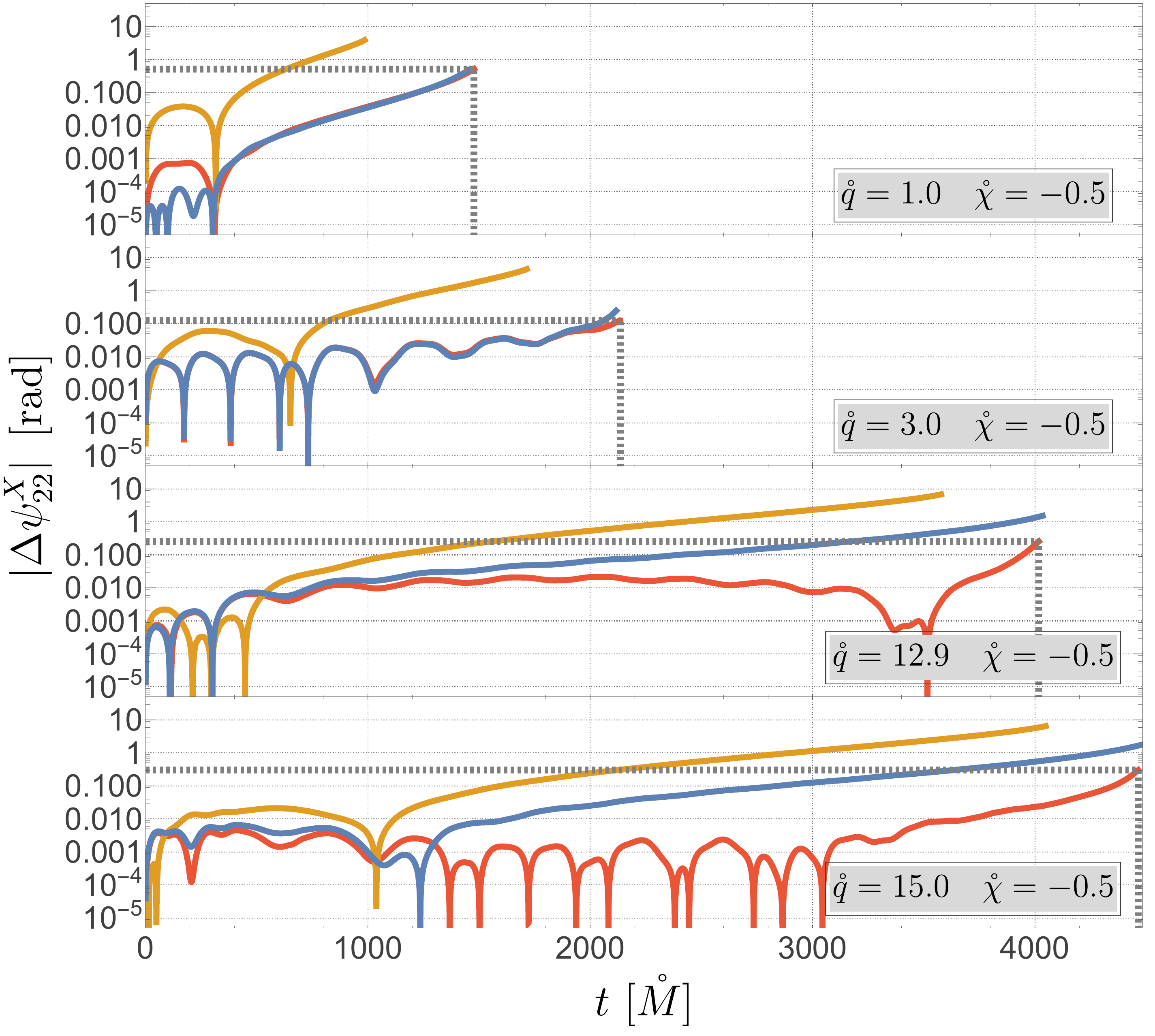}
 \caption{\label{fig:dephasingretrograde}Accumulated phase error of the hybrid (red), PN (blue), and 0PA (orange) waveform models against NR simulations for a set of retrograde quasicircular orbits with primary spin $\mathring \chi=-0.5$, and with other details as in Fig.~\ref{fig:dephasingprograde}. The SXS simulations used in this figure are those with $\mathring \chi=-0.5$ in Table \ref{tab:mismatchdephasing}.
 } 
 \end{figure}
 \begin{figure}[tp]
 \center
 \includegraphics[width=\columnwidth]{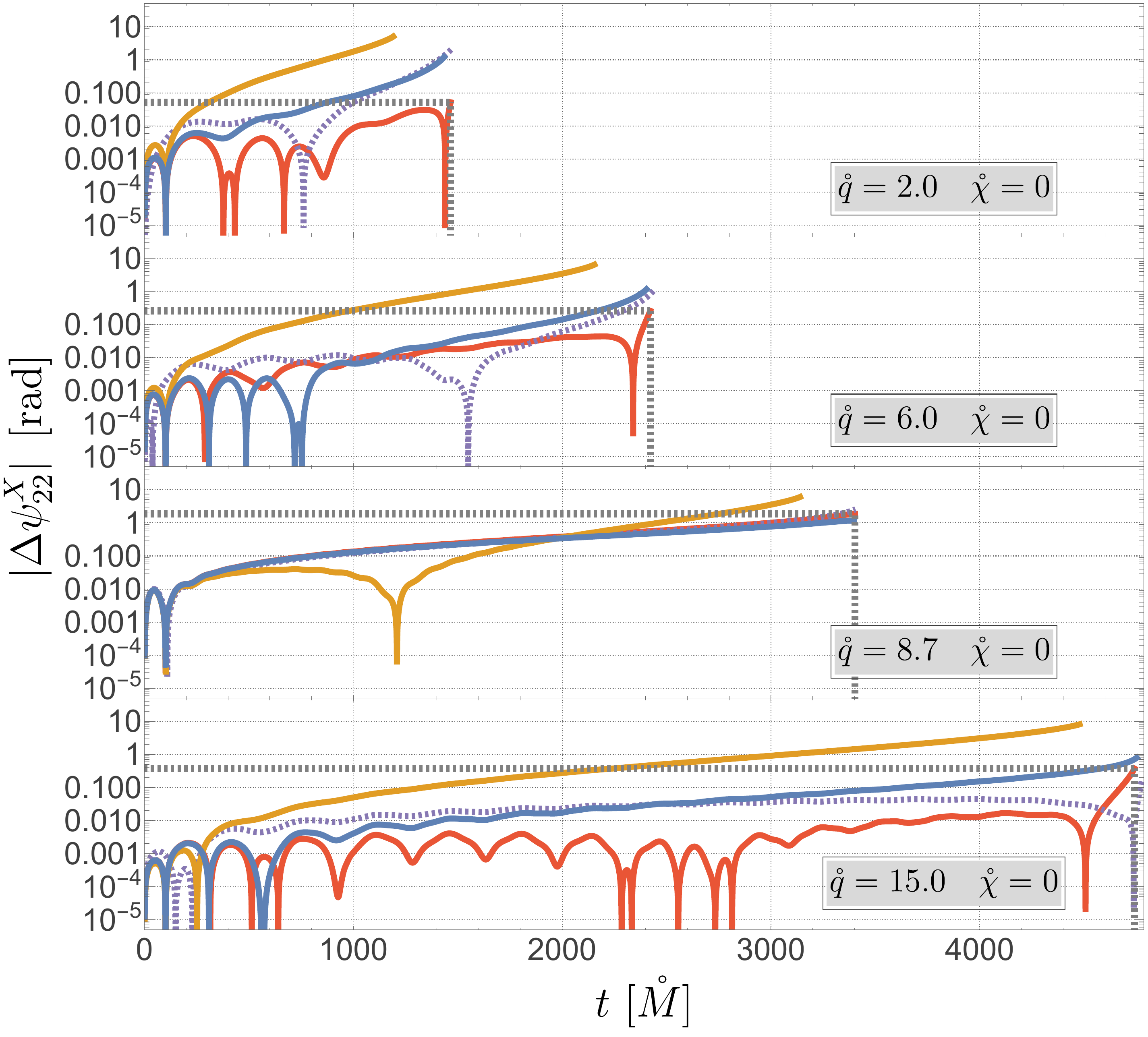}
 \caption{\label{fig:dephasingSchwarzschild}Accumulated phase error of the hybrid (red), PN (blue), and 0PA (orange) waveform models against NR simulations for a set of nonspinning quasicircular binaries, with the dephasing of the \texttt{1PAT1} model (dotted purple) against the same NR simulations also included for comparison. Other details are as in Fig.~\ref{fig:dephasingprograde}.  The SXS simulations used in this figure are those with $\mathring \chi=0$ in Table \ref{tab:mismatchdephasing}.} 
 \end{figure}

Figures~\ref{fig:dephasingprograde}--\ref{fig:dephasingSchwarzschild} illustrate three ``Fixed$\chi$'' comparisons. Figures \ref{fig:dephasingprograde} and~\ref{fig:dephasingretrograde} show the accumulated phase error for a set of simulations with  $\mathring\chi=+0.5$ (prograde orbits) and $\mathring\chi=-0.5$ (retrograde orbits), respectively. We see the hybrid model always has an accumulated phase error of $\lesssim1$ radian over the full inspiral. Figure \ref{fig:dephasingSchwarzschild} shows the accumulated phase error for nonspinning binaries, for which we also display the dephasing that the \texttt{1PAT1} model of~\cite{Wardell:2021fyy} accumulates against the same NR simulations. At least at the mass ratios considered $\mathring q\lesssim 15$, the hybrid model competes very well with the full \texttt{1PAT1} model; as noted above, this appears to indicate the significant increase in accuracy from the inclusion of 4PN terms and/or from our resummed form of the PN waveform generation. The hybrid model dephases less than the \texttt{1PAT1} model for comparable-mass binaries, as it includes high-PA information from PN expansions. However, the \texttt{1PAT1} model gets more and more accurate and becomes slightly more reliable than the hybrid at more asymmetric mass ratios. This is an expected behavior that will remain true for EMRIs, as the \texttt{1PAT1} phase error goes to zero like $1/\mathring q$ (a 2PA error), while the hybrid's phase error goes to a constant (a $1$PA error).\\ \\

A critical question is the size of this 1PA error in the hybrid model, as it can indicate how well the hybrid would perform for EMRIs. From the total phase error of the hybrid models in Figs.~\ref{fig:dephasingprograde}--\ref{fig:dephasingSchwarzschild}, we can extract the following rough estimate for the hybrid's constant 1PA residual: $0.5\text{ rad}$ for $\mathring\chi=0.5$ and $[x_i,x_f]=[0.086,0.228]$; $0.3 \text{ rad}$ for $\mathring\chi=-0.5$ and $[x_i,x_f]=[0.079,0.135]$; and $ 0.4\text{ rad}$ for $\mathring\chi=0$ and $[x_i,x_f]=[0.091,0.167]$. In contrast to these small errors, the $0$PA and PN errors scale, as expected, as a (large) constant and as a $1/\mathring\nu$ power law, respectively.

Finally, we consider ``$\Omega_0$'' comparisons, computing the total dephasing between the model $X$ and the NR waveforms on the entire frequency range of the NR waveform inspiral, starting from the reference frequency all the way up to the primary's ISCO frequency. The values of the dephasing can be found in Table~\ref{tab:mismatchdephasing}. Since the frequency intervals under which we compare the models now differ from one simulation to another, it is hard to meaningfully compare between different configurations, but the values in Table~\ref{tab:mismatchdephasing} offer a notion of the relative performance of the hybrid model against the $0$PA and PN models on the full frequency range of the NR inspiral signal.

\subsubsection{Mismatches}

Next, we compute the mismatch of model $X$ against NR as
\begin{equation}\label{eq:mismatch}
	\mathcal{M}_X=1-\max_{\deltaA t,\deltaA\varphi}\frac{\langle h^X_{\deltaA t,\deltaA \varphi},h^\text{NR}\rangle}{\| h^X\|\| h^\text{NR}\|},
\end{equation}
where~\cite{LIGOScientific:2019hgc}
\begin{multline}
	\langle h^X_{\deltaA t,\deltaA \varphi},h^\text{NR}\rangle \\
    =\frac{2}{\pi} \text{Re}\int_{\omega_i}^{\omega_f}  \frac{\left({\mathcal{F}[h^X_{\deltaA t,\deltaA \varphi}]}(\omega)\right)^* {\mathcal{F}[h^\text{NR}]}(\omega)}{S_n(\omega/(2\pi))}d\omega,
\end{multline}
and $h^X_{\deltaA t,\deltaA \varphi}=h_{22}^X(t+\deltaA t)e^{-i\deltaA\varphi}$ is the model $X$ waveform translated in time and phase by $\deltaA t$ and $\deltaA \varphi$, respectively. $\mathcal{F}[h^X]$ and $\mathcal{F}[h^\text{NR}]$ are the Fourier transforms of the model $X$ and NR waveforms, respectively. In this work we take $S_n(f)$ as the power spectral density of the AdVirgo$+$ detector for Run O5 from~\cite{KAGRA:2013rdx}  defined in the range $10-10000\text{Hz}$ (located on the LIGO DCC~\cite{LVK_PSD}). As we are optimizing different quantities, the optimized values of $\deltaA t$, $\deltaA \varphi$ generically differ from the values obtained from the waveform alignment procedure described earlier. 

As before, we perform two sets of mismatch analyses. In the first set ``Fixed$\chi$'', the model $X$ and NR signals are examined on a fixed waveform frequency range starting from the maximal reference initial frequency $\omega_\text{NR}(t_i)$ among all simulations selected and up to the primary ISCO frequency $\Omega_{\star}(\mathring\chi)$. From a stationary phase approximation, we expect that in the Fourier domain, only frequencies $\omega$ close to the frequencies $m\, \omega_{X}(t)$ will be excited; see, for example, the discussion around Eq.~(3.1) of Ref.~\cite{Arun:2004hn}. This motivates us to choose the bounds $\omega_i$ and $\omega_f$ of the discrete Fourier transform as follows: we choose $\omega_i$ to be a small fraction (for definiteness $1/5$) of the maximal reference initial frequency $\omega_\text{NR}(t_i)$ among all simulations selected and $\omega_f$ to be a multiple (for definiteness $1.1$) of the primary's lightring frequency $\Omega_{l.r}(\mathring\chi)$. We checked that the Fourier-domain waveforms do not depend significantly upon slightly varying the choices of $\omega_i$ and $\omega_f$.

\begin{figure}[!htb]
 \center
  \includegraphics[width=0.47\textwidth]{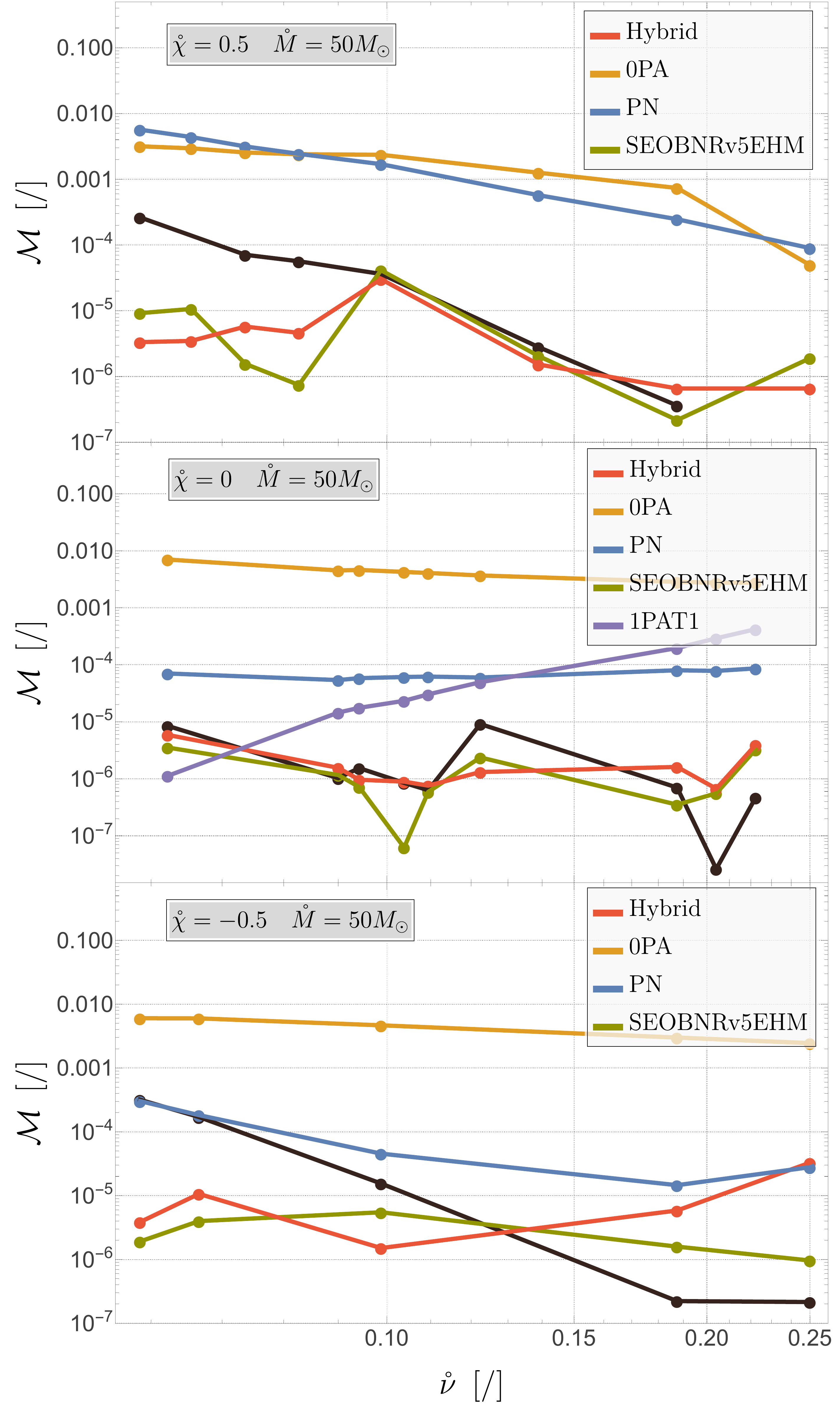}
\caption{\label{fig:MismatchPrograde}Mismatches of the hybrid, 0PA, PN, SEOBNRv5, and \texttt{1PAT1} models against NR waveforms for prograde ($\mathring\chi=+0.5$, top panel), nonspinning (middle panel), and retrograde ($\mathring\chi=-0.5$, bottom panel) binaries. Mass ratio ranges from $\mathring q=1$ ($\mathring\nu=0.25$) to $\mathring q=15$ ($\mathring \nu\sim0.059$), with fixed total mass $50M_\odot$. The sensitivity curve is that of the AdVirgo$+$ detector for Run O5 from~\cite{KAGRA:2013rdx}. The simulations used here are those with IDs ending 2329, 2427, 2385, 0065, 2469, 2467, 2476, 2464 (top panel), 3630, 0259, 0298, 0299, 0300, 0301, 1165, 2265, 2480 (middle panel), 2328, 2358, 3623, 2474, 2463 (bottom panel) in Table \ref{tab:mismatchdephasing}. SEOB waveforms were generated using version 0.3.1 of the \texttt{pyseobnr} package~\cite{Mihaylov:2023bkc,Ramos-Buades:2023ehm,Pompili:2023tna,Khalil:2023kep,vandeMeent:2023ols}.} 
\end{figure}

In the second set of comparisons, ``$\Omega_0$'', for each simulation the waveforms are examined on the maximal frequency range from the initial reference frequency $\Omega_\text{ref}$ to the final ISCO frequency $\Omega_*(\mathring\chi)$ for the given spin of the simulation. The definition of $\omega_i$ and $\omega_f$ are the same as before: a small fraction of $\Omega_\text{ref}$ for $\omega_i$ and a multiple of $\Omega_{l.r}(\mathring\chi)$ for $\omega_f$. The total mass is restricted such that the frequency $\omega_i/(2\pi \MTOT)$ is larger than $10\text{Hz}$. This enforces that the whole simulated inspiral from the initial reference time is in the bandwidth of current ground-based detectors. We define in either case the initial and final times $t_i$ and $t_f$ as the times corresponding to the frequencies $\omega_i$ and $\omega_f$. Our prescription is closely related to but slightly differs from Ref.~\cite{Moore:2016qxz}, which takes 10Hz as $\omega_i/(2\pi \MTOT)$ and the ISCO frequency for $\omega_f/(2\pi \MTOT)$. 

When performing the fast Fourier transform over a finite time window, we first perform a windowing of the functions $h^X_{\deltaA t,\deltaA \varphi}$ and $h^\text{NR}$ using a Tukey window function. The width of the window's walls is fixed by requiring that the power spectrum of the frequency-domain waveform $|h^\text{NR}(\omega)|^2$ has its high-frequency noise suppressed.

To optimize over phase and time shifts, we first optimize the mismatch analytically for the phase shift. Indeed, the bound 
\begin{align}
    1-\frac{\langle h^X_{\deltaA t,\deltaA \varphi},h^\text{NR}\rangle}{\| h^X\|\| h^\text{NR}\|}&\geq\nonumber\\*
    1-\frac{2/\pi}{\| h^X\|\| h^\text{NR}\|}&\left|\int_{\omega_i}^{\omega_f} \frac{\left({\mathcal{F}[h^X_{\deltaA t,\deltaA \varphi}]}(\omega)\right)^* {\mathcal{F}[h^\text{NR}]}(\omega)}{S_n(\omega/(2\pi))}d\omega\right|
\end{align}
can always be saturated by a well-chosen rotation $\deltaA\varphi$ in the complex plane. On the other hand, we perform the optimization over the time shift numerically. We leverage the fact that the function
\begin{equation}
    1-\frac{2/\pi}{\| h^X\|\| h^\text{NR}\|}\left|\int_{\omega_i}^{\omega_f}  \frac{\left({\mathcal{F}[h^X_{\deltaA t,\deltaA \varphi}]}(\omega)\right)^* {\mathcal{F}[h^\text{NR}]}(\omega)}{S_n(\omega/(2\pi))}d\omega\right|
\end{equation}
is typically a parabola with a minimum to once again implement a ternary search algorithm for finding the location of the minimum $\deltaA t_\text{min}$.

\begin{figure*}[!htb]
 \center
 \includegraphics[width=0.48\textwidth]{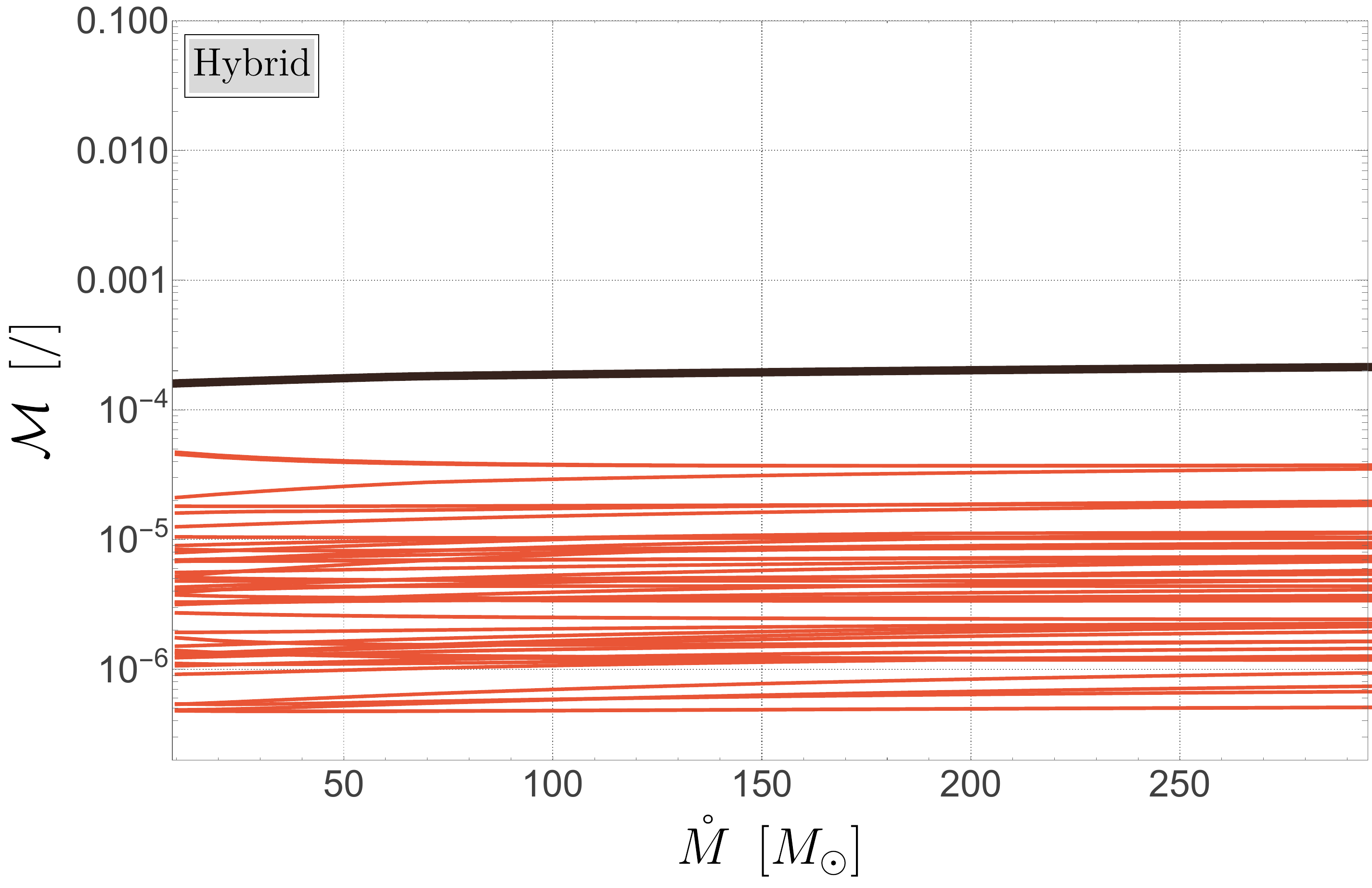}
 \hspace{5 mm}
 \includegraphics[width=0.48\textwidth]{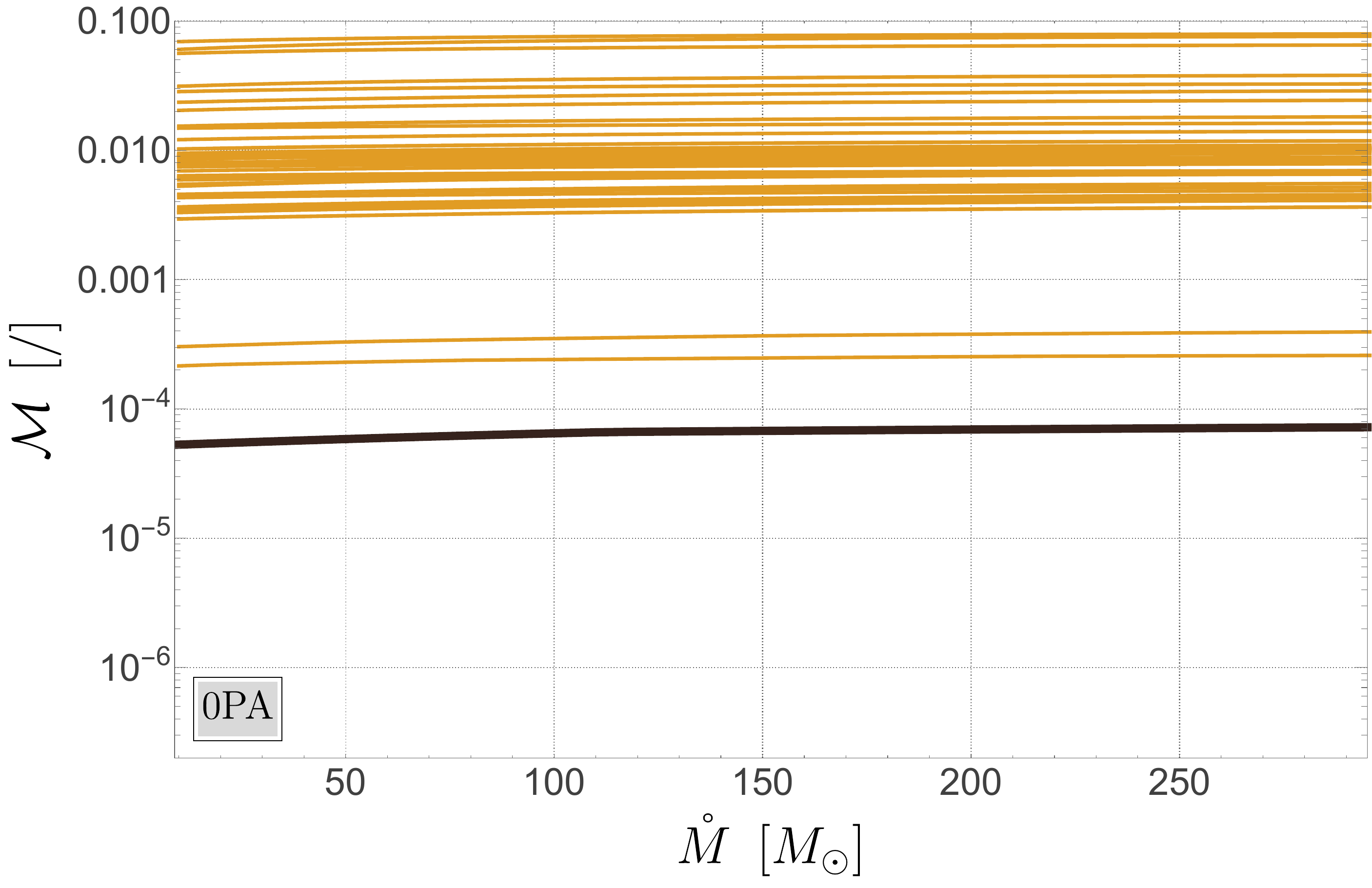}
 \hspace{5 mm}
 \includegraphics[width=0.48\textwidth]{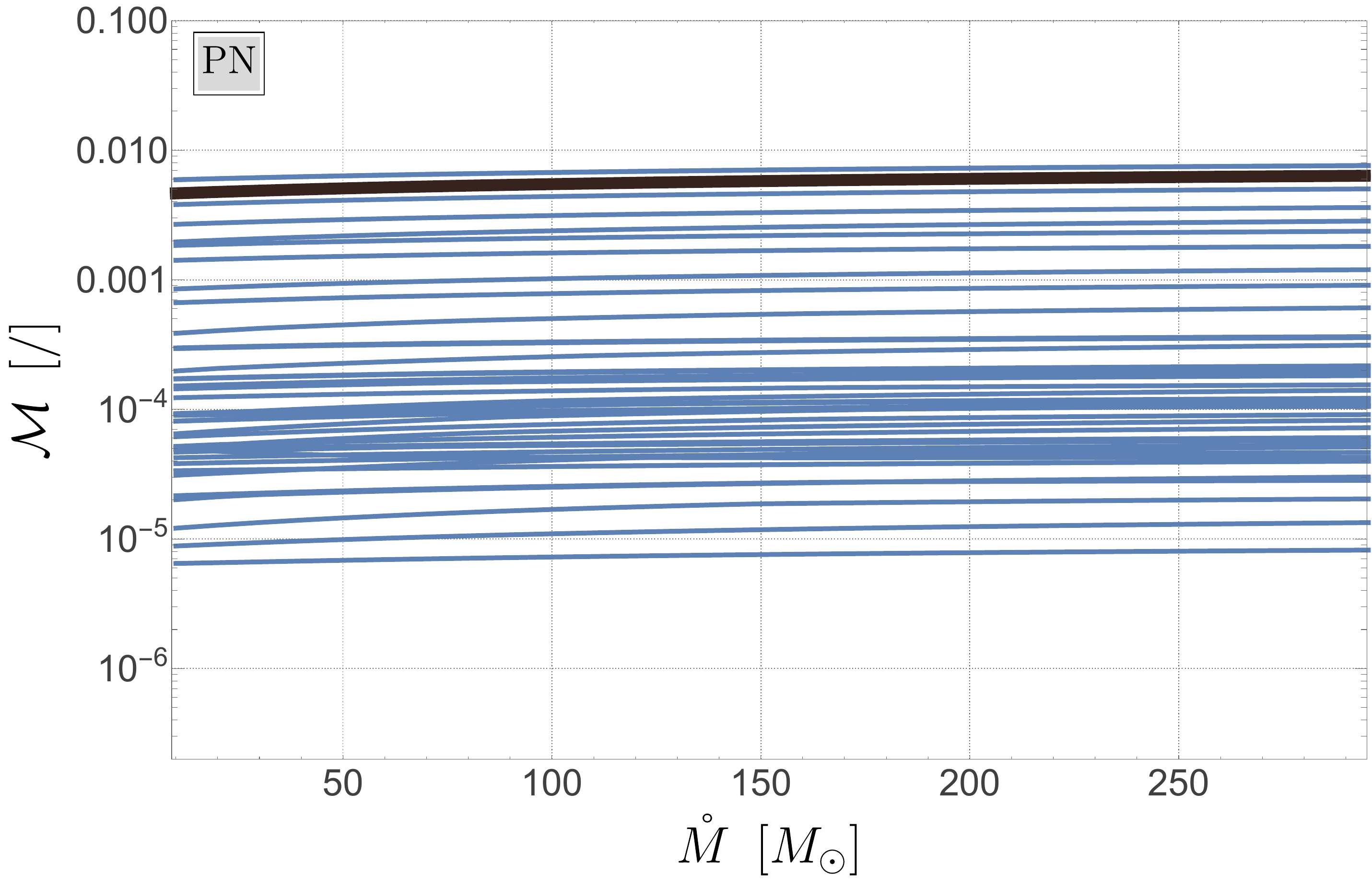} 
 \hspace{5 mm}
 \includegraphics[width=0.48\textwidth]{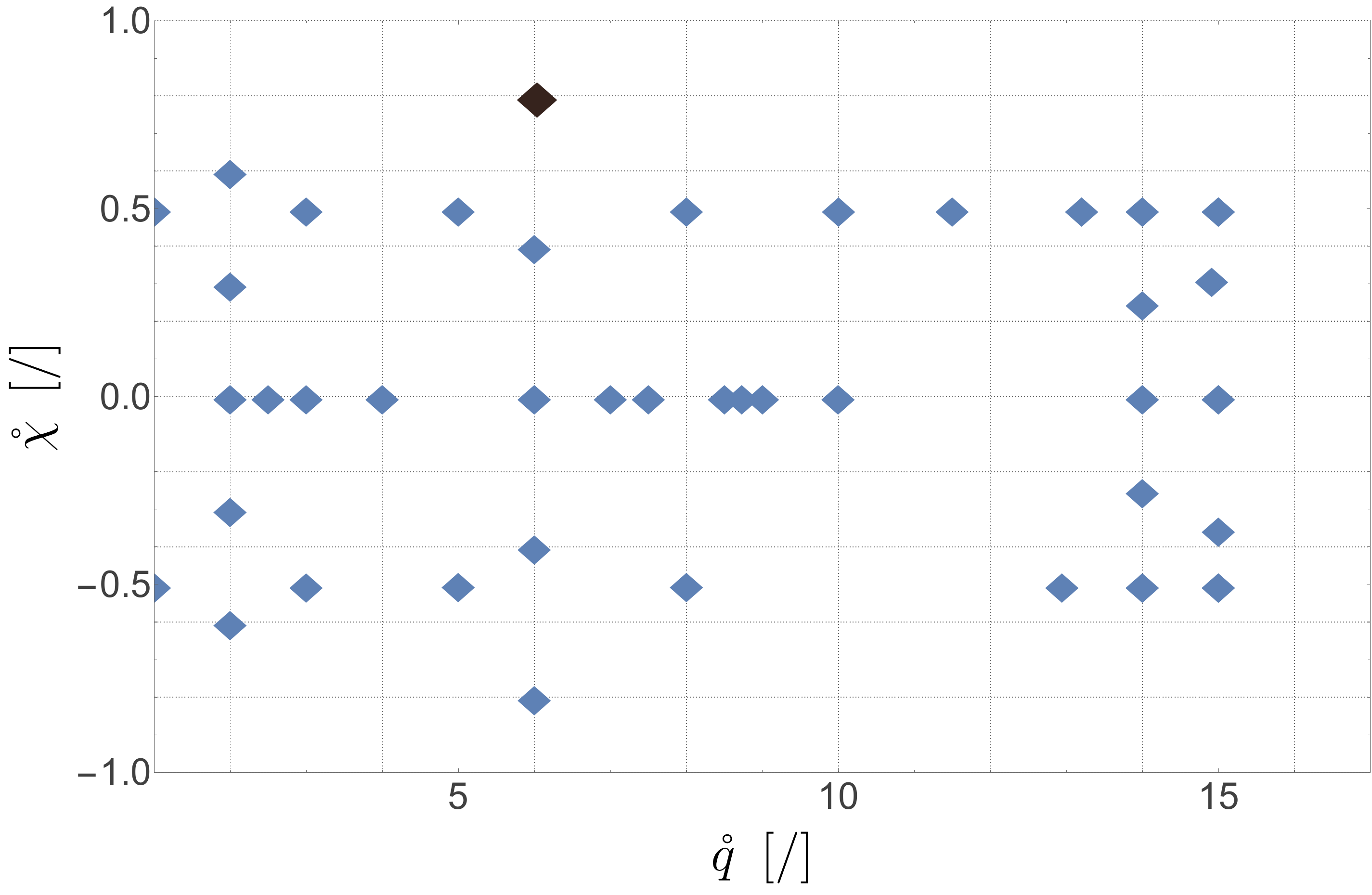} 
 \caption{\label{fig:Spaghet}Inspiral-only mismatches of model $X\in\{H, \text{0PA}, \text{PN}\}$ (top left, top right, and bottom left panels, respectively) against NR simulations as a function of the total mass of the system. Note the mismatches are nearly independent of the total mass. The bottom-right panel shows the population of binary configurations used in this study. The mismatches here are computed over the full range of frequencies that each NR template sweeps through in its inspiral phase (i.e., from $x_i=\Omega_\text{ref}^{2/3}$ to $x_f=\Omega_\star^{2/3}$). The black highlighted curves correspond to the diamond on the bottom-right panel, which is the configuration that yields the worst mismatch between the hybrid model and NR. This corresponds to $\mathring q=6$ and a highly spinning primary with $\mathring\chi=+0.8$. The 0PA model achieves its  \emph{best} match for this configuration, but the reason is accidental: the $1$PA forcing term (as estimated from the hybrid model) omitted by the $0$PA model is very small at this peculiar value of the primary spin. The set of SXS simulations used here are listed in Table~\ref{tab:mismatchdephasing}.} 
 \end{figure*}

Figure~\ref{fig:MismatchPrograde} shows the ``Fixed$\chi$'' comparisons: the evolution of the mismatch of the different models against NR as we move to more asymmetric binaries while keeping the spin of the primary and the frequency band fixed. We also show for reference the mismatch of the SEOBNRv5EHM model~\cite{Liu:2021pkr,Gamboa:2024imd,Gamboa:2024hli} against the same NR simulations. We obtain the SEOBNRv5 waveforms using version $0.3.1$ of the \texttt{pySEOBNR} python package~\cite{Mihaylov:2023bkc}. As another point of comparison, we further compute the mismatch between the highest and next-to-highest NR resolutions, which provides a crude estimate of the NR error. Finally, for the set of nonspinning binaries, we additionally show  the mismatch of the \texttt{1PAT1} model against NR.

As explained in~\cite{Mitman:2025tmj}, the mismatch between two waveforms tends to grow with the square of the dephasing. Therefore, we expect the $0$PA and the hybrid models to have a mismatch that asymptotes to a constant for asymmetric binaries, while the PN mismatch against an NR template should grow like $\sim\left(\deltaE\psi^\text{PN}_{22}\right)^2\sim1/\mathring\nu^2$. In contrast, the \texttt{1PAT1} model's mismatch should go to zero like $\sim\left(\deltaE\psi^{1\text{PAT}1}_{22}\right)^2\sim\mathring\nu^2$. These scalings are consistent with the behaviors that we observe in Fig.~\ref{fig:MismatchPrograde}.  Importantly, at small $\mathring\nu$ the mismatch between the hybrid model and the NR waveform settles down below $\mathcal M \lesssim 10^{-5}$, three orders of magnitude below the value for the 0PA model. However, we still see the superiority of the complete \texttt{1PAT1} model for sufficiently small $\mathring\nu$.

We also observe that in the range of mass ratios we consider, the hybrid model achieves mismatches comparable to those of SEOBNRv5, except in the case of comparable-mass retrograde binaries. In most cases, and particularly for small mass ratios, both the hybrid and SEOBNRv5 achieve lower mismatches than the lower-resolution NR simulations. For that mismatch between the two NR resolutions, the scaling with mass ratio is intricate as it depends both on the scaling of the length of the simulation with $\mathring\nu$ and on the scaling of the resolution with~$\mathring\nu$. 

When using SEOBNRv5 as a benchmark in this way, we stress that our comparisons are restricted to a finite portion of the inspiral, omitting the merger-ringdown regime; unlike EOB models, our model does not attempt to cover this most challenging regime. On the other hand, we also stress that for sufficiently small $\mathring\nu$, the SEOBNRv5 mismatch in the inspiral will grow as $1/\mathring\nu^2$ due to the model's inexact 0PA information; see the companion letter~\cite{PaperIV}.

Our mismatches have been computed for inspirals that sweep the parameters $[x_i,x_f]=[0.086,0.228]$ for the set of prograde orbits, $[x_i,x_f]=[0.079,0.135]$ for the set of retrograde orbits, and $[x_i,x_f]=[0.091,0.167]$ for the set of nonspinning binaries. This corresponds to an orbital separation range of $[11.5\MTOT,4.2 \MTOT]$, $[12.8\MTOT,7.5\MTOT]$, and $[11.0\MTOT,6.0\MTOT]$, respectively. Because a 1PA error is independent of mass ratio, we expect our hybrid to achieve the same small error, $\mathcal M \lesssim 10^{-5}$, for IMRIs and EMRIs in these same frequency windows.

Finally, for the ``$\Omega_0$'' batch of simulations (i.e., computing the mismatch over the full inspiral of each simulation), we display in Fig.~\ref{fig:Spaghet} the mismatch of model $X$ against every NR simulation listed in Table~\ref{tab:mismatchdephasing}. We see that the hybrid model's mismatch is consistently below $10^{-4}$ across all but one simulation; the lone outlier, with mismatch $\approx 2\times 10^{-4}$, is highlighted in black. This level of mismatch is, in general, several orders of magnitude below that of the 0PA model (top right panel) and one or more orders of magnitude below that of the PN model (bottom left).

\subsection{Accuracy in the high-$\mathring q$ regime}
\label{subsec:accuracy}

The comparisons in Sec.~\ref{subsec:comparison} were performed for comparable masses or moderately low mass ratios ($\mathring q\lesssim 15$), as that is the range currently covered by NR. However, our hybrid model is aimed at larger mass ratios. Although the trends we observed above suggest that the model's error remains small all the way to the EMRI regime, we are cognizant that the model is not exact at 1PA order. We also wish to assess how much of an improvement, if any, we make by using the full 1PA binding energy instead of its 4PN approximation. This motivates a more careful estimation of its accuracy at high~$\mathring q$. We carry out that assessment by appraising the accuracy of the model's underlying data and then extrapolating the effects of any inaccuracy to large $\mathring q$. 

\begin{figure}[!htb]
\includegraphics[width=\columnwidth]{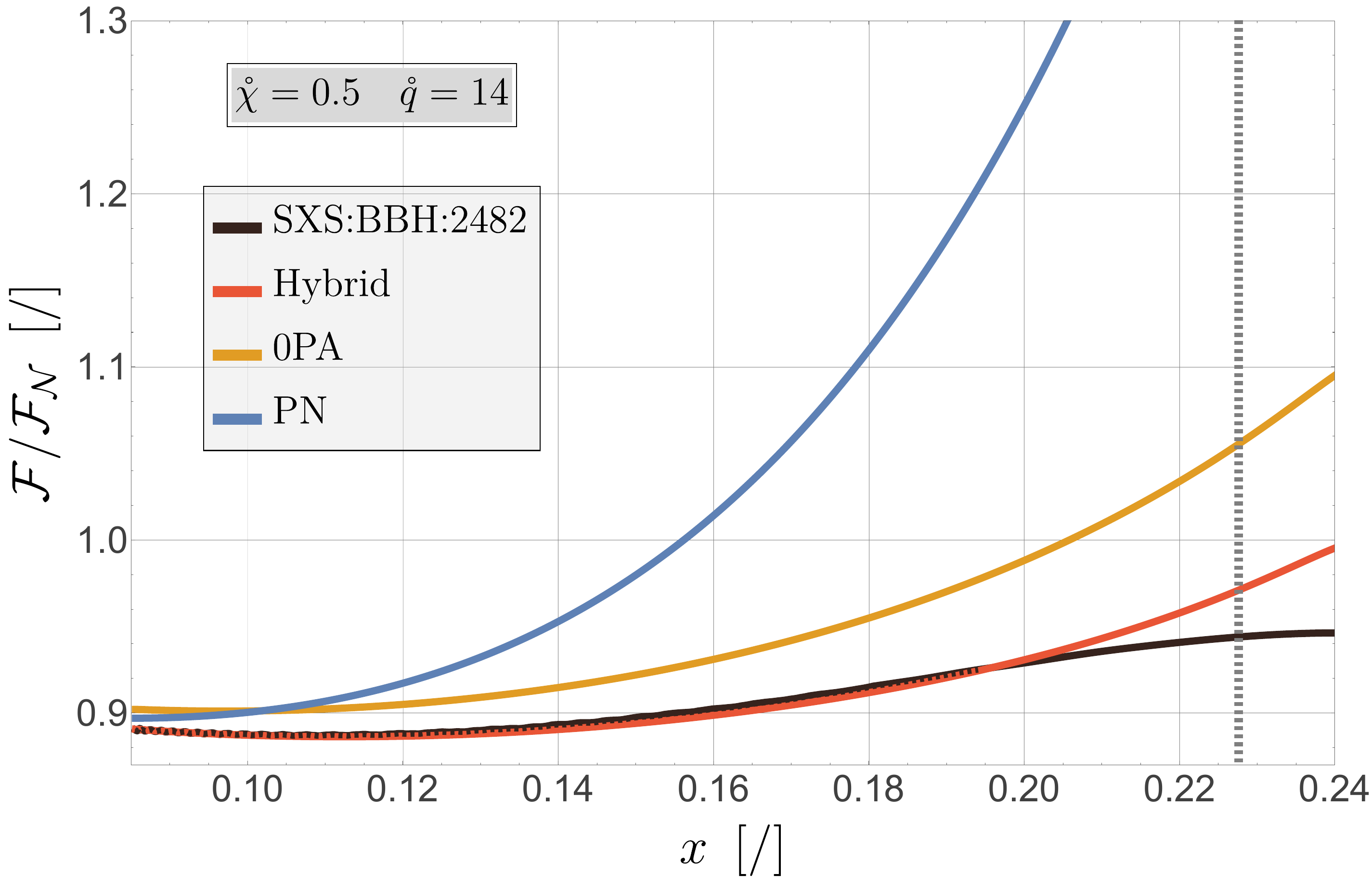}
\includegraphics[width=\columnwidth]{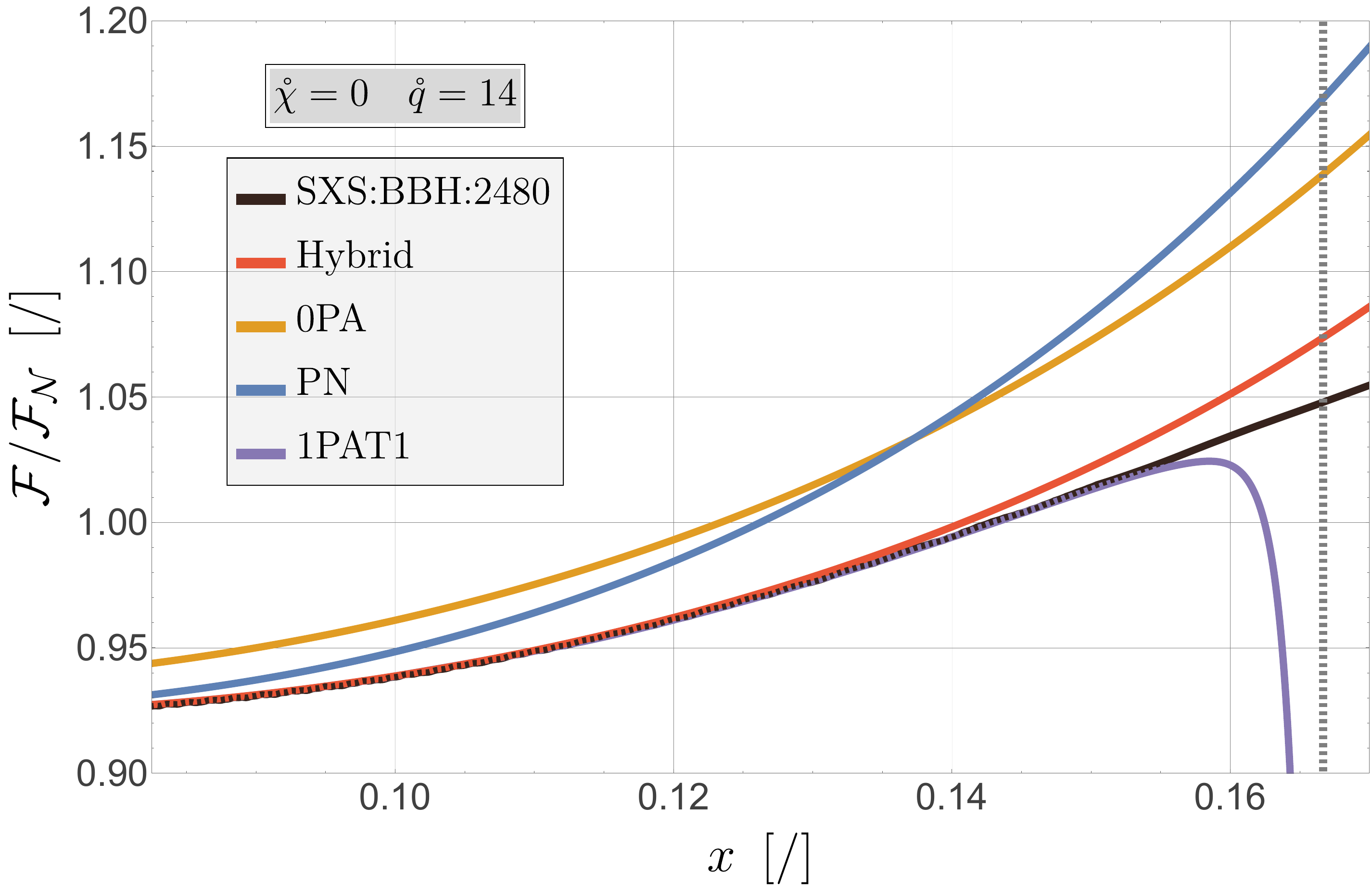}
\includegraphics[width=\columnwidth, trim={0 30pt 0 0}]{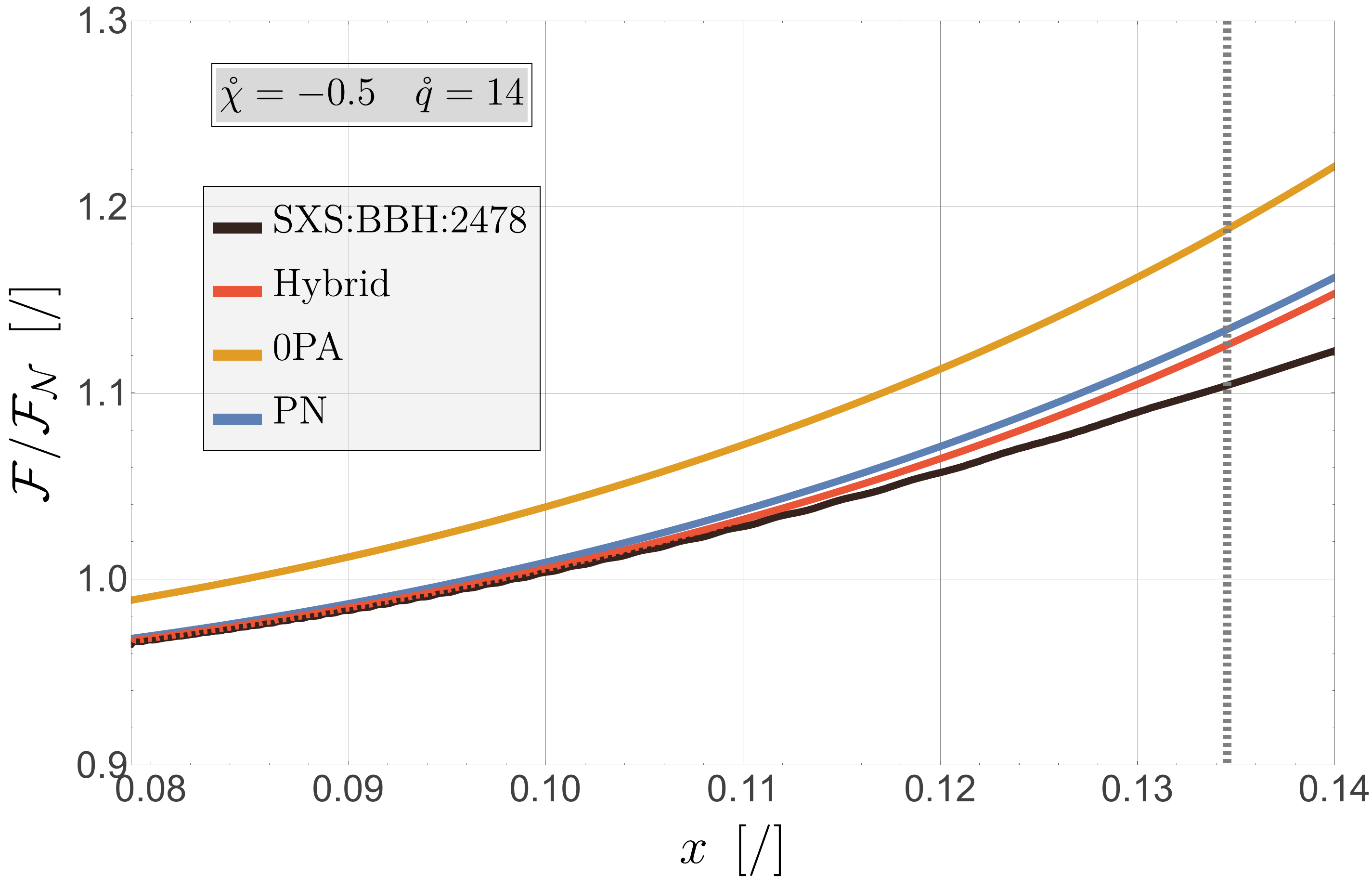}
\caption{\label{fig:fluxqualitative} 
Newtonian-normalized energy fluxes of the three models $H$ (red), $0$PA (orange), and PN (blue) as a function of the PN parameter $x=\omega^{2/3}$. The fluxes are normalized by the leading PN quadrupole formula: $\mathcal{F}_\mathcal{N}=32\nu^2x^5/5$. All three NR simulations (see Table~\ref{tab:mismatchdephasing} for references) have the same large mass ratio $\mathring q=14$ but different values of the primary spin~$\mathring \chi$: $\mathring \chi=+0.5$ (prograde, top panel), $\mathring \chi=0$ (nonspinning, middle panel), and  $\mathring \chi=-0.5$ (retrograde, bottom panel). In the nonspinning case, we have also included (in purple) the $2$SF energy flux from~\cite{Warburton:2021kwk}, used in the \texttt{1PAT1} model. The gray vertical lines indicate the location of the ISCO.}
\end{figure}

\subsubsection{Energy fluxes and binding energy}

The binding energy of the hybrid model is complete to $1$PA order, while the energy fluxes are only exact to $0$PA order. Hence, we expect mismodeling of the energy fluxes to be the hybrid model's main source of error. 

To estimate the impact of an error in the flux and binding energy on the waveform phase, we compare them against NR. This necessitates extracting the flux and binding energy from an NR waveform, which we do in the following way: differentiating the total energy $\ETOT=\MTOT+E$ with respect to time $t$ and using the relations $\frac{d}{dt}\ETOT=-\mathcal F^\infty$, $\frac{d}{dt}\MTOT=\mathcal F^{\mathcal H}$, we can write the binding energy as 
\begin{align}\label{eq:NRbinding1}
   E(t) = E(t_i)-\int_{t_i}^{t} dt \, \left( \mathcal F^\infty + \mathcal F^{\mathcal H} \right). 
\end{align}
The NR flux is computed from a given NR simulation as~\cite{Warburton:2024xnr}
\begin{equation}\label{eq:NRflux}
    \mathcal{F}_\text{NR}^\infty=\sum_{\ell=2}^{\ell_\text{max}}\sum_{m=-\ell}^{\ell}\frac{1}{16\pi}|\dot{h}^\text{NR}_{\ell m}|^2,
\end{equation}
where $\ell_\text{max}=8$ for a typical SXS simulation. For the NR flux down the horizon of the primary $\mathcal{F}^\mathcal{H}$, one could in principle extract from an NR simulation the time-dependent value of the masses of the primary $m_1(t)=\MATHRINGm1+ \int_{t_0}^t \mathcal{F}^\mathcal{H}dt$. However, we found the errors of the NR values of $m_1(t)$ were too large and spoiled the comparison. In practice, we discard the NR contribution to the horizon flux and approximate it by the $0$PA horizon flux. The horizon flux is very small compared to the flux at infinity: had we completely dismissed the horizon flux or approximated it by the $0$PA horizon flux, almost no difference could be seen in practice. 

\begin{figure}[!htb]
\includegraphics[width=\columnwidth]{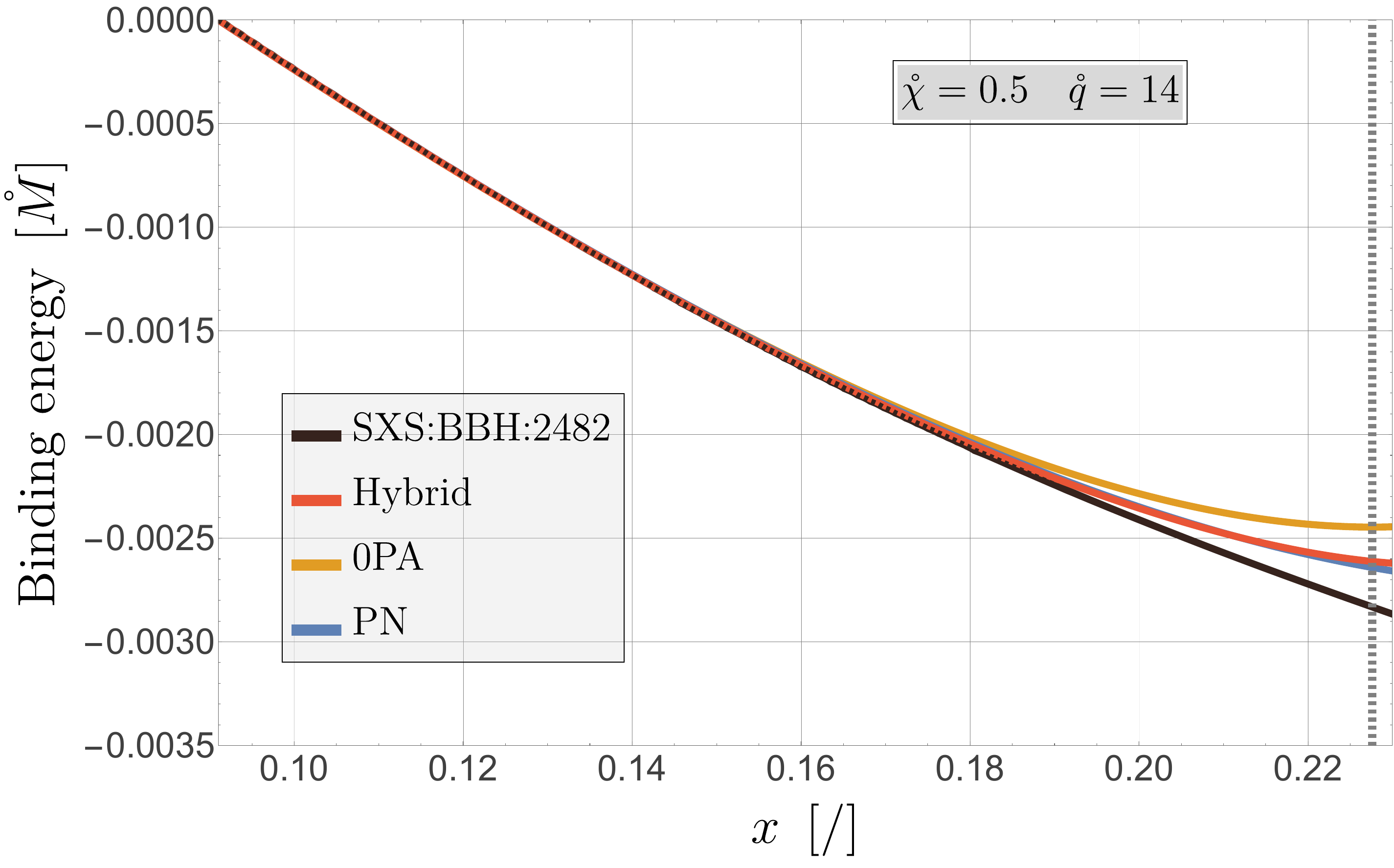}
\includegraphics[width=\columnwidth]{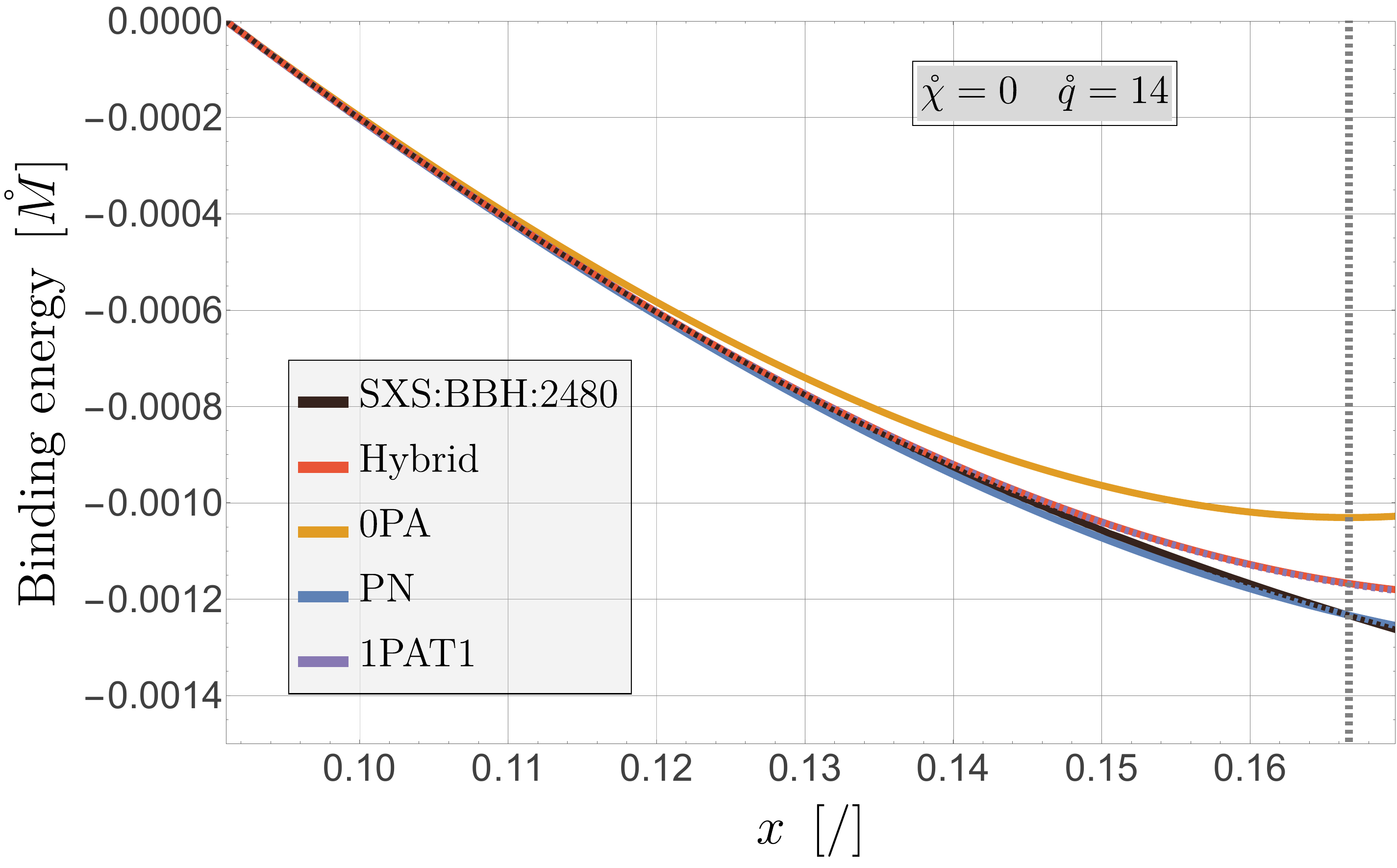}
\includegraphics[width=\columnwidth]{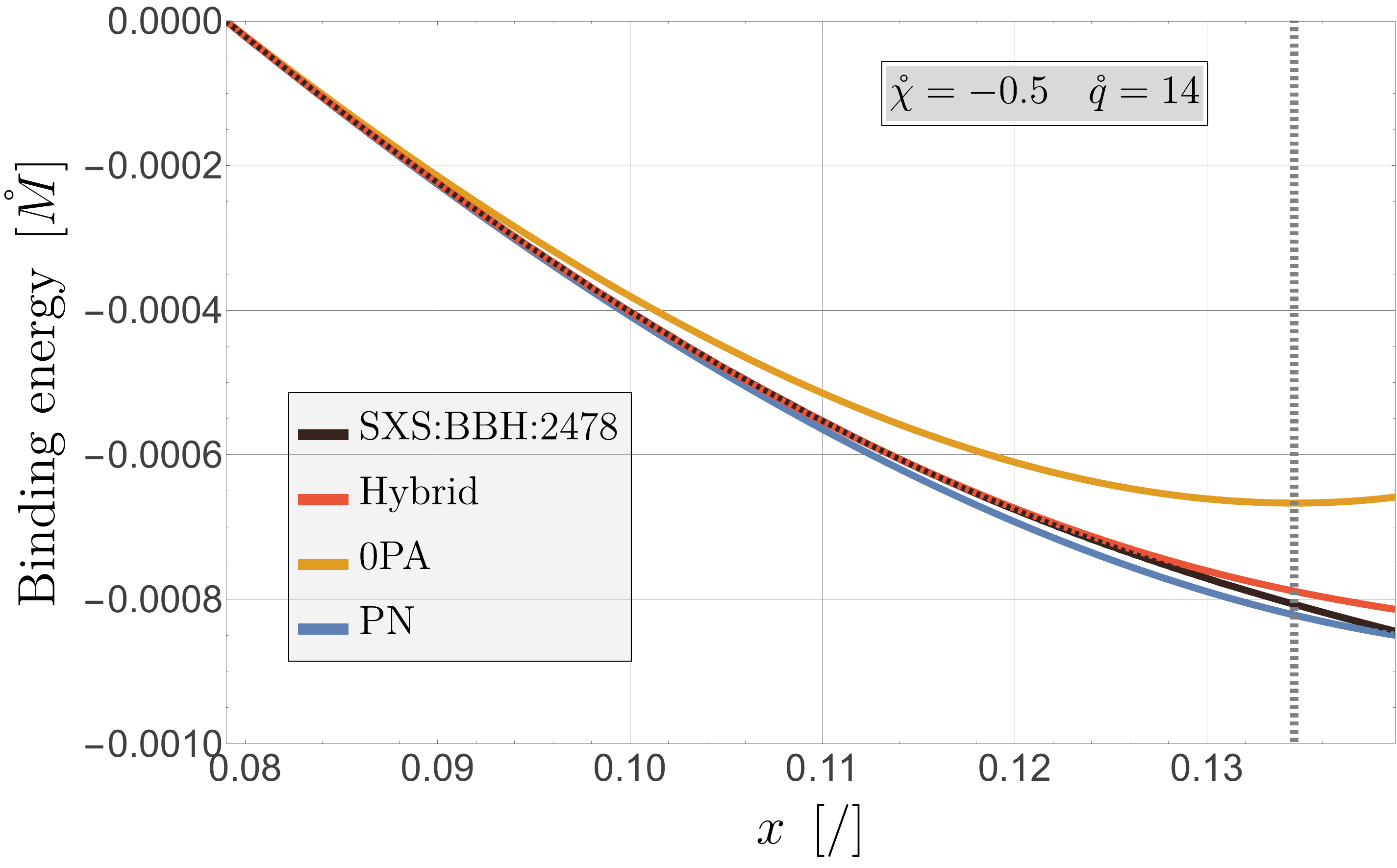}
\caption{\label{fig:energyqualitative} 
Binding energies for the three models $H$ (red), $0$PA (orange), and PN (blue) as a function of the PN parameter $x=\omega^{2/3}$. We have set the integration constant of the NR sbinding energy to $E(t_i)=0$ in Eq.~\eqref{eq:NRbinding1}, setting the binding energy to vanish at the left boundary of each plot. The binding energies in the other models are correspondingly shifted by constants to bring them to zero there. Other details are as in Fig.~\ref{fig:fluxqualitative}.
}
\end{figure}

We show in Fig.~\ref{fig:fluxqualitative} a comparison of the energy fluxes in the hybrid, 0PA, and PN models with respect to NR. The $0$PA fluxes badly reproduce the NR curves for any value of the primary spin $\mathring\chi$. The PN flux curve, on the other hand, converges well to the NR curves when the PN parameter $x$ decreases, but it starts to deviate badly from the NR curve as we move to the strong-field regime. It is worth noting that the large discrepancy of the PN flux with NR close to the ISCO is drastically less pronounced for retrograde orbits, where the ISCO radius is located further away from the primary black hole. 

The figure illustrates that the hybrid fluxes are always more accurate than the $0$PA and PN fluxes, for both retrograde and prograde orbits. However, this accuracy improvement is less pronounced for retrograde orbits, where the PN approximant already reproduces the NR flux very well. It is also worth noting that, for nonspinning binaries, both $2$SF\footnote{Reference~\cite{Warburton:2021kwk} contained a small error in the $2$SF energy flux. In this paper we use the corrected data.} and hybrid fluxes are very similar to each other, except very close to the ISCO where both curves start to deviate from NR, corroborating the outcomes of~\cite{Burke:2023lno}. 

Next, in Fig.~\ref{fig:energyqualitative} we show a comparison between the binding energies of all three models and NR. Note we use the input binding-energy data for each model except NR; we do not evaluate the flux integral~\eqref{eq:NRbinding1} for the other models. The discrepancy between the hybrid and PN models is more subtle here than in the flux, both of them reproducing quite well the NR binding energy. Only the $0$PA binding energy is notably inaccurate as compared to NR. For the case of a nonspinning binary, we were also able to compare against the \texttt{1PAT1} model. As we see, for $\mathring q=14$ the hybrid and \texttt{1PAT1} binding energies are almost indistinguishable. This is to be expected because both use the same 1PA binding energy; they differ only by the 2PA and higher terms included in the hybrid model.

\begin{figure}[!hbt]
\includegraphics[width=\columnwidth]{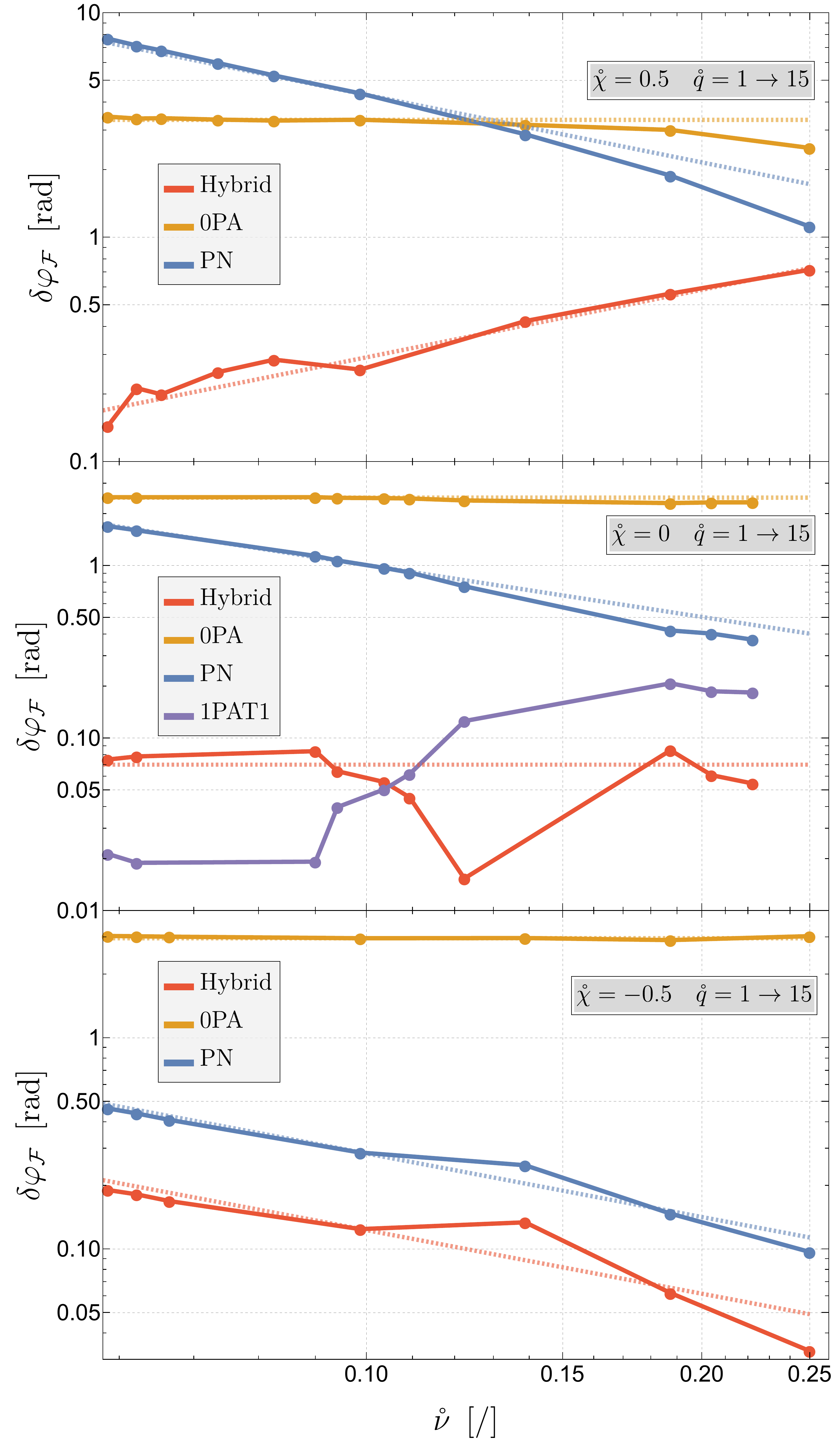}
\caption{\label{fig:energyquantitative} 
Estimate of the waveform phase error due to a mismodeling of the energy flux $\deltaE\varphi_\mathcal{F}$. The integration in Eq.~\eqref{eq:dephasingflux2} is performed on the maximal range of PN parameter $[x_i,x_f]$ that is shared between all simulations of a given panel. Explicitly, $[x_i,x_f]=[0.086,0.228]$ for the prograde orbits (top panel), $[x_i,x_f]=[0.091,0.167]$ for the nonspinning binaries (central panel), and $[x_i,x_f]=[0.079,0.135]$ for the retrograde orbits (bottom panel). We also show  reference power laws in dashed lines: $\mathring\nu^{-1}$ in dashed blue, $\mathring\nu^{0}$ in dashed orange. The red dashed line is a $\mathring\nu^1$ reference (top panel), $\mathring\nu^0$ (middle), or $\mathring\nu^{-1}$ (bottom). For the nonspinning binaries, we have also included (in purple) the estimate of the phase error due to the error in the \texttt{1PAT1} flux. The simulations used in these figures are the same as in Fig.~\ref{fig:MismatchPrograde}; see Table \ref{tab:mismatchdephasing}.}
\end{figure}

\subsubsection{Phase error due to  energy flux error}

\begin{figure}[!hbt]
\includegraphics[width=\columnwidth]{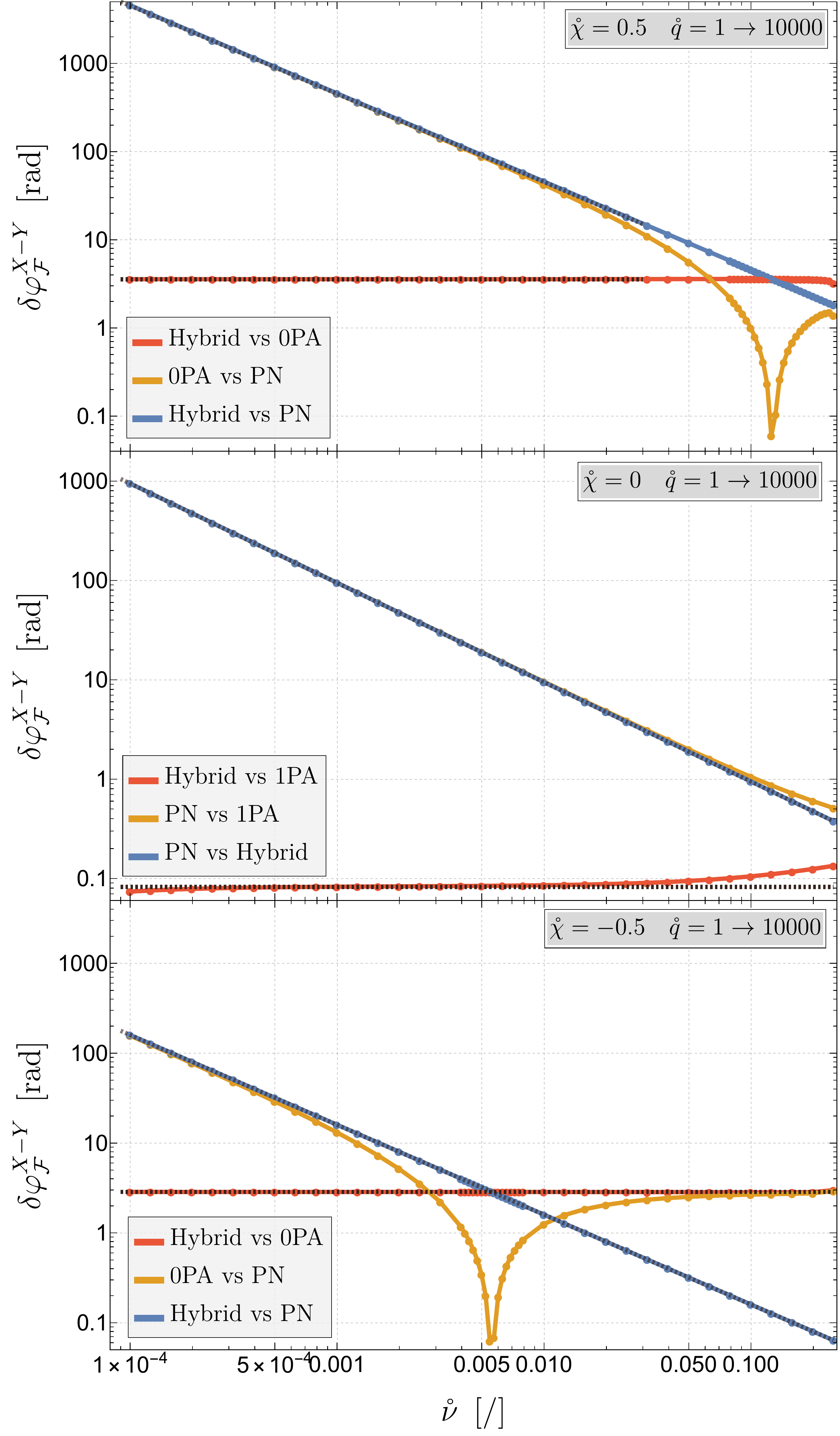}
\caption{\label{fig:energyquantitative2}Estimate of the accumulated dephasing between two models $X,Y\in\{H,0\text{PA},\text{PN},1\text{PA(T}1)\}$ over a fixed frequency window due to the discrepancy in their energy flux; see Eq.~\eqref{eq:dephasingflux3}. In the top panel, we set the primary spin to $\mathring\chi=+0.5$ (prograde orbits); in the central panel, to $\mathring\chi=0$ (nonspinning binaries); and in the bottom panel, to $\mathring\chi=-0.5$ (retrograde orbits). We vary the mass ratio from $\mathring q=1$ to $\mathring q=10^4$. Notice that, for the central panel, we do not compare the hybrid and PN models against a $0$PA energy flux but rather against the $2$SF ($1$PA) flux from~\cite{Warburton:2021kwk}.}
\end{figure}

Our model is exact at 1PA order \emph{except} in its use of a PN approximation to the 2SF energy flux. We now assess how an error in the energy flux to infinity, $\mathcal{F}^\infty$, propagates to a total accumulated phase error $\deltaE\varphi^X_\mathcal{F}$ over a fixed waveform frequency interval $[\omega_i,\omega_f]$. The chosen frequency window depends upon the chosen batch of simulations and is made explicit in Figs.~\ref{fig:energyquantitative}--\ref{fig:bindingquantitative}. 

We make use of the chain rule to write the total accumulated phase error in model X on the interval $[\omega_i,\omega_f]$ as
\begin{align}
\deltaE\varphi^X&=\int_{\omega_i}^{\omega_f}\left(\frac{\omega }{F_\omega^X}-\frac{\omega }{F_\omega^\text{NR}}\right)d\omega \label{eq:phaseerror1}.
\end{align}
Here, we have used for the integration variable the waveform frequency $\omega$. We now derive, at leading order in $\nu$, how a mismodeling in the energy flux contributes to this dephasing. To that end, we write the waveform frequency forcing term as a functional $F_\omega\left(\mathcal{F}^\infty,\mathcal{F}^\mathcal{H},\mathcal{G}^\mathcal{H},E\right)$ and expand Eq. \eqref{eq:phaseerror1} at leading order in $\nu$ while keeping $\mathcal{F}^\mathcal{H}$, $\mathcal{G}^\mathcal{H}$, and $E$ fixed. At leading order, the linear functional $\deltaE\varphi_\mathcal{F}^X$ can be written as
\begin{align}\label{eq:dephasingflux}
    \deltaE\varphi_\mathcal{F}^X&=\nu \int_{\omega_i}^{\omega_f}\frac{\omega \partial E_{geo}/\partial\omega }{\mathcal{F}_{0\text{PA}}^2}\left(\mathcal{F}^\infty_\text{NR}-\mathcal{F}^\infty_X\right)d\omega ,
\end{align}
where $\mathcal{F}_{0\text{PA}}=\nu^2 \mathcal{F}^\text{SF}_{(0)}(x,\chi)$. Changing the integration variable to $x$, and using the explicit expression~\eqref{Egeo} for the geodesic binding energy, one can rewrite the integral as
\begin{align}\label{eq:dephasingflux2}
\deltaE\varphi^X_{\mathcal{F}}=\int_{x(\omega_i)}^{x(\omega_f)}\frac{\nu\omega^{7/3}U_{(0)}^3D_{}}{2\mathcal{F}_\text{0PA}^2}\left(\mathcal{F}_X-\mathcal{F}_\text{NR}\right)dx,
\end{align}
where the functions $D$ and $U_{(0)}$ have been defined in Eqs.~\eqref{U0} and \eqref{eqD}.

The first factor in the integrand only includes 0PA terms, which are analytically known for $U_{(0)}$ and $D$, and numerically computed on a Chebyshev grid for $\mathcal{F}_{0\text{PA}}$. The second factor is the error between the model $X$ energy flux and the NR energy flux computed with Eq.~\eqref{eq:NRflux}. The first factor comes as a weight that encodes how the model phase error responds to an error in the flux. It is interesting to observe that large errors on the energy flux close to the ISCO are suppressed by the weight function, as the factor $D$ vanishes at the ISCO. Instead, the leading PN behavior of the weight function in Eq.~\eqref{eq:dephasingflux} diverges as
\begin{align}
\frac{\nu\omega^{7/3}U_{(0)}^3D_{}}{2\mathcal{F}_\text{0PA}^2}=\frac{25}{2048\nu^3}\frac{1}{x^{17/2}}+\mathcal{O}(1/x^8)
\end{align}
at small $x$, hence amplifying any small deviation from the energy flux in the early inspiral. 

Equation~\eqref{eq:dephasingflux2} serves as a proxy for estimating the impact of the error of the energy flux on the performance of our model. We compute this estimate for the ``Fixed$\chi$'' batch of comparisons and show the results in Fig.~\ref{fig:energyquantitative}.

In all the comparisons in Fig.~\ref{fig:energyquantitative}, we observe that the dephasing estimate $\deltaE\varphi_\mathcal F$ grows as $\sim\mathring \nu^{-1}$ for the PN model and as $\sim\mathring \nu^{0}$ for the $0$PA model. This is to be expected from the expansion of the waveform phase in Eq.~\eqref{eq:accphase}. Indeed, the PN flux is not complete at 0PA order, which means that its phase error grows as $\sim\mathring \nu^{-1}$ (though with a visibly small coefficient). On the other hand, the $0$PA model omits 1PA flux terms, causing a phase error that scales as $\sim\mathring \nu^{0}$ (but with a visibly large coefficient). 

For the hybrid model, though, the behavior of the dephasing estimate is now more delicate: the hybrid model is $0$PA complete but lacks high-order PN information at $1$PA and higher. It is not clear beforehand whether the $1$PA or $2$PA and higher errors will dominate in the range of mass ratios we can consider. For prograde orbits (top panel), the estimate $\deltaE\varphi^H_\mathcal{F}$ scales like $\mathring\nu^1$, showing that, at these mass ratios and at $\mathring\chi=+0.5$, the error in the hybrid model's energy flux to infinity is dominated by the lack of high-PN information at $2$PA order. This behavior cannot last forever as we go to smaller and smaller mass ratio; the $2$PA errors are suppressed by a factor $\mathring\nu$ relative to the $1$PA errors and will eventually become negligible. Hence, the hybrid error $\deltaE\varphi^H_\mathcal{F}$ in the top panel of Fig.~\ref{fig:energyquantitative} will eventually hit a plateau. This nevertheless gives us a conservative estimate of the $1$PA residual of the hybrid model at $\mathring\chi=+0.5$ and $[x_i,x_f]=[0.086,0.228]$ to be $\varphi_{(1)}(\mathring\nu t)\sim0.2\ \text{rad}$. 
For nonspinning binaries (middle panel), the hybrid has settled down to the plateau where $1$PA errors numerically dominate $2$PA errors already at $\mathring q\sim15$. This gives us the estimate $\varphi_{(1)}(\mathring\nu t)\sim0.1\ \text{rad}$ for $\mathring\chi=0$ and $[x_i,x_f]=[0.091,0.167]$. These estimates are roughly comparable to the full waveform dephasing computed in Sec.~\ref{sec:wfdephasing}. As expected, this points towards the fact that, at large $\mathring q$, the main source of error in our model comes from not including the $2$SF (1PA) correction to the energy flux. However, this error appears to be small. 

Contrastingly, for retrograde orbits the error estimate $\deltaE\varphi^H_\mathcal{F}$ of the hybrid appears to grow as  a $0$PA error, scaling as $\nu^{-1}$. We have thoroughly checked our implementation and found no error. One possible conclusion for the behavior of $\deltaE\varphi^H_\mathcal{F}$ is that one would need to reach smaller mass ratios to see the expected plateau scaling as $\nu^0$. Based on the results of~\cite{vandeMeent:2023ols}, one could argue that the numerical coefficients of the residual 2SF errors are higher for retrograde orbits than for prograde orbits; see Fig.~10 of that reference. One would therefore need to reach lower values of the mass ratio to identify the ${\cal O}(\nu^0)$ scaling. In that interpretation, the scaling observed in the range $q=[1,15]$ does not indicate a 0PA error. 

Finally, we also define the difference between the phase errors $\deltaE\varphi_\mathcal{F}^X$ and $\deltaE\varphi_\mathcal{F}^Y$ as
\begin{align}\label{eq:dephasingflux3}
    \deltaE\varphi_\mathcal{F}^{X-Y}&=\deltaE\varphi_\mathcal{F}^X-\deltaE\varphi_\mathcal{F}^Y\nonumber\\
&=\int_{\omega_i}^{\omega_f}\frac{\nu\omega^{7/3}U_{(0)}^3D_{}}{2\mathcal{F}_\text{0PA}^2}\left(\mathcal{F}_X-\mathcal{F}_\text{Y}\right)dx,
\end{align}
which is now independent of the NR template. In Fig.~\ref{fig:energyquantitative2} we display the estimate~\eqref{eq:dephasingflux3} for more extreme mass ratios  where NR data is unavailable. Notice that the PN model's lack of an exact 0PA flux leads it to accumulate hundreds to thousands of radians of dephasing as the mass ratio moves to the EMRI regime $\mathring\nu\lesssim10^{-4}$, pinpointing the need to include at least the complete $0$PA dynamics in an EMRI model. Importantly, the estimated dephasing between the PN and the hybrid models is very small for retrograde orbits at comparable mass ratios, showing a great improvement on the quality of the hybrid model compared to a simple 0PA waveform in this part of the parameter space. The same holds for nonspinning binaries: the hybrid model compares slightly better to the PN model at equal mass ratio than does the \texttt{1PAT1} model. Finally, for nonspinning binaries, the dephasing due to the discrepancy between the hybrid and $2$SF flux is estimated to be $\sim 0.09\text{ rad}$. That is to say, the effect of neglecting the high-PN $1$PA terms in the energy flux to infinity translates to a total waveform dephasing of less than $0.1\text{ rad}$ over the PN parameter range $[x_i,x_f]=[0.091,0.167]$.

\subsubsection{Phase error due to binding energy error}

As our final assessment, we explore how much accuracy the hybrid model would sacrifice if it used a 4PN approximation to the complete 1PA binding energy.  

We again start from 
\begin{align}
\deltaE\varphi^X&=\int_{\omega_i}^{\omega_f}\left(\frac{\omega }{F_\omega^X}-\frac{\omega }{F_\omega^\text{NR}}\right)d\omega \label{eq:phaseerror2} 
\end{align}
and now consider a small perturbation of the binding energy $E$, while keeping the fluxes fixed. At leading order in the mass ratio, Eq. \eqref{eq:phaseerror2} yields
\begin{align}\label{eq:dephasingbinding}
   \deltaE\varphi_{E}^X=-\int_{x_i}^{x_f}\frac{\omega}{\mathcal{F}_\text{0PA}}\frac{d}{dx}\left(E_X-E_\text{NR}\right)dx,
\end{align}
where the NR binding energy is computed with Eq. \eqref{eq:NRbinding1}. Eq. \eqref{eq:dephasingbinding} serves as a proxy for estimating the impact of the error of the binding energy on the performance of our model. In this case, the numerical errors on the NR binding energy are the main source of error in Eq. \eqref{eq:dephasingbinding}. It is therefore not possible to use Eq. \eqref{eq:dephasingbinding} with the NR binding energy as a benchmark. Only the $0$PA binding energy is imprecise enough to be located above the NR noise.

\begin{figure}[!htb]
\includegraphics[width=\columnwidth]{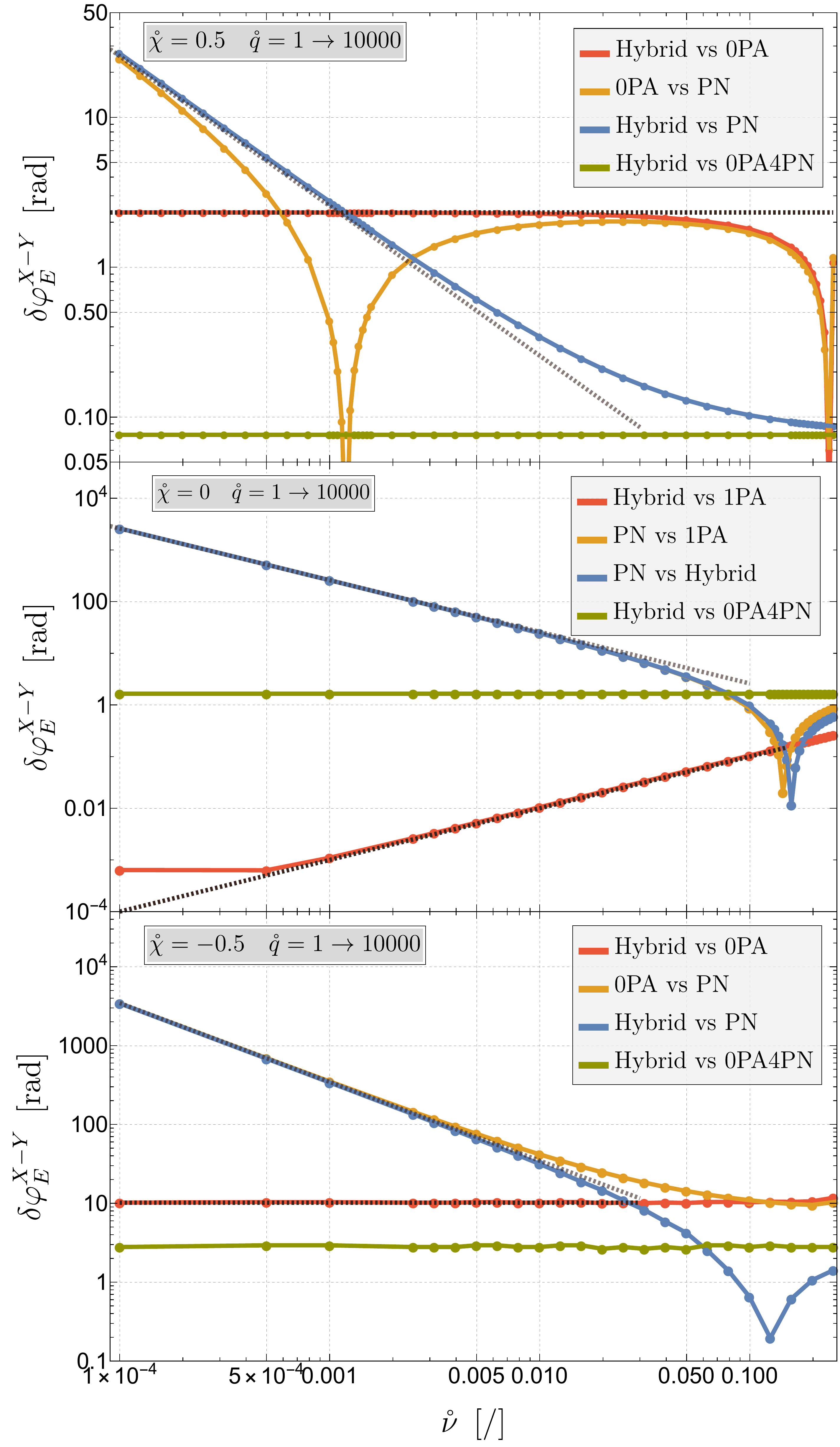}
\caption{\label{fig:bindingquantitative} Estimate of the accumulated dephasing between two models $X,Y\in\{H,0\text{PA},\text{PN},1\text{PA(T}1),\text{0PA4PN}\}$  over a fixed frequency window due to the discrepancy in their binding energy; see Eq. \eqref{eq:dephasingbinding2}. Other details are as in Fig.~\ref{fig:energyquantitative2}. For the central panel, we do not compare the hybrid and PN models against the geodesic binding energy but rather against the $1$SF ($1$PA) binding energy computed from the Detweiler redshift (as used in both \texttt{1PAT1} and the hybrid model).
}
\end{figure}

Therefore, rather than comparing the binding energy to an NR template, we investigate how two different models, $X$ and $Y$, dephase from one another due to their difference in binding energy. We define
\begin{align}\label{eq:dephasingbinding2}
    \deltaE\varphi_{E}^{X-Y}&=\deltaE\varphi_{E}^X-\deltaE\varphi_{E}^Y\nonumber\\
    &=\int_{x_i}^{x_f}\frac{\omega}{\mathcal{F}_\text{0PA}}\frac{d}{dx}\left(E_X-E_Y\right)dx,
\end{align}
which is now independent of the template $E_\text{NR}$. $ \deltaE\varphi_{E}^{X-Y}$ is shown in Fig.~\ref{fig:bindingquantitative} for the same set of primary spins as in the ``Fixed$\chi$'' batch of comparisons: $\mathring\chi=0.5$, $\mathring\chi=0$, and $\mathring\chi=-0.5$. In contrast to the energy flux, the binding energy of the hybrid model is complete at $1$PA. Therefore, when compared to the \texttt{1PAT1} binding energy (i.e., 1PA and no higher), the residual $E^H-E^{1\text{PAT}1}\sim\mathcal{O}(\nu^3)$, and the dephasing estimate is $ \deltaE\varphi_{E}^{H-1\text{PAT}1}\sim\mathcal{O}(\nu)$. This is indeed the observed behavior in Fig.~\ref{fig:bindingquantitative}. $ \deltaE\varphi_{E}^{H-1\text{PAT}1}$ reaches a plateau at mass ratio $\mathring\nu\sim10^{-4}$ due to the limited numerical precision of the $1$PA binding energy of the hybrid model. This unwanted behavior can easily be overcome by increasing the numerical precision of the interpolated binding energy.

The impact of the $1$PA correction to the binding energy on the dephasing is more pronounced for retrograde orbits than for prograde orbits: for $\mathring\chi=-0.5$, the accumulated phase difference between the $0$PA and hybrid models over the range $[x_i,x_f]=[0.079,0.135]$ settles down to $\sim10.2\ \text{rad}$. As the hybrid binding energy is $1$PA complete, this estimate also serves as the $1$PA contribution to the accumulated phase in Eq. \eqref{eq:accphase}: $\varphi_{(1)}(\mathring\nu t)\sim 10.2\ \text{rad}$ for $\mathring\chi=-0.5$ and $[x_i,x_f]=[0.079,0.135]$. Similarly, for prograde orbits, $\varphi_{(1)}(\mathring\nu t)\sim 2.32\ \text{rad}$ for $\mathring\chi=0.5$ and $[x_i,x_f]=[0.086,0.228]$. 

Since the PN binding energy has not been resummed to encapsulate the geodesic binding energy, it is  not $0$PA complete and $ \deltaE\varphi_{E}^{\text{PN}-X}\sim\mathcal{O}(1/\mathring\nu)$ for any model $X$ that is complete at $0$PA. This is indeed what is observed in Fig. \ref{fig:bindingquantitative}. However, we note that EOB models encode the geodesic dynamics by construction~\cite{Buonanno:1998gg}, meaning they are not susceptible to this error. 

Finally, we examine whether our use of an exact 1SF (1PA) binding energy has a significant impact on our phasing. To answer this question, we measure the phase difference~\eqref{eq:dephasingbinding2} between our hybrid model and the 0PA4PN model, which is exact at 0PA but uses the 4PN binding energy in place of the full 1SF binding energy. As we see in the green curves in Fig.~\ref{fig:bindingquantitative}, the impact can be large: the 0PA4PN model suffers dephases from the hybrid model by between 0.07 radians ($\mathring\chi=0.5$) and 3 radians ($\mathring\chi=-0.5$). This reinforces a similar conclusion reached for the nonspinning case in Ref.~\cite{Leather:2025nhu}.

\subsection{Comparison with \texttt{InspiralESIGMAHM}}

Our comparisons have focused on benchmarking our model against 0PA, PN, and SEOBNRv5 models. In the companion letter~\cite{PaperIV}, we also compare against TEOBResumS. However, in addition to EOB, there are other notable hybrid models that incorporate SF and PN information, such as \texttt{ENIGMA} and its successor \texttt{ESIGMAHM}~\cite{Huerta:2016rwp,Huerta:2017kez,Paul:2024ujx}. Here we briefly compare our hybrid model to this family of models. The original \texttt{ENIGMA} model for nonspinning binaries~\cite{Huerta:2016rwp,Huerta:2017kez} utilized 6PN approximations to the 0PA (1SF) energy flux at infinity and to the 1PA (1SF) binding energy~\cite{Barausse:2011dq} to build hybrid PN-SF expressions for these quantities. \texttt{ESIGMAHM} extended the model to incorporate spin effects (while reducing the use of SF information)~\cite{Paul:2024ujx}. This construction is  similar in spirit to our model, but it differs in important ways. As stressed in the Introduction, our model prioritizes ``exact'' SF results, using the complete 0PA energy fluxes, 0PA waveform mode amplitudes, and 1PA binding energy, while \texttt{ENIGMA}/\texttt{ESIGMAHM} uses relatively low-order PN expansions of the 0PA flux and 1PA binding energy and uses strictly PN waveform amplitudes. Moreover, while we leave our evolution equations exact in terms of the fluxes and binding energy, the \texttt{ENIGMA}/\texttt{ESIGMAHM} models fully expand $dx/dt$ in a PN series. 

Part of this difference in approach stems from us targeting high mass ratios, which  \texttt{ENIGMA} and \texttt{ESIGMAHM} do not. However, we find a significant accuracy advantage in our ``SF forward'' approach even for more comparable masses. The inspiral-only model \texttt{InspiralESIGMAHM} exhibits mismatches with \texttt{SEOBNRv4}~\cite{Bohe:2016gbl}, \texttt{IMRPhenomXAS}~\cite{Pratten:2020fqn}, and \texttt{SEOBNRv5\_ROM}~\cite{Pompili:2023tna} in the range $10^{-2}-10^{-3}$ for prograde orbits (see Fig.~3 of~\cite{Paul:2024ujx}). For an equivalent configuration, our model exhibits mismatches of order $10^{-4}-10^{-6}$ against a set of NR simulations (in a frequency window that starts at the minimal reference frequency of each NR simulation and ends at the ISCO for each spin, which is a larger frequency window than the one considered in Fig.~3 of~\cite{Paul:2024ujx}); see Fig.~\ref{fig:Spaghet}. On the other hand, our results also suggest the importance of including the most accurate PN information: our pure PN model exhibits mismatches in the range $10^{-2}-10^{-5}$; while less accurate than our hybrid model, this PN model appears to significantly outperform \texttt{InspiralESIGMAHM} in many cases.

\section{Outlook}\label{sec:outlook}

Building faithful and fast waveform models for IMRIs and EMRIs is an open challenge that needs to be addressed to maximize the science output of current and future GW detectors. Due to the lack of NR simulations for mass ratios higher than $\approx$15--20, only SF, PN, and PM theory give us first-principles descriptions of such systems. While SF theory alone should suffice for modelling strong-field inspirals, it is currently limited to leading, 0PA order for all but the simplest systems; moreover, it loses accuracy in the weak field (unless carried to 2PA order or beyond~\cite{Albertini:2022rfe}). As we outlined in the Introduction, the best strategy to overcome these limitations is to use all the first-principles information that is available, as espoused in the ``tutti frutti'' philosophy~\cite{Bini:2019nra,Bini:2020wpo,Bini:2020nsb,Bini:2020hmy,Bini:2020rzn,Bini:2023byt}, for example. 

To this end, we have laid down a new framework that allows us to incorporate information from both PN and SF expansions into a first-principles, accurate, efficient scheme for asymmetric compact binary inspirals. The efficiency of our scheme  leverages the multiscale framework, in which necessary data on parameter space is pre-computed in an offline stage, and the online waveform generation then only involves a rapid evolution through phase space using this pre-computed data.

We constructed the first such hybrid waveform model by consistently assembling state-of-the-art numerical 1SF and analytical 4PN (4.5PN in the nonspinning sector) results for the quasicircular inspiral of compact binaries with a spinning primary. Our hybridization scheme is based upon the consistent truncated hybridization of four defining quantities of the flux-balance laws governing the physical gravitational waveform evolution: the fluxes of energy at infinity and at the primary black hole's horizon, the flux of angular momentum at the horizon, and the binding energy. Our model takes into account the hybridization of the  amplitudes of all harmonic modes beyond the 22 mode (up to the highest values of $\ell$ for which the PN modes are nonzero). Our quasicircular hybrid model can be straightforwardly updated as new SF and PN results are obtained. 

We tested the faithfulness of our hybrid waveform model against 50 NR simulations from the SXS catalog for the power spectral density of the AdVirgo+ detector for Run O5. For comparison, we also tested the faithfulness of the separate purely PN and 0PA SF models. The resulting mismatches between the hybrid model and the NR templates are in the range $10^{-4} - 10^{-6}$, while the PN and 0PA models have much worse mismatches in the range $10^{-2} - 10^{-5}$ and $10^{-1} - 10^{-4}$, respectively. More precisely, the hybrid model improves the median mismatch by a factor of 40 or 2000 with respect to, respectively, PN or 0PA SF  models taken separately.

We also assessed the dephasing between the hybrid model and the NR templates, which is generally below $\deltaE\psi_{22}^H\lesssim1\text{ rad}$ during the frequency interval considered, a strong improvement over the 0PA SF and purely PN models. The hybrid model performed also remarkably well against other state-of-the-art waveform models such as the 2SF model \texttt{1PAT1} in the case of nonspinning binaries and the EOB model \texttt{SEOBNRv5EHM} restricted to quasicircular inspirals. Our analysis showed the significant impact of using high-precision SF data without PN truncation. It is known that 0PA terms must be exact to at least 6 digits~\cite{Khalvati:2025znb} for EMRIs since they contribute an amount of order $1/\mathring\nu$ to the waveform phase. Requirements on 1PA terms are less severe because they contribute a mass-ratio-independent amount to the phase, but our analysis suggests that truncating the 1PA (1SF) binding energy at 4PN order can lead to several radians of dephasing in the frequency intervals we consider. This importance of avoiding PN truncations in certain terms is also supported by the analysis in Ref.~\cite{Leather:2025nhu}. It is additionally supported by comparison with another PN-SF hybrid model, \texttt{InspiralESIGMAHM}; our model performs significantly better against NR, a result which we ascribe to our use of exact SF data.

In the EMRI regime, our model's limiting factor is its use of a 4PN approximation to the 1PA (2SF) energy flux to infinity. By comparing our hybrid model with the 2SF-exact \texttt{1PAT1} model, we found that this truncation results in a dephasing of around $0.09\, \text{rad}$  during the frequency interval considered for nonspinning quasicircular inspirals, dramatically lower than the dephasing due to PN truncation of the 1PA binding energy. This agrees with previous studies~\cite{Isoyama:2012bx,Albertini:2022dmc}, which likewise indicated that PN truncation of the 2SF flux has a relatively small effect. However, it is less clear that this extends to the spinning case; through comparisons with NR, we arrived at a significantly larger (though still subradian) estimated dephasing due to our inexact 2SF flux, particularly for retrograde systems. Unlike in the non-spinning case, our comparisons with NR also did not go deep enough into the small-$\mathring\nu$ regime to safely extrapolate to the EMRI case. Hence, fully numerical 2SF flux results will likely be needed to assess the accuracy of our model for spinning EMRIs.

We have made our model publicly available as \texttt{WaSABI-C} v0.9 (Waveform Simulations of Asymmetric Compact Binary Inspirals - Circular) within the \href{https://bhptoolkit.org/WaSABI/}{\textsc{WaSABI}} Mathematica package~\cite{BHPT_WaSABI}. In the companion Letter~\cite{PaperIV}, we extend the model to \texttt{WaSABI-C} v1.0, which includes the complete 1PA secondary-spin information for generic primary spin, all available PN secondary-spin information (excluding quartic terms), and the complete 1PA (2SF) energy flux to infinity in the nonspinning sector, utilizing SF results from Refs.~\cite{Warburton:2021kwk,Piovano:2024yks,PaperII}. The positive outcome of our faithfulness analysis here and in the companion Letter encourages us to further extend our baseline hybrid model to include additional physical parameters in the hybridization framework, particularly eccentricity and inclination (i.e., planar precession). Such extensions will likewise be made publicly available for community use in \href{https://bhptoolkit.org/WaSABI/}{\textsc{WaSABI}}. These models are already fast enough to use in accuracy comparisons between models, but to make them usable in data analysis we will implement them in the FEW infrastructure~\cite{Katz:2021yft,Chapman-Bird:2025xtd}. By design, such implementation poses no barriers.

This straightforward integration into FEW, and ultimately into EMRI data analysis pipelines that utilize FEW, is a major part of our motivation. Relative to other types of hybrid and composite models, such as variants of EOB, our hybrid model is ``self-force forward'', not just in using as much exact SF information as is available, but also in structuring our model in the multiscale form amenable to implementation in FEW. This ultimately ensures that waveforms can be generated in tens of milliseconds, even for  eccentric binaries with spin precession~\cite{Katz:2021yft}. Otherwise, our hybrid model is broadly similar to EOB's inspiral waveform. Both our model and EOB treat the inspiral waveform as a function on the binary's phase space and generate waveforms by evolving trajectories through that phase space; but in the multiscale expansion for quasicircular orbits this phase space reduces to an angle and a frequency (the orbital $\phi_p$ and $d\phi_p/dt$ or the waveform's $\psi$ and $\omega)$, together with the primary's parameters, while in EOB one would typically solve for the trajectory in the full phase space $(r,\phi,p_r,p_\phi)$. Both frameworks allow the inclusion of PN and SF results, and both can add physical parameters and higher-order perturbative results in a modular way; but we do gain in simplicity by using ``exact'', fully relativistic SF data, including SF amplitudes, as we do not require Pad\'e resummations or the factorized, resummed form of the waveform itself that is employed in EOB models. Important future work will be to compare our model with versions of EOB that specifically target high mass ratios~\cite{Nagar:2022fep,Albanesi:2023bgi,Leather:2025nhu}.

Given our model's accuracy and efficiency for almost all mass ratios, we anticipate it will be directly relevant for analysis of high-$\mathring q$ systems detectable with ground-based detectors such as LVK and ET, in addition to IMRIs and EMRIs observable with LISA. However, for systems detectable by ground-based detectors, we expect that our model must be extended to the merger and ringdown phases. References~\cite{Kuchler:2024esj,Kuchler:2025hwx,Roy:2025kra,Honet:2025dho} have laid out a method of performing this extension while preserving the capability of rapid waveform generation.   

\begin{acknowledgments}

We gratefully acknowledge Maarten van de Meent for providing first-order self-force data. We also thank Chris Kavanagh, Josh Mathews, Zach Nasipak, David Trestini, Maarten van de Meent, Niels Warburton, and Barry Wardell for helpful discussions. AP acknowledges the support of a Royal Society University Research Fellowship and the ERC Consolidator/UKRI Frontier Research Grant GWModels (selected by the ERC and funded by UKRI [grant number EP/Y008251/1]). LH acknowledges the support of the Fonds National pour la Recherche Scientifique through a FRIA doctoral grant. G.C. acknowledges financial support from the Win4Project ETLOG grant of the Walloon Region.  G.C. is Research Director of the FNRS. This work makes use of the Black Hole Perturbation Toolkit and PNpedia. Part of the numerical computations supporting this work were performed on the Lyra cluster hosted at ULB.

\end{acknowledgments}

\appendix
\section{Chebyshev interpolation of self-force data}
\label{sec:ChebyshevInterpolation}

In this Appendix we describe our interpolation of the 0PA energy flux, 1PA binding energy, and 0PA waveform mode amplitudes.

\subsection{Interpolation of the energy flux}\label{sec:ECheb}

The 0PA energy fluxes through the primary's horizon and to null infinity are found in the standard way~\cite{Barack:2018yvs,Pound:2021qin} by solving the Teukolsky equation sourced by a point mass on a circular geodesic. Explicitly, the Weyl scalar $\Psi_4$ is decomposed in spin-weighted tensor spheroidal harmonics as~\cite{Pound:2021qin} 
\begin{equation}
\zeta^4\Psi_4=\sum_{\ell' =2}^\infty\sum_{m'=-\ell'}^{\ell'} \prescript{}{-2}{\Psi}_{\ell' m'}(\tilde 
 r)\prescript{}{-2}{S}_{\ell' m'}(\theta,\phi)
\end{equation}
with $\hat r=r/\MATHRINGm1$,  $\zeta=\hat r-i\mathring\chi\cos{\theta}$, and $\prescript{}{-2}{S}_{\ell' m'}$ satisfying the spin-weight $-2$ spheroidal harmonic equation. The radial coefficients $\prescript{}{-2}{\Psi}_{\ell' m'}(\hat r)$ satisfy the Teukolsky radial equation sourced by $\prescript{}{-2}{T}_{\ell' m'}$, which is constructed from the stress-energy tensor of the point mass. Primed indices $(\ell',m')$ refer to a decomposition in spheroidal harmonics, while unprimed indices $(\ell,m)$ (as used in the body of the paper) refer to spherical harmonics. The radial equation for $\prescript{}{-2}{\Psi}_{\ell' m'}(\hat r)$ is solved through the method of variation of constants~\cite{Pound:2021qin}:
\begin{equation}
    \prescript{}{-2}{\Psi}_{\ell' m'}=\prescript{}{-2}{C}^{\rm in}_{\ell' m'}\prescript{}{-2}R^{\rm in}_{\ell' m'}+\prescript{}{-2}{C}^{\rm up}_{\ell' m'}\prescript{}{-2}R^{\rm up}_{\ell' m'},\label{eq:varconstant}
\end{equation}
where $\prescript{}{-2}R^{\rm in}_{\ell' m'}(\hat r)$ and $\prescript{}{-2}R^{\rm up}_{\ell' m'}(\hat r)$ are two linearly independent homogeneous solutions representing an ingoing wave at the horizon and an outgoing wave at infinity, respectively. The coefficients are given by
\begin{align}
    \prescript{}{-2}{C}^{\rm in}_{\ell' m'}(\hat r)&=\int_{\hat r}^\infty \frac{\prescript{}{-2}R^{\rm up}_{\ell' m'}(p)}{W(p)\Delta(p)}\prescript{}{-2}{T}_{\ell' m'}
(p) dp,\\
    \prescript{}{-2}{C}^{\rm up}_{\ell' m'}(\hat r)&=\int_{\hat r_+}^{\hat r} \frac{\prescript{}{-2}R^{\rm in}_{\ell' m'}(p)}{W(p)\Delta(p)}\prescript{}{-2}{T}_{\ell' m'}
(p) dp,
\end{align}
with $\hat r_+=1+\sqrt{1-\mathring\chi^2}$ the outer event horizon radius, $W$ the Wronskian between $\prescript{}{-2}R^{\rm in}_{\ell' m'}(\hat r)$ and $\prescript{}{-2}R^{\rm up}_{\ell' m'}(\hat r)$, and $\Delta(r)=\hat r^2-2\hat r+\mathring\chi^2$.

The $(\ell',m')$ mode contribution to the 0PA energy fluxes can be written as
\begin{align}
    \mathcal{F}^\infty_{(0)\ell'm'}&=\frac{2\pi}{(m'\hat\Omega)^2}\left|\prescript{}{-2}{C}^{\rm up}_{\ell' m'}\right|^2\\
    \mathcal{F}^\mathcal{H}_{(0)\ell'm'}&=\frac{2\pi\alpha_{\ell'm'}}{(m'\hat\Omega)^2}\left|\prescript{}{-2}{C}^{\rm in}_{\ell' m'}\right|^2,
\end{align}
with the coefficients $\alpha_{\ell'm'}$ given in Eq. (97) of~\cite{Pound:2021qin}, $\hat\Omega=\MATHRINGm1 \Omega$, and 
\begin{align}
\prescript{}{-2}{C}^{\rm up}_{\ell' m'} &=\lim_{\hat r\rightarrow\infty}\prescript{}{-2}{C}^{\rm up}_{\ell' m'}(\hat r),\label{Cup}\\
\prescript{}{-2}{C}^{\rm in}_{\ell' m'}&=\lim_{\hat r\rightarrow \hat r_+}\prescript{}{-2}{C}^{\rm in}_{\ell' m'}(\hat r). 
\end{align}
The $0$PA fluxes $\mathcal{F}^{\infty\, }_{(0)}\left(\hat{\Omega}^{2/3},\mathring\chi\right)$ and $\mathcal{F}^{\mathcal{H}\, }_{(0)}\left(\hat{\Omega}^{2/3},\mathring\chi\right)$ are then computed through a sum over $(\ell' ,m')$ modes.

Points in the two-dimensional parameter space of quasicircular trajectories around a primary Kerr black hole can be labeled by the 0PA orbital radius $r_{(0)}\in [r_\star(\mathring \chi),\infty)$, ranging from the ISCO radius $r_\star(\mathring \chi)$ up to infinity, and by the primary black hole's spin per unit mass ${\mathring\chi} = \mathring a/\MATHRINGm1\in[-1,1]$. In order to perform a Chebyshev interpolation, we first compactify the parameter space into a square using the transformation 
\begin{subequations}\label{parametertransfo}
\begin{align}
r_\text{C}&=2r_\star(\mathring\chi)/r_p-1,\\
\mathring\chi_\text{C}&=\mathring\chi.
\end{align}
\end{subequations}
The parameters $(r_\text{C},\mathring\chi_\text{C})$ now lie in the range $(r_\text{C},\mathring\chi_\text{C})\in[-1,1]\times[-1,1]$ and a Chebyshev interpolation can be performed. We use data provided by Maarten van de Meent~\cite{MaartenPrivate}, which consists of the horizon and null infinity energy fluxes $\mathscr{F}^a_{ij}=\mathscr{F}^a(r_{\text{C},i},\mathring\chi_{\text{C},j})$ ($a\in\{\infty,\mathcal{H}\}$) including all modes up to $\ell' =50$ evaluated on Chebyshev nodes located at
\begin{equation}\label{ChebyshevNodes} (r_{\text{C},i},\mathring\chi_{\text{C},j})=\left(\cos{\left(\frac{2i+1}{n_R}\frac{\pi}{2}\right)},\cos{\left(\frac{2j+1}{n_A}\frac{\pi}{2}\right)}\right)
\end{equation}
with $n_R=18$ and $n_A=27$. In the companion letter~\cite{PaperIV}, we upgrade to a refined grid \eqref{ChebyshevNodes} with $n_R=36$, $n_A=36$, where the compactification \eqref{parametertransfo} is upgraded to 
\begin{align}\label{parametertransfo2}
r_\text{C}&=\sqrt{r_\star(\mathring\chi)/r_{(0)}},\\
\mathring\chi_\text{C}&=\frac{\log(1-\mathring \chi)-\log(1-\mathring \chi_\text{max})}{\log 2 - \log (1-\mathring \chi_\text{max})},
\end{align}
defined in terms of the Thorne limit $\chi_\text{max}=0.998$~\cite{1974ApJ...191..507T}. The parameters $(r_\text{C},\mathring\chi_\text{C})$ lie in the range $[0,1] \times [0,1]$. The logarithmic scaling of the primary spin in Eq. \eqref{parametertransfo2} increases the convergence of the interpolation. With this refined grid, the estimated relative interpolation error is $10^{-10}$~\cite{PaperIV}.

\begin{figure}[tp]
\includegraphics[width=.49\textwidth]{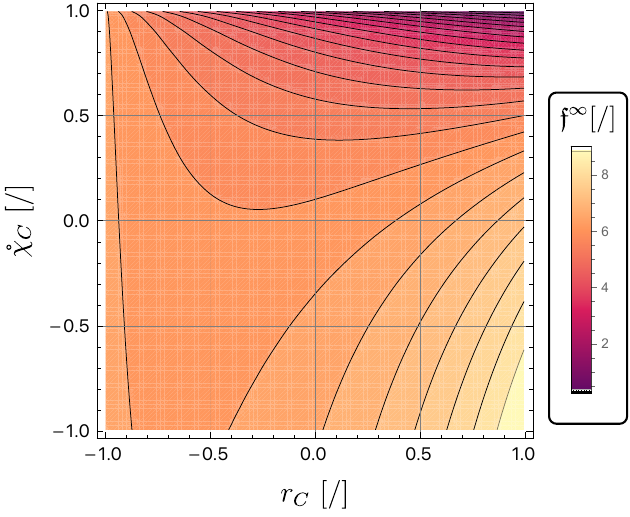}
\caption{\label{fig:eChebyshevGrid}Interpolating function $\mathfrak{f}^\infty(r_\text{C},\mathring\chi_\text{C})$ on a $(r_\text{C},\mathring\chi_\text{C})$ grid as defined in Eqs. \eqref{parametertransfo} and \eqref{ChebyshevNodes}. The right boundary $r_\text{C}=1$ corresponds to the particle being located at the ISCO $r_{(0)}=r_\star(\mathring\chi)$, while the left boundary $r_\text{C}=-1$ corresponds to the particle being located at infinity. The top boundary corresponds to prograde orbits around an extremal Kerr primary, while the bottom boundary corresponds to retrograde orbits around an extremal Kerr primary.}
\end{figure}

Before performing the Chebyshev interpolation, we factor out from $\mathscr{F}^a_{ij}$ its leading-order PN behavior as $r_{(0)}\rightarrow\infty$. That is to say, we rather interpolate the data $\mathfrak{f}^a_{ij}$ defined on each node through
\begin{align}
    \mathscr{F}^\infty_{ij}(r_{\text{C},i},\mathring\chi_{\text{C},j})&=\mathfrak{f}^a_{ij}(r_{\text{C},i},\mathring\chi_{\text{C},j})/r_{(0)}^5(r_{\text{C},i},\mathring\chi_{\text{C},j}),\\ 
    \mathscr{F}^{\mathcal H}_{ij}(r_{\text{C},i},\mathring\chi_{\text{C},j})&=\mathfrak{f}^a_{ij}(r_{\text{C},i},\mathring\chi_{\text{C},j})/r_{(0)}^{\frac{15}{2}}(r_{\text{C},i},\mathring\chi_{\text{C},j}). 
\end{align}
The resulting functions $\mathfrak{f}^a(r_{\text{C}},\mathring\chi_{\text{C}})$ are of order unity over the parameter space and therefore have a more rapidly convergent Chebyshev interpolation. We deduce the functions $\mathcal{F}_{(0)}^{\infty \, }(\hat\Omega^{2/3},\mathring\chi)$ and $\mathcal{F}_{(0)}^{\mathcal{H} \, }(\hat\Omega^{2/3},\mathring\chi)$ from
\begin{align}
\mathcal{F}_{(0)}^{\infty\, }\left(\hat\Omega^{2/3},\mathring\chi\right) &=\frac{\hat\Omega^{10/3}}{(1-\mathring\chi \hat\Omega)^{10/3}}\mathfrak{f}^\infty\left(\frac{2r_\star\hat\Omega^{2/3}}{(1-\mathring\chi \hat\Omega)^{2/3}}-1,{\mathring\chi}\right),\\
\mathcal{F}_{(0)}^{\mathcal{H}\, }\left(\hat\Omega^{2/3},\mathring\chi\right) &=\frac{\hat\Omega^{15/3}}{(1-\mathring\chi \hat\Omega)^{15/3}}\mathfrak{f}^\mathcal{H}\left(\frac{2r_\star\hat\Omega^{2/3}}{(1-\mathring\chi \hat\Omega)^{2/3}}-1,{\mathring\chi}\right).
\end{align}
As an example, we display the interpolating function  of $\mathfrak{f}^\infty(r_\text{C},\mathring\chi_\text{C})$ in Fig. \ref{fig:eChebyshevGrid}. 

In Eqs. \eqref{fluxSF}, we reexpanded the energy fluxes in terms of the dynamical variables $\nu$, $\chi$, and $x=\omega^{2/3}$, where $\omega$ is the 22 mode waveform frequency. Since $\mathring\nu=\nu+\mathcal{O}(\nu^2)$, $\mathring\chi=\chi+\mathcal{O(\nu)}$, and $\hat\Omega^{2/3}=x+\mathcal{O}(\nu)$, the reexpansion of the $0$PA fluxes solely reduces to replacing $\hat\Omega^{2/3}$ by $x$ and $\mathring\chi$ by $\chi$:
\begin{align}
\mathcal{F}_{(0)}^{\infty \text{SF}}\left(x,\chi\right) &=\mathcal{F}_{(0)}^{\infty  }(x,\chi),\\
\mathcal{F}_{(0)}^{\mathcal{H} \text{SF}}\left(x,\chi\right) &=\mathcal{F}_{(0)}^{\mathcal H }(x,\chi).
\end{align}

\subsection{Interpolation of the binding energy}
\label{app:1b}

The next-to-leading order correction to the binding energy, $E_{(1)}$, can be obtained from 1SF computations of the Detweiler redshift~\cite{Shah:2012gu,Isoyama:2014mja,van_de_Meent_2015}. We again make use of data provided by Maarten van de Meent~\cite{MaartenPrivate} for the $1$SF correction to the  redshift, $z_{(1)}(\hat\Omega^{2/3},\mathring\chi)=Z_{(1)}(r_{\text{C}},\mathring\chi_{\text{C}})$.

We use the same Chebyshev grid as for the energy fluxes, factoring out the leading PN behavior of $Z_{(1)}$ as
\begin{align}
    Z_{(1)}(r_{\text{C},i},\mathring\chi_{\text{C},j})=\mathfrak{Z}_{(1)}(r_{\text{C},i},\mathring\chi_{\text{C},j})/r_{(0)}(r_{\text{C},i},\mathring\chi_{\text{C},j})
\end{align}
and interpolating the function $\mathfrak{Z}_{(1)}(r_{\text{C}},\mathring\chi_{\text{C}})$. We can then write the redshift factor as
\begin{align}
    z_{(1)}(\hat\Omega^{2/3},\mathring\chi)&=Z_{(1)}(r_{\text{C}},\mathring\chi_{\text{C}})\\
    &=\frac{\hat\Omega^{2/3}}{(1-\mathring\chi\hat\Omega)^{2/3}}\mathfrak{Z}_{(1)}\left(\frac{2r_\star \hat\Omega^{2/3}}{(1-\mathring\chi \hat\Omega)^{2/3}}-1,{\mathring\chi}\right).
\end{align}

The function $ z_{(1)}(\hat\Omega^{2/3},\mathring\chi)$ is related to the $1$SF correction to the binding energy as~\cite{Isoyama:2014mja}
\begin{align}
    E_{(1)}\!\left(\hat{\Omega}^{2/3},\mathring\chi\right)=\frac{1}{2}\left(z_{(1)}-\hat{\Omega}\frac{\partial z_{(1)}}{\partial \hat\Omega}\right).
\end{align}
In contrast to the energy fluxes, this correction occurs at next-to-leading order,  the leading-order term being the geodesic binding energy $\hat E^{\rm geo}\left(\hat\Omega^{2/3},\mathring\chi\right)$ as appears in Eq.~\eqref{Egeo}. Therefore, a special care needs to be taken when reexpanding the binding energy to the set of variables $(x,\chi,\nu,\MTOT)$. Indeed, starting with the expansion
\begin{equation}
    E=m_2 \left[\hat{E}^\text{geo}\left(\hat{\Omega}^{2/3},\mathring\chi\right)+\mathring\varepsilon E_{(1)}\left(\hat{\Omega}^{2/3},\mathring\chi\right)\right]+\mathcal{O}(\mathring\varepsilon^3)
\end{equation}
and reexpanding to $\nu$ at fixed $(x,\chi,\MTOT)$ yields 
\begin{align}
    E&=\nu \MTOT \hat{E}^\text{geo}\!\left(x,\chi\right)\nonumber +\nu^2 \MTOT E^{\rm SF}_{(1)}(x,\chi)
+\mathcal{O}(\nu^3),
\end{align}
with $ E^{\rm SF}_{(1)}$ given by Eq. \eqref{eq:E1reexp}.

\subsection{Mode amplitudes in spherical basis}\label{sec:AmpChebyshev}

The mode amplitudes $\tilde h_{\ell' m'}$, $m'\neq 0$,  written in the basis of spin-weighted spheroidal harmonics, are related to the Teukolsky coefficients $\prescript{}{-2}{C}^{\rm up}_{\ell' m'}$ from Eq.~\eqref{Cup} according to~\cite{Pound:2021qin}
\begin{equation}\label{defom}
    \tilde h_{\ell' m'}=\frac{2}{(m'\Omega)^2}\prescript{}{-2}{C}^{\rm up}_{\ell' m'}.
\end{equation}
Since PN waveforms are decomposed in spin-weighted spherical harmonics, hybridizing the amplitudes requires us to first reexpress the Teukolsky amplitudes in the basis of spin-weighted spherical harmonics, following Refs.~\cite{Hughes:1999bq,Cook:2014cta,Stein:2019mop}:
\begin{equation}
    \prescript{}{-2}{C}^{\rm up}_{\ell m}= \delta_{mm'}\sum_{\ell'=2}^\infty \mu_{\ell,\ell'}(\gamma_{m}) \prescript{}{-2}{C}^{\rm up}_{\ell' m'},
\end{equation}
where $\mu_{\ell,\ell'}(\gamma_{m})$ are the coefficients of the change of basis, which depend upon the spheroidicity parameter $\gamma_{m} = m\mathring{\chi}\hat \Omega$.

Using the \texttt{Teukolsky} Mathematica package~\cite{TeukolskyPackage} available on the Black Hole Perturbation Toolkit~\cite{BHPToolkit}, we can analytically find the coefficients $\mu_{\ell,\ell'}(\gamma_{m})$ as a Taylor expansion in the spheroidicity parameter. For example, for the $(\ell,m)=(2,2)$ mode we have 
\begin{align}
    &\prescript{}{-2}{C}^{\rm up}_{\ell m} = \prescript{}{-2}{C}^{\rm up}_{2'2'}-\gamma_{2'} \left(\frac{2}{9} \sqrt{\frac{5}{7}}\prescript{}{-2}{C}^{\rm up}_{3'2'}\right)\nonumber\\
    &\ \ +\gamma_{2'}^2\left(\frac{\sqrt{5}}{294}\prescript{}{-2}{C}^{\rm up}_{4'2'}-\frac{1}{162}\sqrt{\frac{5}{7}}\prescript{}{-2}{C}^{\rm up}_{3'2'}-\frac{10}{567}\prescript{}{-2}{C}^{\rm up}_{2'2'}\right)\nonumber\\
    &\ \ +\gamma_{2'}^3 \left(\frac{1}{270 \sqrt{55}}\prescript{}{-2}{C}^{\rm up}_{5'2'}+\frac{\sqrt{5}}{588}C^{\text{SF}}_{4'2'}+\frac{\sqrt{35}}{729}\prescript{}{-2}{C}^{\rm up}_{3'2'}\right.\nonumber\\
    &\left.\phantom{000000}-\frac{{5}}{5103}\prescript{}{-2}{C}^{\rm up}_{2'2'}\right)+\mathcal{O}(\gamma_{2'}^4),
\end{align}
with $\gamma_{2'}=2\mathring\chi\hat\Omega$.

Rather than interpolating the mode amplitudes in spheroidal basis and then performing the change of basis, we first perform the change of basis to spherical harmonics using the spectral method of Ref.~\cite{Cook:2014cta} and then perform a Chebyshev interpolation. Explicitly, we use the same Chebyshev grids as for the energy fluxes and binding energy, Eq. \eqref{ChebyshevNodes} with $n_R=18$ and $n_A=27$. On each node $(r_{\text{C},i},\mathring\chi_{\text{C},j})$, we use the \texttt{Teukolsky} Mathematica package~\cite{TeukolskyPackage} to compute the Teukolsky amplitudes $c^{\text{SF}}_{\ell' m',(0)}\left(r_{\text{C},i},\mathring\chi_{\text{C},j}\right)$  in the basis of spheroidal harmonics. Here, we have defined $c^{\text{SF}}_{\ell' m',(0)}\left(r_{\text{C}},\mathring\chi_{\text{C}}\right)\equiv \prescript{}{-2}{C}^{\rm up}_{\ell' m'}\left(\hat\Omega^{2/3},\mathring\chi\right)$. Using the spectral method of~\cite{Cook:2014cta}, we perform the change of basis to spherical harmonics at each node of the Chebyshev grid:
\begin{align}
    c^{\text{SF}}_{\ell m,(0)}\left(r_{\text{C},i},\mathring\chi_{\text{C},j}\right)&=\sum_{\ell'=2}^\infty \mu_{\ell,\ell'}\left(m\mathring\chi_{\text{C},j}\hat\Omega_{i,j}\right) \nonumber\\
    &\quad \times c^{\text{SF}}_{\ell' m,(0)}\left(r_{\text{C},i},\mathring\chi_{\text{C},j}\right),
\end{align}
where $\hat{\Omega}_{i,j}=\left[\mathring\chi_{\text{C},j}+\left(\frac{2r_{\star}(\mathring \chi_{C,j})}{r_{\text{C},i}+1}\right)^{3/2}\right]^{-1}$. Afterwards, we factor out the leading PN behavior of the mode amplitudes as well as a conventional factor $1/\hat \Omega^2$ originating from the definition \eqref{defom}. Explicitly, we define $\mathfrak{c}^{\text{SF}}_{\ell m,(0)}$ as 
\begin{align}
   \mathfrak{c}^{\text{SF}}_{\ell m,(0)} 
   &=\frac{r_{p,ij}^{\ell/2}}{\hat\Omega_{i,j}^2}c^{\text{SF}}_{\ell m,(0)}
   &\text{for $\ell+m$ even},\\
   \mathfrak{c}^{\text{SF}}_{\ell m,(0)}
   &=\frac{r_{p,ij}^{(1+\ell)/2}}{\hat\Omega_{i,j}^2}c^{\text{SF}}_{\ell m,(0)}
   &\text{for $\ell+m$ odd},
\end{align}
and perform the Chebyshev interpolation on the functions $\mathfrak{c}^{\text{SF}}_{\ell m,(0)}\left(r_{\text{C},i},\mathring\chi_{\text{C},j}\right)$. Here, $r_{p,ij}=2r_{\star,j}/\left(r_{\text{C},i}+1\right)$. We then deduce the mode amplitudes written in the basis of spherical harmonics to be
\begin{widetext}
\begin{align}
    C^{\text{SF}}_{\ell m,(0)}\left(\hat\Omega^{2/3},\mathring\chi\right)&=\frac{\hat\Omega^{2+\ell/3}}{(1-\mathring\chi\hat\Omega)^{\ell/3}}\mathfrak{c}^{\text{SF}}_{\ell m,(0)}\left(\frac{2r_\star \hat\Omega^{2/3}}{(1-\mathring\chi \hat\Omega)^{2/3}}-1,{\mathring\chi}\right)&\text{for}\ \  \ell+m\ \  \text{even},\\
    C^{\text{SF}}_{\ell m,(0)}\left(\hat\Omega^{2/3},\mathring\chi\right)&=\frac{\hat\Omega^{2+(1+\ell)/3}}{(1-\mathring\chi\hat\Omega)^{(1+\ell)/3}}\mathfrak{c}^{\text{SF}}_{\ell m,(0)}\left(\frac{2r_\star \hat\Omega^{2/3}}{(1-\mathring\chi \hat\Omega)^{2/3}}-1,{\mathring\chi}\right)&\text{for}\ \  \ell+m\ \  \text{odd}.
\end{align}
\end{widetext}
At leading SF order, reexpanding the mode amplitudes to fixed $x$ and evolving spin $\chi$ only amounts to replacing $\hat\Omega^{2/3}\rightarrow x$ and $\mathring\chi\rightarrow\chi$ in the above expressions.

\section{Post-Newtonian data}
\label{app:PN}

In this appendix we summarize the PN expressions we use for the energy flux to null infinity, the binding energy, the energy flux through the primary's horizon, and the waveform amplitudes.

\subsection{$\mathcal F^\infty_{\text{PN}}$}

First, the energy flux at null infinity is decomposed as the sum
\begin{align}
\mathcal F^\infty_{\text{PN}} &=\mathcal F^{\infty}_{4.5\text{PN}}+\mathcal F^{ \infty}_{SO\, 4\text{PN}}+\mathcal F^{ \infty}_{SS\, 4\text{PN}}+\mathcal F^{\infty}_{SSS\, 4\text{PN}}.     
\end{align}
We summarize each term in turn.

The energy flux at null infinity for nonspinning compact binaries was obtained at 4.5PN order, $\mathcal F^{\infty}_{4.5\text{PN}}$~\cite{Blanchet:2023bwj}, following~\cite{Marchand:2016vox,Marchand:2020fpt,Larrouturou:2021dma,Larrouturou:2021gqo}; see also~\cite{Blanchet:2023sbv,Trestini:2023wwg,Warburton:2024xnr}. We use Eq.~(483) of Blanchet's review~\cite{Blanchet:2013haa}, 
\begin{align}\label{F4.5PN}
\mathcal F^{\infty}_{4.5\text{PN}}(x,\nu) =\frac{32}{5}\nu^2 x^5 \left( 1+ \sum_{n=2}^9 \mathcal F^{\infty}_{\frac{n}{2}\text{PN}}(\nu) x^{n/2}\right). 
\end{align}

The spin-orbit contribution  $\mathcal F^{ \infty}_{SO\, 4\text{PN}}$ is known at 4PN precision~\cite{Blanchet:2006gy,Bohe:2013cla,Marsat:2013caa,Cho:2022syn}. We use Eq. (612) of Blanchet's review~\cite{Blanchet:2013haa}, 
\begin{align}
\mathcal F^{\infty}_{SO\, 4\text{PN}}(x,\nu ; \MTOT,a)&= \frac{32\nu^2 x^{13/2}}{5\MTOT^2}\Biggl[-4S_\ell-\frac{5}{4}\Delta \Sigma_\ell \nonumber\\
&+ \sum_{n=2}^5 x^{n/2} F^{\infty}_{SO, \frac{n}{2}}(\nu ;\MTOT,a) \Biggr],    
\end{align}
 with $S_\ell = m_1 a$, $\Sigma_\ell=- \MTOT a$ and $\Delta=\sqrt{1-4\nu}$.  
 
The spin-spin interaction term  $\mathcal F^{ \infty}_{SS}$ is known to 3PN order from Refs.~\cite{Bohe:2015ana,Cho:2021mqw}. At that order we use Eq.~(618) of Blanchet's review~\cite{Blanchet:2013haa} with $\kappa_+=2$ and $\kappa_-=0$ (the values for a Kerr black hole). The 3.5PN terms are derived in Eq.~(41) of Ref.~\cite{Cunningham:2024dog}. We complement this expression with the 4PN result from Eq.~(13) of~\cite{Cho:2022syn} to obtain
 \begin{align}
\mathcal F^{\infty}_{SS\, 4\text{PN}}(x,\nu ; \MTOT,a)&= \frac{32\nu^2 x^{7}}{5M_{\text{tot} }^4}\Biggl[\mathcal F^{\infty}_{SS,0}(\nu,S_\ell,\Sigma_\ell)\nonumber\\
&\hspace{-2cm}+  x \mathcal F^{\infty}_{SS,1}(\nu,S_\ell,\Sigma_\ell)+ x^{3/2} \mathcal F^{\infty}_{SS,3/2}(\nu,S_\ell,\Sigma_\ell)\nonumber \\
&\hspace{-2cm} +  x^2 \mathcal F^{\infty}_{SS,2}(\nu,S_\ell,\Sigma_\ell) \biggr]   . 
\end{align}

The cubic-in-spin interaction term through  4PN order is known from Ref.~\cite{Marsat:2014xea}. From its Eq. (6.18) we have
 \begin{align}
\mathcal F^{\infty}_{SSS\, 4\text{PN}}(x,\nu ; \MTOT,a)= \frac{32 \nu^2 x^{5+\frac{7}{2}}}{5\MTOT^6}f_{SSS}[S_\ell,\Sigma_\ell , \nu].
\end{align}
Note that the coefficient of $x^9$ in ${\cal F}^{\infty}_{SSS}$ identically vanishes, meaning the above 3.5PN term is the complete ${\cal F}^{\infty}_{SSS}$ through 4PN order.

These expressions can be written in terms of the variables $(x,\MTOT,\nu,\chi)$ after using the substitution 
\begin{equation}
    a=m_1\chi = \frac{\MTOT}{2}\left(1+\sqrt{1-4\nu}\right)\chi.\label{subsa}
\end{equation}
This substitution does not affect the PN counting (i.e., which terms are $n$PN), but it does affect the SF counting (i.e., which terms are $n$PA). 

\subsection{$E_{\text{PN}}$}

The binding energy is decomposed as the sum 
\begin{align}
E_{\text{PN}} &=E_{4\text{PN}}+E_{SO\, 4\text{PN}}+E_{SS\, 4\text{PN}}+E_{SSS\, 4\text{PN}}.     
\end{align}

For the contribution for nonspinning binaries, $E_{4\text{PN}}(x , \nu; \MTOT)$, we use the 4PN-accurate energy  written in Eq. (5.5) of~\cite{Damour:2014jta},
 \begin{multline}\label{E4PN}
E_{4\text{PN}} = -\frac{\nu \MTOT}{2} x \Biggl[ 1+\sum_{n=1}^3 e_{n\text{PN}}(\nu) x^n   
 + e_{4\text{PN}}(\nu,\log x) x^4 \Biggr] . 
\end{multline}

We emphasize that the expressions for $E_{\rm 4PN}$ in the literature are for the binary's local mechanical energy, written as a function of the orbital parameter we have denoted $y_{\rm PN}=(M\Omega_{\rm PN})^{2/3}$. As we stressed in Sec.~\ref{sec:binding energy}, the local mechanical energy differs from the Bondi mass beginning at 4PN order. Trestini recently showed that once one accounts for the Schott term relating the local mechanical energy to the Bondi mass, the 4PN binding energy (as defined from the Bondi mass) is simply given by the 4PN mechanical energy with the direct replacement $y_{\rm PN}\to x$~\cite{Trestini:2025nzr}; the Schott term relating the energies exactly cancels the correction relating the frequencies in Eq.~\eqref{xxorb}. This is what we use in Eq.~\eqref{E4PN}.

Since we do not account for the (not-yet-computed) Schott term in the 1SF (1PA) binding energy, this introduces a small inconsistency in our hybridization. However, the inconsistency is comparable to our already existing error in omitting the 1SF Schott term. 

In the spin sectors below, this subtlety does not arise. The contributions from the spin first enter the energy at subleading PN orders, meaning the difference between $x$ and $y_{\text{PN}}$ would only affect them at orders higher than 4PN. Hence, in those expressions we can use $y_{\text{PN}}$ and $x$ interchangeably.

The 4PN-accurate expression for the spin-orbit coupling term $E_{SO\, 4\text{PN}}(x , \nu; \MTOT,a)$ can be found in Eq. (592) of Blanchet's review~\cite{Blanchet:2013haa}. It takes the form
\begin{align}
E_{SO\, 4\text{PN}} &=-\frac{\nu x^{5/2}}{2\MTOT} \sum_{n=0}^2 e_{SO\, \left(n+1.5\right)\text{PN}}(S_\ell,\Sigma_\ell,\nu) x^n. 
\end{align}

We take the expression for the spin-spin coupling term $E_{SS\, 4\text{PN}}(x , \nu; \MTOT,a)$ from Eq.~(617) of~\cite{Blanchet:2013haa} supplemented by the 4PN results from Eq.~(12) of~\cite{Cho:2022syn}: 
 \begin{align}
E_{SS\, 4\text{PN}} &=-\frac{\nu x^{3}}{2\MTOT^3} \sum_{n=0}^4 e_{SS\, \left(\frac{n}{2}+2\right)\text{PN}}(S_\ell,\Sigma_\ell,\nu) x^{n/2} . 
\end{align}

Finally, the cubic-in-spin contribution through 4PN, $E_{SSS\, 4\text{PN}} (x , \nu; \MTOT,a)$, is taken from Eq. (6.16) of~\cite{Marsat:2014xea},
\begin{equation}
E_{SSS\, 4\text{PN}} = -\frac{\nu}{2\MTOT^5} x^{1+\frac{7}{2}}e_{SSS}(S_\ell,\Sigma_\ell , \nu). 
\end{equation}
Note that the 4PN terms ($\propto x^5$) in $E_{SO}$ and $E_{SSS}$ identically vanish. 

Again, we rewrite these expressions in terms of the variables $(x,\MTOT,\nu,\chi)$ using Eq.~\eqref{subsa}, which does not affect the PN counting but does affect the SF counting.

\subsection{$\mathcal F^{\mathcal H}_{\text{PN}}$}  

The energy flux through the horizon for spinning compact binaries starts at 2.5PN order (relative to the leading, quadrupole formula for the flux to infinity), while for nonspinning black holes it starts at 4PN order~\cite{Tagoshi:1997jy,Alvi:2001mx,Porto:2007qi,Chatziioannou:2012gq,Saketh:2022xjb}. More explicitly, the energy flux $\mathcal F^{\mathcal H}_{\text{PN}}$ is given by the 4PN expression $\mathcal F^{\mathcal  H}_{4 \text{PN}} $ provided in Eqs.~(4.23)--(4.25) of~\cite{Saketh:2022xjb}, 
 \begin{align} 
\mathcal F^{\mathcal H}_{4 \text{PN}} &= -\frac{16m_1^2\nu^2 (1+\kappa) x^{6}}{5}\Omega_\text{tidal}(\Omega_{\mathcal H}-\Omega_\text{tidal})\biggl[1+3{\chi}^2 \nonumber \\
&\quad +\sum_{n=2}^3 x^\frac{n}{2} C_{n}(\nu;{\chi})\biggr],\label{FH4PN}
\end{align}
with typographical corrections from Ref.~\cite{Cunningham:2024dog} for the coefficients $C_n$. 
Here $\kappa=\sqrt{1-\chi^2}$ and $\Omega_{\mathcal H} = \chi/[2 m_1(1+\kappa)]$ are the primary black hole's extremality parameter and horizon angular velocity, respectively. The quantity 
\begin{equation}
\Omega_\text{tidal} = \frac{x^{3/2}}{M}\left[1-\nu x +\nu\chi x^{3/2} + {\cal O}(x^2)\right]
\end{equation}
is the angular velocity of the tidal field exerted on the primary by the secondary. Note that the coefficient  $C_{3}(\nu;{\chi})$ depends upon $B_2({\chi}):=\text{Im}[\text{PolyGamma}(0,3+2i {\chi}/\kappa)]$. 

As above, we consistently rewrite $\mathcal F^{\mathcal H}_{4 \text{PN}}$ in terms of the variables $(x,\MTOT,\nu,\chi)$.

\subsection{$h^{\text{PN}}_{\ell m}$}

The real-valued $\hat h_{22}$ is provided in Eq. (20) of~\cite{Warburton:2024xnr} up to 4PN, extracted from the earlier results of~\cite{Henry:2022ccf,Blanchet:2023bwj,Blanchet:2023sbv}. The expressions of the amplitudes $h^{\text{PN}}_{\ell m}$ for all spherical harmonic modes have been provided up to 3.5PN order in~\cite{Henry:2022ccf}, including spinning effects for non-precessing quasicircular binaries, which summarizes and completes earlier partial results~\cite{Favata:2008yd,Blanchet:2008je,Faye:2012we,Faye:2014fra,Henry:2021cek,Henry:2022dzx}. They vanish at 3.5PN order for $\ell \geq 10$. 

\bibliography{ThisBib}

\end{document}